\documentclass{andp2012}
\usepackage[english]{babel}
\usepackage{epstopdf}
\usepackage{setspace}
\usepackage{graphicx}
\usepackage{amssymb,amsmath}

\newcommand{\ket}[1]{{| #1 \rangle}\xspace}

\title{A 3D Tight-Binding Model for La-Based Cuprate
  Superconductors} 
\author[R. Photopoulos]{Rapha\"el Photopoulos\inst{1}}
\author[R. Fr\'esard]{Raymond Fr\'esard\inst{1,}\footnote{Corresponding
    author\quad 
			E-mail:~\textsf{Raymond.Fresard@ensicaen.fr},
			Phone: +33\,231\,45\,26\,09,
			Fax: +33\,231\,95\,16\,00}}
\address[1]{Normandie Universit\'e, ENSICAEN, UNICAEN, CNRS, CRISMAT, 14000
  Caen, FRANCE} 
\shortauthors{R. Photopoulos et al.}
\begin{abstract}
Motivated by the recent experimental determination of the three-dimensional
Fermi surface of overdoped La-based cuprate superconductors [Horio $\textit{et al.}$, Phys. Rev. Lett. ${\bf 2018}$, 121, 077004], we revisit the
tight-binding parameterization of their conduction band. We construct a
minimal tight-binding model entailing eight orbitals, two of them involving
apical oxygen ions. Parameter optimization allows to almost perfectly
reproduce the three-dimensional conduction band as obtained from density functional theory (DFT). We
discuss how each parameter entering this multiband model influences it, and
show that the peculiar form of its dispersion severely constraints the
parameter values. We then evidence that standard perturbative derivation of an
effective one-band model is poorly converging because of the comparatively
small value of the charge transfer gap. Yet, this allows us to unravel the
microscopical origin of the in-plane and out-of-plane hopping amplitudes. An
alternative approach to the computation of the tight-binding parameters of the
effective model is presented and worked out. It results that the agreement
with DFT is preserved provided longer-ranged hopping amplitudes are
retained. A comparison with existing models is also performed. Finally, the Fermi
surface, showing staggered pieces alternating in size and shape, is compared
to experiment, with the density of states also being calculated. 

\end{abstract}
\shortabstract
\begin{document}
\maketitle

\section{Introduction}\label{sec:intro}
Transition metal oxides entail a large diversity of functional oriented
properties: high-T$_c$ superconductivity
\cite{bednor86,dagot94,leereview06,fradkin12,keimer15,proust19}, colossal
magnetoresistance observed in the
manganites \cite{Hel93,Tom95,Rav95,Mai95}, and transparent conducting oxides
\cite{Kaw97,Har18}. In addition, a whole series of 
promising materials for thermoelectric applications has been discovered 
\cite{Ter97,Mas00,Mat01,Mic07,Ohta07,Wang13,Gui11}. Furthermore, they also
harbor fascinating phenomena such as superconductivity at the interface of
two insulators \cite{Rey07}, and peculiar metal-to-insulator transitions in
vanadium sesquioxide \cite{McW73,Hel01,Lim03,Gry07}, all of them providing a
strong challenge to study these systems from the theory side. Such a task
implies the derivation of an appropriate microscopical model containing the
relevant degrees of freedom. To some extend, the least degree of complexity
arises when modeling the superconducting cuprates as, in that case, only one
band crosses the Fermi energy. Accordingly, the Hubbard model on the square
lattice is often considered as the proper low energy effective model of the
superconducting cuprates\cite{anders87}. It has been studied by a whole range
of approaches, but it remains fair to say that a widely accepted theory of the
cuprates, and more generally of strongly correlated oxides, is still to come. 

In an effort to set up a generic model applicable to most superconducting
cuprates, P. W. Anderson suggested to focus on the CuO$_2$ layers that are
common to these materials, and, accordingly, to simplify the one-body term of
the Hamiltonian down to nearest-neighbor hopping on the square lattice formed
by the Cu ions \cite{anders87}. He further suggested to assume a fully
screened Coulomb interaction that only retains the local (Hubbard) interaction
term among the $d_{x^2-y^2}$ electrons that form the only band crossing the
Fermi energy. Intense efforts devoted to study the resulting two-dimensional
(2D) Hubbard model has shown that it captures the basic phenomenology of the
cuprates\cite{tremblay06,leprev15,zheng17}. However, these endeavors did not
allow to systematically yield long-ranged pure d-wave pairing order in
the hole doped region of interest (from the underdoped $\delta \simeq$ 5$\%$
region to the overdoped $\delta \simeq$ 27$\%$ one with $\delta$ the hole
doping in a half-filled band). In addition, the so-called 1/8 anomaly is found
to be related to a magnetic stripe. Yet, the period of the calculated one is
twice larger than experimentally observed \cite{tranqua95,Racz06,leblanc15}. Eventually, this
failure may originate from an oversimplification of the model, and it has been
suggested that including hopping to the next-nearest neighbors (with hopping
amplitude $t'$) leads to the correct stripe
periodicity\cite{marcin06,Racz06,imada18,dever18}. Therefore, the physics
entailed in the Hubbard model is sensitive to the very form of its one-body
term. This is even more so since recent calculations indicate that the role of
$t'$ is to suppress d-wave pairing correlations\cite{imada18}. This, however,
is in contrast to the DFT study by Pavarini $\textit{et al}$. \cite{pavar01},
who showed that a larger $t'$ is empirically correlated to a higher maximal
T$_c$ in each family of cuprate compounds. The role of possible longer-ranged
hopping terms, and their relationship with oxygen orbitals, received a lesser
degree of attention. 

Irrespective of the form of the interaction ultimately leading to
superconductivity, the one-body term of the Hamiltonian is of interest on its
own, for several reasons. First, from the experimental point of view it has
been established long ago that the superconducting transition temperature
T$_c$ is quite sensitive to the very structure of a sample, that is primarily
reflected in the hopping term. For instance, the critical temperature is 25 K
in bulk La$_{1.9}$Sr$_{0.1}$CuO$_4$\cite{radaelli94}, while T$_c$ reaches 49 K
in thin film form\cite{perret98,sato00}. Second, knowing the proper parameter
set entering the kinetic energy term of the Hamiltonian might be helpful when
performing numerical simulations, in particular when tackling effective low
energy Hamiltonians such as the one-band Hubbard model. Alternatively, one may
wonder whether the rather weak exotic superconductivity found in the repulsive
$t-U$ Hubbard model\cite{corboz14,imada14,leprev15,imada18} persists when
further hopping terms are taken into account. Third, it is of interest to
understand the microscopic origin of the various hopping amplitudes entering
the kinetic energy of the effective model. How do effective models based on
``in layer only'' CuO$_{2}$ orbitals compare to models taking the third
dimension into account? Fourth, the reduction of multiband models down to an
effective one-band model often results from a perturbative treatment. Yet, the
relative proximity of the Cu:3d$_{x^2-y^2}$ and O:2p energy levels prevents
this treatment from being rapidly converging, and an alternative procedure to
this downfolding procedure might be precious. Fifth, the
Fermi surface of La$_{1.78}$Sr$_{0.22}$CuO$_4$ as measured by Horio
$\textit{et al.}$ \cite{orio18} is not to be interpreted within a Hubbard
model on the square lattice.

In this work, we therefore propose to re-examine the influence of the oxygen
ions, including the apical ones, on the electronic structure with a particular
emphasis on the form of the inter-layer couplings. More specifically, we take
the single-layer La-based cuprate superconductors La$_{2-x}$Sr$_x$CuO$_4$ (LSCO) as
examples \cite{bednor86}. Their layered body-centered tetragonal structure
(BCT) \cite{raveau80} is explicitely taken into account. In this context, we
neglect all Cu:3d orbitals but the 3d$_{x^2-y^2}$ one, so that the underlying
effective low-energy model still consists of a one-band Hubbard model, but
because of the BCT structure, we retain the six relevant O:2p orbitals that
shape the conduction band. The often advocated Cu:4s orbital is incorporated
in our model, too \cite{ander95,pavar01,mishonov03,mishonov10,bansil05}. 

The paper is organized as follows: we shortly review existing microscopical
models in Section~\ref{sec:motiv}. We set up a three-dimensional eight-band
tight-binding model in Section~\ref{sec:model} where we show that all included
orbitals have a sizeable influence on the dispersion of the conduction
band. We also argue that the other oxygen orbitals may be safely neglected. In
Section~\ref{sec:discus}, we start the discussion by giving a set of optimized
parameters which yields an almost perfect agreement with DFT results by
Markiewicz $\textit{et al.}$ \cite{marki05}. We then address downfolding
procedures to derive an effective one-band model. We first show that the perturbative treatment to fifth order does
not show sign of convergence to the exact result for realistic values of the charge transfer gap. However, this allows us to qualitatively
discuss the different higher order superexchange processes contributing to the
effective hopping integrals. The latter are then numerically computed by means
of the Fourier transform of the exact conduction band. Finally, we discuss the
peculiar calculated Fermi surface of the 3D model which is compared to the
recent experimental results by Horio $\textit{et al.}$ \cite{orio18}. The
density of states is addressed as well. The paper is summarized and concluded
in Section~\ref{sec:conclusion}. 

\section{Emery and related microscopical models}\label{sec:motiv}

Below, we propose a starting point to the description of the low-energy
physics of superconducting La-based cuprates (La$_2$CuO$_4$ as parent
compound). This model goes beyond the popular three-band tight-binding Emery
model \cite{emer87,varm87,loktev88} which we summarize below, as it will serve
as a basis for comparison. Since it is based on a single CuO$_2$ layer, it
entails no dispersion in the direction perpendicular to them, in contrast to
DFT calculations that are performed for the 3D compounds
\cite{marki05,lindroo06}. 

In real space, the one-body term describing an ideal CuO$_2$ layer reads:
\begin{equation}
\hat{H}_{\rm E}^{(0)} = \epsilon_d\sum_{{\bf i},\sigma} \hat{n}_{{\bf i},\sigma}^d + 
\epsilon_p \left(\sum_{{\bf j},\sigma}\hat{n}^{(X)}_{{\bf j},\sigma}
+ \sum_{{\bf l},\sigma}\hat{n}^{(Y)}_{{\bf l},\sigma} \right)
+ \hat{T}_{pd} + \hat{T}_{pp}\,, 
\label{eq.emery_simple}
\end{equation}
with
\begin{equation}
\begin{aligned}
\hat{T}_{pd} &= t_{pd}\sum_{{\bf i},\sigma}\hat{d}_{{\bf i}, \sigma}^\dagger\left(\hat{p}_{x,{\bf i}+\frac{a}{2}{\bf e_x},\sigma}^{(X)} - \hat{p}_{x,{\bf i}-\frac{a}{2}{\bf e_x},\sigma}^{(X)}- \hat{p}_{y,{\bf i}+\frac{a}{2}{\bf e_y},\sigma}^{(Y)}\right. \\&
\left.+\hat{p}_{y,{\bf i}-\frac{a}{2}{\bf e_y},\sigma}^{(Y)}\right) + h.c. \,\,
{\rm and}\,,\\
\hat{T}_{pp} &= t_{pp}\sum_{{\bf i},\sigma}\hat{p}_{x,{\bf i}+\frac{a}{2}{\bf e_x},\sigma}^{(X)\dagger}\left(\hat{p}_{y,{\bf i}-\frac{a}{2}{\bf e_y},\sigma}^{(Y)} -\hat{p}_{y,{\bf i}+\frac{a}{2}{\bf e_y},\sigma}^{(Y)}\right) \\&
+ \hat{p}_{x,{\bf i}-\frac{a}{2}{\bf e_x},\sigma}^{(X)\dagger}\left(\hat{p}_{y,{\bf i}+\frac{a}{2}{\bf e_y},\sigma}^{(Y)} -\hat{p}_{y,{\bf i}-\frac{a}{2}{\bf e_y},\sigma}^{(Y)}\right) + h.c.
\,,
\end{aligned}
\label{eq.kinetic_emery}
\end{equation}
where ${\bf i}\equiv{\bf R}_i$ is the position of the Cu site in the CuO$_2$
plaquette of the Bravais lattice. The lattice parameter $a$ is the shortest
Cu-Cu distance, and $\hat{d}^\dagger_{{\bf i},\sigma}$ is the creation
operator of an electron with the spin $\sigma$ =
$\left(\uparrow,\downarrow\right)$ on the Cu:3d$_{x^2-y^2}$ orbital on site
${\bf i}$ with on-site energy $\epsilon_d$ and occupation number operator
$\hat{n}_{{\bf i},\sigma}^d$. Two O:2p oxygen orbitals are embodied in the
model through the $\hat{p}_{x,{\bf j},\sigma}^{(X)\dagger}$ (${\bf j} \equiv
{\bf R}_i +a{\bf e}_x/2$), and
$\hat{p}_{y,{\bf l},\sigma}^{(Y)\dagger}$ (${\bf l} \equiv
{\bf R}_i +a{\bf e}_y/2$) creation operators. The on-site
oxygen energy is $\epsilon_p$ with the occupation number operators
$\hat{n}^{(X)}_{{\bf j},\sigma}$ and $\hat{n}^{(Y)}_{{\bf l},\sigma}$. The
kinetic part of the Hamiltonian 
Eq.~(\ref{eq.kinetic_emery}) refers to the hopping amplitude $t_{pd} = \langle
3d_{x^2-y^2,{\bf i}}|\hat{T}_{pd}|p_{x,{\bf i}+a{\bf e}_x/2}^{(X)}\rangle$ =
$-\langle 3d_{x^2-y^2}|\hat{T}_{pd}|p_{y,{\bf i}+a{\bf e}_y/2}^{(Y)}\rangle$
between the 2p$_{x}^{(X)}$ and 2p$_{y}^{(Y)}$ oxygen orbitals and the
3d$_{x^2-y^2}$ orbital, and the hopping amplitude between both oxygen orbitals
given by $t_{pp} = -\langle p_{x,{\bf i}+a{\bf
    e}_x/2}^{(X)}|\hat{T}_{pp}|p_{y,{\bf i}+a{\bf 
    e}_y/2}^{(Y)}\rangle$. The form of the kinetic part of the Emery model is
better known in momentum space. Applying the Fourier transform on
Eqs.~(\ref{eq.emery_simple}) and (\ref{eq.kinetic_emery}), the non-interacting
Emery three-band Hamiltonian in momentum space is given by:  
\begin{equation}
\hat{H}_{\rm E}^{(0)}  = \sum_{{\bf k},\sigma}\hat{\Phi}_{{\bf k},\sigma}^\dagger\begin{pmatrix}
\epsilon_d & 2it_{pd}p_x & -2it_{pd}p_y \\
-2it_{pd}p_x & \epsilon_p & -4t_{pp}p_xp_y \\
2it_{pd}p_y & -4t_{pp}p_xp_y & \epsilon_p
\end{pmatrix}\hat{\Phi}_{{\bf k},\sigma}
\label{eq.emery_matrix}
\end{equation}
with
\begin{equation}
p_x \equiv \sin{\left(\frac{k_x a}{2}\right)}\:\:,\:\:p_y \equiv \sin{\left(\frac{k_y a}{2}\right)}\,.
\label{eq.pxpy}
\end{equation}
Here, all creation operators of electrons on Cu:3d and O:2p orbitals with the
momentum ${\bf k}$ and spin $\sigma$ are gathered in the three-component
operator $\hat{\Phi}_{{\bf k},\sigma}^\dagger = \left(\hat{d}_{{\bf
      k},\sigma}^\dagger,\hat{p}_{x,{\bf k},\sigma}^{(X)\dagger},\right.$
$\left.\hat{p}_{y,{\bf k},\sigma}^{(Y)\dagger}\right)$. Consensus about the
value of the parameters does not seem to have been reached but typical values
in t$_{pd}$ unit are given by: $\epsilon_d - \epsilon_p \simeq 2.5-3.5 t_{pd}$
and $t_{pp} \simeq 0.5-0.6t_{pd}$, with $t_{pd}\simeq 1.2-1.5$
eV\cite{macma87,emer87,mila88,hyberts89,kampf94,dagot94,dopf92,pavar01,kent08}.  

Since a realistic non-interacting Hamiltonian which can be implemented in
quantum correlated treatments is crucial in order to correctly describe the
unconventional properties of the superconducting cuprates
\cite{fradkin12,jang15}, other models have been put forward. For instance, a
generic 2D four-band model for CuO$_2$ planes was suggested by Labb\'e and
Bok\cite{labbe87}. Later, Andersen \textit{et al.} \cite{ander95} derived a
2D 8-band model extending the Emery model which interpolates the LDA band
structure of stoichiometric YBa$_2$Cu$_3$O$_7$. They thereby showed that the
shape of the Fermi surface is entirely characterized by the in-plane hopping
parameters $t$, $t' \simeq -0.30t$ and $t'' \simeq 0.20t$. In order to shed
light on their DFT results for LSCO, Markiewicz \textit{et al.}
\cite{marki05} introduced a phenomenological effective model. Using,
$E_{3D}(k_x,k_y,k_z) = \epsilon_M + E_{2D}(k_x,k_y) + E_z(k_x,k_y,k_z)$ the
purely 2D contribution reads: 
\begin{equation}
\begin{split}
E_{2D}(k_x,k_y) = -2t\left[\cos{(k_xa)} + \cos{(k_ya)}\right]\\
-4t'\cos{(k_xa)}\cos{(k_ya)} \\
-2t''\left[\cos{(2k_xa)}+\cos{(2k_ya)}\right]\\
-4t'''\left[\cos{(k_xa)}\cos{(2k_ya)} + \cos{(k_ya)}\cos{(2k_xa)}\right]\,,
\end{split}  
\label{eq.marki_eparall}
\end{equation}
where $t$, $t'$, $t''$ and $t'''$ represent first, second, third and fourth
nearest-neighbor in-plane hopping integrals. Regarding the out-of-plane
dispersion Markiewicz \textit{et al.} entangle all three space dimensions using: 
\begin{equation}
 E_z(k_x,k_y,k_z) = -2t_z\pi_x\pi_y\pi_z\left[\cos{(k_xa)} - \cos{(k_ya)}\right]^2
 \label{eq.marki_ez}
\end{equation}
where 
\begin{equation}
\pi_x \equiv \cos{(k_xa/2)}\:\:,\:\: \pi_y \equiv \cos{(k_ya/2)} \:\:,\:\:
\pi_z \equiv \cos{(k_zc/2)}\,, 
\end{equation}
and $t_z$ denotes an inter-layer hopping parameter. The offset
by (a/2, a/2) of the successive CuO$_2$ layers is taken into account through
the factor $\pi_x\pi_y$ \cite{marki05}. The constant $\epsilon_M$ makes
$E_{3D}(0,0,0)$ vanishing. This model has also been used in order to fit the
experimental out-of-plane Fermi surface of overdoped LSCO ($\delta$ = 0.22)
obtained with ARPES \cite{orio18}. In addition, a three-dimensional (3D)
four-band tight-binding model was derived by Mishonov $\textit{et
  al.}$\cite{mishonov10} in order to explain the observation of the 3D
Fermi surface in Tl$_2$Ba$_2$CuO$_{6+\delta}$\cite{hussey03}. 

\section{Extended three-dimensional model}\label{sec:model}
The goal of this section is to set up a tight-binding model for La-based
cuprates. More specifically we attempt to reproduce the dispersion of the band
based on the Cu:3d$_{x^2-y^2}$ orbital along the main symmetry lines of the
Brillouin zone, including the ouf-of-plane ones, as obtained by DFT
calculations by Markiewicz $\textit{et al.}$\cite{marki05}. There is an
extensive literature on copper orbitals that may play an important role to
high-T$_c$ superconductivity (HTSC) \cite{dagot94,leereview06,damasc03}. It
has predominantly been focused on the 3d$_{x^2-y^2}$, 3d$_{3z^2-r^2}$, and 4s
orbitals that are closest to the Fermi energy
\cite{ander95,scalap95,pavar01,jang15,matt18}. The assumption that some form of
the one-band Hubbard model harbors the key ingredient to HTSC leads to
neglect the 3d$_{3z^2-r^2}$ (filled) orbital altogether, as it may not be
simply integrated out because of the strong interaction of these electrons
with the ones populating the 3d$_{x^2-y^2}$ orbitals. In other words, keeping
both e$_g$ orbitals unavoidably results in a two-band Hubbard model
\cite{sakakib10,matt18}. Since the interaction of the latter electrons with
the 4s electrons is much weaker, integrating out the latter is better
justified and we explicitly take them into account in our tight-binding model,
as proposed in Refs.~\cite{ander95,pavar01}.

We start the construction of our model by considering the four inequivalent
oxygen ions building octahedra surrounding a given copper atom: there are two
in-plane oxygen ions O$^{(X)}$ and O$^{(Y)}$ along the x and y directions
respectively and two apical oxygens O$^{(a)}$ and O$^{(b)}$, located above and
below each copper ion, respectively. This leads to twelve nearly degenerate
O:2p orbitals to be considered. Yet, as will be addressed below, only six of
them significantly contribute to the dispersion of the Cu:3d$_{x^2-y^2}$
band. Structure wise, the numerical value of the lattice parameters is given
by: $a = b = 3.78~$\AA~and $c = 13.18~$\AA. The position of
the copper site and the oxygen sites in a unit-cell $i$ are given by ${\bf
  R}_{Cu} = {\bf R}_i$, ${\bf R}_{O^{(X)}} = {\bf R}_i + a{\bf e}_x/2$, ${\bf
  R}_{O^{(Y)}} = {\bf R}_i + a{\bf e}_y/2$, ${\bf R}_{O^{(a)}} = {\bf R}_i +
d_{\rm Cu-O_{ap}}{\bf e}_z$ and ${\bf R}_{O^{(b)}} = {\bf R}_i - d_{\rm
  Cu-O_{ap}}{\bf e}_z$. The distance $r \equiv d_{\rm Cu-O_{ap}} = 2.42~$\AA$ = 0.64a$
characterizes the elongation of the octahedra. Let us
observe here that symmetry forces a whole series of hopping amplitudes to
vanish. In particular one has: 
\begin{equation}
\langle 3d_{x^2-y^2}|\hat{T}| p_z^{(X,Y)}\rangle = \langle 3d_{x^2-y^2}|\hat{T}| p_{x,y}^{(a,b)}\rangle = 0 
\label{eq.orb_vanish_1}
\end{equation}
and
\begin{equation}
\langle 4s|\hat{T}| p_z^{(X,Y)}\rangle = \langle 4s |\hat{T}| p_{x,y}^{(a,b)}\rangle = 0.
\label{eq.orb_vanish_2}
\end{equation}
Hence those six oxygen orbitals do at best play a minor role and are therefore
neglected. Accordingly, all along this work we consider the eight orbitals:
Cu:3d$_{x^2-y^2}$, Cu:4s, $O^{(X)}:2p_x^{(X)}$, $O^{(Y)}:2p_y^{(Y)}$,
$O^{(X)}:2p_y^{(X)}$, $O^{(Y)}:2p_x^{(Y)}$, $O^{(a)}:2p_z^{(a)}$,
$O^{(b)}:2p_z^{(b)}$, in this order.  Further arguments for this choice can be
found in
Refs.~\cite{pavar01,ander95,raimondi96,feiner92,lau11,hansman14,weber10,xiang96,peng17}. These
eight orbitals are labeled by an index $\mu$ that runs from 1 to 8. Subsets
of orbitals involving all of them but the Cu:3d$_{x^2-y^2}$ one are labeled
by an index $\nu$ that runs from 2 to 8, subsets of orbitals involving the
in-plane oxygen orbitals are labeled by an index $\kappa$ running from 3 to
6, while the subset of apical oxygen orbitals is labeled by $\rho$ that runs
from 7 to 8. Our eight-band tight-binding Hamiltonian may be expressed as:  
\begin{equation} 
\mathcal{\hat{H}}_{\rm (8)} = \hat{H}_0 + \hat{T} + \hat{H}_d,
\label{eq.ham8b_1}
\end{equation}
where $\hat{H}_0$ stands for the on-site orbital energies relative to
$\epsilon_d$ which denotes the on-site energy of the 3d$_{x^2-y^2}$ orbital
and $\hat{T}$ is the kinetic energy term. Introducing the Fourier transform
$\hat{d}_{\bf k,\sigma}$ of the annihilation operator of an electron on site
$i$ with spin $\sigma$ on the 3d$_{x^2-y^2}$ orbital: 
\begin{equation}
\hat{d}_{{\bf k},\sigma} = \frac{1}{\sqrt{L}}\sum_{{\bf k},\sigma}e^{-i{\bf k}\cdot{\bf R}_i}\hat{d}_{i,\sigma}\,, 
\end{equation}
and analogous expressions for the other orbitals, $\hat{H}_0$ reads:
\begin{equation}
\hat{H}_0 = \sum_{{\bf k},\sigma}\left[-\Delta_{pd}\sum_{\kappa}\hat{n}_{{\bf
      k},\sigma,\kappa}^p -\Delta_{z}\sum_{\rho}\hat{n}_{{\bf
      k},\sigma,\rho}^{p_z} 
 + \Delta_s\hat{n}_{{\bf k},\sigma}^s\right] \,,
\end{equation}
where $\Delta_{pd} = \epsilon_d - \epsilon_p$, $\Delta_{z} =
\epsilon_d-\epsilon_z$, $\Delta_{s} = \epsilon_s-\epsilon_d$. Additionally,
$\epsilon_p$, $\epsilon_z$ and $\epsilon_s$, denote the on-site energies of the
2p$_{x,y}^{(X,Y)}$, 2p$_z$ and Cu:4s orbitals, respectively. The various
$\hat{n}_{{\bf k},\sigma, \mu}$ operators represent the occupation number operators
of a given orbital with momentum ${\bf k}$ and spin $\sigma$. L is the size of
the lattice. Gathering all creation operators in the eight-component operator
$\hat{\Psi}_{{\bf k},\sigma, \mu}^\dagger$ = $\left(\hat{d}_{{\bf
      k},\sigma}^\dagger\right.$, $\hat{s}_{{\bf k},\sigma}^\dagger$,
$\hat{p}_{x,{\bf k},\sigma}^{(X)\dagger}$, $\hat{p}_{y,{\bf
    k},\sigma}^{(Y)\dagger}$, $\hat{p}_{y,{\bf k},\sigma}^{(X)\dagger}$,
$\hat{p}_{x,{\bf k},\sigma}^{(Y)\dagger}$, $\hat{p}_{z,{\bf
    k},\sigma}^{(a)\dagger}$, $\left.\hat{p}_{z,{\bf
      k},\sigma}^{(b)\dagger}\right)$, the kinetic energy may be written as: 
\begin{equation}
\hat{T} = \sum_{{\bf k},\sigma}\sum_{\mu,\mu'}t_{\bf k}^{\mu,\mu'}
\hat{\Psi}_{{\bf k},\sigma,\mu}^\dagger\hat{\Psi}_{{\bf k},\sigma,\mu'}^{\phantom{\dagger}} \,. 
\end{equation}
Here $t_{\bf k}^{\mu,\mu'}$ is the hopping integral in momentum space between
orbital $\mu$ and orbital $\mu'$. Finally, we have 
\begin{equation}
\hat{H}_d = \epsilon_d\sum_{{\bf k},\sigma}\sum_{\mu}
\hat{\Psi}_{{\bf k},\sigma,\mu}^\dagger\hat{\Psi}_{{\bf k},\sigma, \mu}^{\phantom{\dagger}}
\,.
\end{equation}
Altogether, we focus on the one-body Hamiltonian 
\begin{equation}
\hat{\mathcal{H}}=\hat{H}_0+\hat{T}
\label{eq.modelref}
\end{equation}
expressed as:
\begin{equation}
\mathcal{\hat{H}} = \sum_{{\bf k},\sigma}\sum_{\mu,\mu'}\hat{\Psi}_{{\bf k},\sigma,\mu}^\dagger H_{\bf k}^{\mu,\mu'} \hat{\Psi}_{{\bf k},\sigma, \mu'}^{\phantom{\dagger}}\,,
\label{eq:modeltot}
\end{equation}
with
\begin{equation}
H_{{\bf k}} = \begin{pmatrix}
A_{{\bf k}_\parallel} & B_{{\bf k}_\parallel}& C_{k_z}   \\
B_{{\bf k}_\parallel}^\dagger  & D_{{\bf k}_\parallel} & E_{{\bf k}} \\
C_{k_z}^\dagger & E_{{\bf k}}^\dagger & F_{{\bf k}} 
\end{pmatrix}\,.
\:\:\:\:\:\:\:\:\
\label{eq.matrixmodel}
\end{equation}
Here, the Hamiltonian matrix is expressed in terms of the sub-matrices
A$_{{\bf k}_\parallel}$, D$_{{\bf k}_\parallel}$ and F$_{\bf k}$ entailing the
tight-binding Hamiltonians in the copper, in-plane oxygen, and out-of-plane
oxygen orbital subspaces, respectively. The coupling between these subspaces
is accounted for by the submatrices B$_{{\bf k}_\parallel}$, C$_{k_z}$ and
E$_{\bf k}$. While the Hamiltonian matrix Eq.~(\ref{eq.matrixmodel}) depends
on the momentum ${\bf k} = ({\bf k}_\parallel,k_z)$, we clarified the momentum
dependence of the sub-matrices, that are derived below. 

\subsection{In-plane hopping integrals}
In this sub-section, we set up the contribution to the Hamiltonian arising
from the copper and in-plane oxygen orbitals. This leads to a two-dimensional
model. The positions in the unit-cell $i$ of the O$^{(X)}$ and O$^{(Y)}$ ions
are respectively labeled by ${\bf j} \equiv {\bf R}_i +a{\bf e}_x/2$ and
${\bf l}\equiv {\bf R}_i +a{\bf e}_y/2$. 

\subsubsection{Oxygen - copper and direct copper - copper hopping integrals}
The strong d-p $\sigma$-hybridization has already been presented in
Section~\ref{sec:motiv}. Then, in agreement with arguments given in
Refs.~\cite{pavar01,ander95} we include in the model the effect of the Cu:4s
orbital. It strongly hybridizes with the nearest neighboring
O$^{(X)}$:2p$_{x}^{(X)}$ and O$^{(Y)}$:2p$_{y}^{(Y)}$ orbitals. They are
located at relative positions $\pm a{\bf e}_x/2$ and $\pm a{\bf e}_y/2$ and
the corresponding hopping matrix elements are $\pm t_{sp} = \langle 4s_{\bf i}
|\hat{T}|p_{x, {\bf i}\pm a{\bf e}_x/2}^{(X)}\rangle$. Symmetry implies that
$\pm t_{sp} = \langle 4s_{\bf i} |\hat{T}|p_{y, {\bf i}\pm a{\bf
    e}_y/2}^{(Y)}\rangle$. Note that the matrix elements of the kinetic energy
between the Cu:4s orbital and the remaining O$^{(X)}$ and O$^{(Y)}$ orbitals
vanish. Since the Cu:4s orbital is more extended than the 3d one we also
include direct nearest neighbor 4s-4s hopping amplitude $-t_{ss}$ = $\langle
4s_{\bf i}|\hat{T}|4s_{{\bf i}\pm a{\bf e}_x}\rangle$ and the next nearest
neighbor one $-t_{ss}'$ = $\langle 4s_{\bf i}|\hat{T}|4s_{{\bf i}\pm a({\bf
    e}_x+{\bf e}_y)}\rangle$, together with their symmetry related
counterparts. The matrices $A_{{\bf k}_\parallel}$ and $B_{{\bf k}_\parallel}$
then follow as: 
\begin{equation}
A_{{\bf k}_\parallel} =
 \bordermatrix{~& \ket{3d_{x^2-y^2}} & \ket{4s} \cr
& 0 & 0   \cr
 & 0 & \tilde{\Delta}_{{\bf k}_\parallel}   \cr}\,,
\end{equation}
where we introduced the short-hand notation:
\begin{equation}
\begin{aligned}
\tilde{\Delta}_{{\bf k}_\parallel} = \Delta_s &- 2t_{ss}(\cos{(k_x a)} +
\cos{(k_y a}))\\ 
 &- 4t_{ss}'\cos{(k_x a)}\cos{(k_y a)}\,,
\end{aligned}
\end{equation}
and
\begin{equation}
B_{{\bf k}_\parallel} =
\bordermatrix{~& \ket{p_x^{(X)}} & \ket{p_y^{(Y)}} & \ket{p_y^{(X)}} & \ket{p_x^{(Y)}} \cr
 & 2it_{pd}p_x & -2it_{pd}p_y & 0 & 0   \cr
  & 2it_{sp}p_x & 2it_{sp}p_y & 0 & 0 \cr }\,.
 \end{equation}
The matrix B$_{{\bf k}_\parallel}$ accounts for the coupling between the Cu
(3d and 4s) orbitals and the involved in-plane oxygen one. With the above sign
convention $t_{pd}$, $t_{sp}$, $t_{ss}$ and $t'_{ss}$ are all positive. 

\subsubsection{Oxygen-Oxygen hopping integrals}
\paragraph{a) Neighboring oxygen ions along the axes of the square lattice}
$ $

\begin{figure}[h!]
     \centering
     \includegraphics[width=0.7\columnwidth]{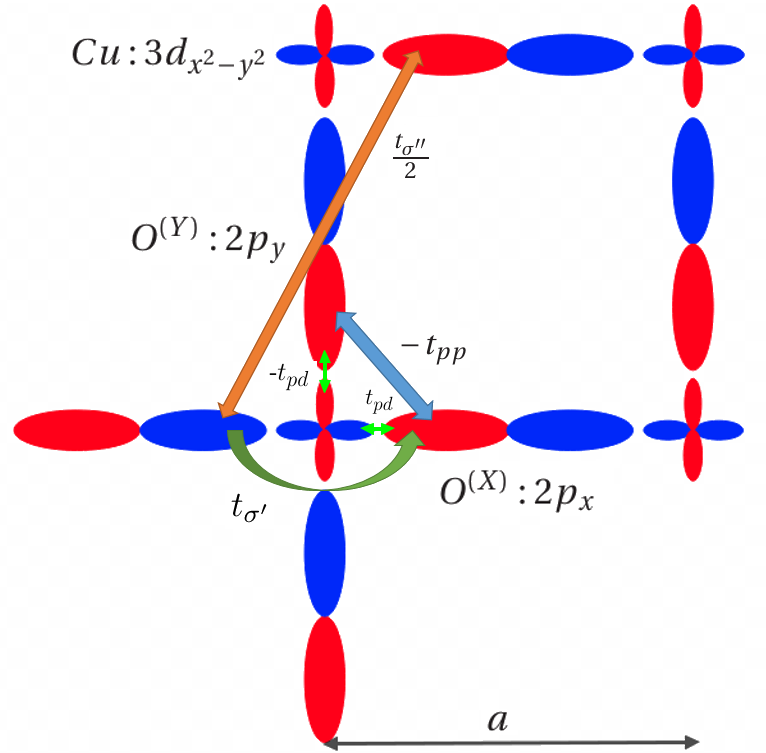}
     \caption{(Color online) Illustration of $t_{pd}$, $t_{\sigma'}$, $t_{pp}$
       and $t_{\sigma''}$ in-plane hopping amplitudes. Note that t$_{pp}$ =
       $(t_\sigma+t_\pi)/2$ and $t_{\sigma''}$ are introduced using the
       rotated orbital basis (2p$_\xi^{(X,Y)}$, 2p$_\eta^{(X,Y)}$).} 
     \label{fig:dessin_1}
\end{figure}

We first consider the hopping integrals between the in-plane
2p$_{x,y}^{(X,Y)}$ oxygen orbitals. While the ionic radius of
Cu$^{2+}$  is commonly accepted to be close to $0.75~$\AA, the one of O$^{2-}$
is $1.35 ~$\AA, which results into a strong hybridization of the oxygen orbitals
among themselves. Indeed, the diameter of the O$^{2-}$ ions is comparable to
the distance between nearest-neighbors O$^{(X)}$ and O$^{(Y)}$ oxygen ions on
a CuO$_2$ plaquette given by $a/\sqrt{2}$ = 2.68~\AA. We thus introduce the
$\sigma$-type 
hopping integral between nearest-neighbor 2p$_x^{(X)}$ orbitals, $t_{\sigma'}$
= $\langle p_{x,{\bf j}}^{(X)}|\hat{T}|p_{x,{\bf j}\pm a{\bf
    e_x}}^{(X)}\rangle$ as depicted in Fig.~\ref{fig:dessin_1}. Similarly,
$t_{\sigma'}$ = $\langle p_{y,{\bf j}}^{(X)}|\hat{T}|p_{y,{\bf j}\pm a{\bf
    e_y}}^{(X)}\rangle$ is the hopping amplitude between nearest-neighbor
2p$_{y}^{(X)}$ orbitals. Likewise, for symmetry reasons, $\langle p_{y,{\bf
    l}}^{(Y)}|\hat{T}|p_{y,{\bf l}\pm a{\bf e_y}}^{(Y)}\rangle$ and $\langle
p_{x,{\bf l}}^{(Y)}|\hat{T}|p_{x,{\bf l}\pm a{\bf e_x}}^{(Y)}\rangle$ will be
given by $t_{\sigma'}$ as well. In addition, $\pi$-type coupling between two
O$^{(X)}$ neighbors linked along $\pm a {\bf e_y}$ and $\pm a {\bf e_x}$ are
also considered. They are both accounted for by the hopping integral
$-t_{\pi'}$ = $\langle p_{x,{\bf j}}^{(X)}|\hat{T}|p_{x,{\bf j}\pm a{\bf
    e_y}}^{(X)}\rangle$ = $\langle p_{y,{\bf j}}^{(X)}|\hat{T}|p_{y,{\bf
    j}\pm a{\bf e_x}}^{(X)}\rangle$. Symmetry implies that the $\pi$-type
hopping integrals involving the oxygen O$^{(Y)}$ ions are again given by
$-t_{\pi'}$ = $\langle p_{x,{\bf l}}^{(Y)}|\hat{T}|p_{x,{\bf l}\pm a{\bf
    e_y}}^{(Y)}\rangle$ = $\langle p_{y,{\bf l}}^{(Y)}|\hat{T}|p_{y,{\bf
    l}\pm a{\bf e_x}}^{(Y)}\rangle$. With the above used sign conventions,
$t_{\sigma'}$ and $t_{\pi'}$ are positive. Furthermore symmetry forces all
matrix elements $\langle p^{(X,Y)}_{x,y,{\bf j}}|\hat{T}|p^{(X,Y)}_{z,{\bf
    j}}\rangle$ to vanish. 

$ $
 
\paragraph{b) Neighboring oxygen ions along the diagonal axis of the square
  lattice}

$ $

In order to introduce the $\sigma$- and $\pi$-type hopping matrix elements
between nearest-neighbor O$^{(X)}$ and O$^{(Y)}$ ions we take advantage of a
transformation consisting in a rotation by $\pi/4$ around ${\bf e}_z$ of the
2$p_{x,y}^{(X,Y)}$ orbitals for each given oxygen site. It leads to define
$\hat{p}_{\xi}^{(\beta)} = (\hat{p}_x^{(\beta)} +
\hat{p}_y^{(\beta)})/\sqrt{2}$ and $\hat{p}_{\eta}^{(\beta)} =
(-\hat{p}_x^{(\beta)} + \hat{p}_y^{(\beta)})/\sqrt{2}$ where
$\beta\in(X,Y)$. With this transformation we obtain an alternative (or
rotated) 2p oxygen orbital basis in which $\sigma$- and $\pi$-hybridization
along the diagonal directions between all oxygen ions of the lattice is easily
taken into account. We thus consider hybridization between 2p$_{\xi}^{(X)}$
and 2p$_{\xi}^{(Y)}$ along $({\bf e}_x+{\bf e}_y)/\sqrt{2}$ with the
associated hopping integral $t_\sigma$ = $\langle p_{\xi, {\bf
    j}}^{(X)}|\hat{T}|p_{\xi, {\bf j}\pm a ({\bf e}_x+{\bf
    e}_y)/2}^{(Y)}\rangle$. The same hopping amplitude $t_{\sigma}$ is
obtained when the 2p$_{\eta}^{(X,Y)}$ orbitals are considered along the
orthogonal direction: $t_\sigma$ = $\langle p_{\eta, {\bf
    j}}^{(X)}|\hat{T}|p_{\eta, {\bf j} \pm a ({\bf e}_x-{\bf
    e}_y)/2}^{(Y)}\rangle$. Furthermore, concerning the $\pi$-type
hybridization, we have $-t_\pi$ = $\langle p_{\eta, {\bf
    j}}^{(X)}|\hat{T}|p_{\eta, {\bf j}\pm a ({\bf e}_x+{\bf
    e}_y)/2}^{(Y)}\rangle$ along $({\bf e}_x + {\bf e}_y)/\sqrt{2}$ and
$-t_\pi$ = $\langle p_{\xi, {\bf j}}^{(X)}|\hat{T}|p_{\xi, {\bf j}\pm a ({\bf
    e}_x-{\bf e}_y)/2}^{(Y)}\rangle$ along $({\bf e}_x - {\bf e}_y)/\sqrt{2}$.

We also take into account, in this basis, the $\sigma$-coupling between
next-nearest neighbors O$^{(X)}$ and O$^{(Y)}$ ions. This leads to the hopping
integral $t_{\sigma''}$ = $\langle p_{\xi, {\bf w}}^{(\beta)}|\hat{T}|p_{\xi,
  {\bf w} \pm a({\bf e}_x + {\bf e}_y)}^{(\beta)}\rangle$. Symmetry implies
$t_{\sigma''}$ = $\langle p_{\eta, {\bf w}}^{(\beta)}|\hat{T}|p_{\eta, {\bf
    w}\pm a({\bf e}_x - {\bf e}_y)}^{(\beta)}\rangle$, with ${\bf w}\in({\bf j,l})$. The corresponding $\pi$-type hopping amplitudes are clearly smaller than the $\sigma$-type ones, and may be neglected (see Fig.$~\ref{fig:dessin_1}$). Then the hopping amplitude between $\sigma$-type 2p:O$^{(X,Y)}$ next-nearest neighbor orbitals is $t_{\sigma''}/2$ in the natural basis. The matrix D$_{{\bf k}_\parallel}$ embodying the tight-binding model for the four in-plane 2p oxygen orbitals therefore results as:
\begin{equation}
\begin{split}
D_{{\bf k}_\parallel} =\:\:\:\:\:\:\:\:\:\:\:\:\:\:\:\:\:\:\:\:\:\:\:\:\:\:\:\:\:\:\:\:\:\:\:\:\:\:\:\:\:\:\:\:\:\:\:\:\:\:\:\:\:\:\:\:\:\:\:\:\:\:\:\:\:\:\:\:\:\:\:\:\:\:\:\:\:\:\:\:\:\:\:\:\:\:\:\:\:\:\:\:\:\:\:\:\:\:\:\:\:\:\:\:\:\:\:\:\:\:\:\:\:\:\:\:\:\:\:\:\:\:\:\:\:\:\:\:\:\:\:\:\:\:\:\:\:\:\:\:\:\:\:\:\\
\resizebox{.89\columnwidth}{!} { \[
 \bordermatrix{~& \ket{p_x^{(X)}} & \ket{p_y^{(Y)}} & \ket{p_y^{(X)}} & \ket{p_x^{(Y)}} \cr
 & \bar{\Delta}_{{\bf k}_\parallel} & -4t_{pp}p_xp_y & -2t_{\sigma''}p_{2x}p_{2y} & 4t_{pp}^{(2)}\pi_x\pi_y   \cr
  & -4t_{pp}p_xp_y & \bar{\Delta}_{{\bf k}_\parallel}' & 4t_{pp}^{(2)}\pi_x\pi_y & -2t_{\sigma''}p_{2x}p_{2y}  \cr
 & -2t_{\sigma''}p_{2x}p_{2y}  & 4t_{pp}^{(2)}\pi_x\pi_y & \bar{\Delta}_{{\bf k}_\parallel}'& -4t_{pp}p_xp_y    \cr
 & 4t_{pp}^{(2)}\pi_x\pi_y &  -2t_{\sigma''}p_{2x}p_{2y}  & -4t_{pp}p_xp_y& \bar{\Delta}_{{\bf k}_\parallel}  \cr}\] }
~
~
~
~
~
\end{split}
\label{eq.matrixd}
 \end{equation}
where we have defined:
\begin{equation}
\begin{split}
\bar{\Delta }_{{\bf k}_\parallel}= -\Delta_{pd} +2(t_{\sigma'}\cos{(k_xa)}-t_{\pi'}\cos{(k_ya)}) \\
+ 2t_{\sigma''}\cos{(k_xa)}\cos{(k_ya)} \\
\bar{\Delta}_{{\bf k}_\parallel}' = -\Delta_{pd} + 2(t_{\sigma'}\cos{(k_ya)}-t_{\pi'}\cos{(k_xa)})\\
+ 2t_{\sigma''}\cos{(k_xa)}\cos{(k_ya)}\:,
\end{split}
\end{equation}
together with
\begin{equation}
t_{pp} = \frac{t_{\sigma}+t_{\pi}}{2}\:\:,\:\:t_{pp}^{(2)} = \frac{t_{\sigma}-t_{\pi}}{2}\,.
\end{equation}
In addition, we  introduced
\begin{equation}
p_{2x} \equiv \sin{(k_xa)}\:\:\:\:,\:\:\:\:p_{2y} \equiv \sin{(k_ya)}\:.
\end{equation}
Furthermore, we note that $t_{pp}$ is identical to the O$^{(X)}$-O$^{(Y)}$
hopping integral involved in the Emery model Eq.~(\ref{eq.emery_matrix})
\cite{emer87}. However, the smaller hopping amplitude $t_{pp}^{(2)}$ is
neglected in the Emery model. While the alternative orbital basis
(2p$_{\xi}^{(X,Y)}$, 2p$_{\eta}^{(X,Y)}$) eases the derivation of D$_{{\bf
    k}_\parallel}$, the latter is written in the natural basis. This completes
the derivation of a two-dimensional model.

\subsection{Out-of-plane hopping integrals}
In this subsection we derive the main contributions leading to dispersion
perpendicular to the CuO$_2$ layers. Because of the body-centered tetragonal
structure there is no obvious leading term describing the hopping of an
electron on a Cu:3d$_{x^2-y^2}$ orbital in one layer to the same orbital on a
neighboring layer. Such processes involve at least the apical oxygens through
their 2p$_z$ orbitals, which themselves couple to the Cu:4s
orbital\cite{feiner92,pavar01,weber10,peng17} and to in-plane O:2p
orbitals. With this, one can attempt to model the rather broad dispersion along
$k_z$ found in DFT calculations\cite{marki05}.  
Below, the position in the unit-cell of apical oxygens O$^{(a)}$ and O$^{(b)}$
are labeled by ${\bf m}\equiv {\bf R}_i + {\rm d_{Cu-O_{ap}}}{\bf e}_z$ and
${\bf n}\equiv {\bf R}_i - {\rm d_{Cu-O_{ap}}}{\bf e}_z$, respectively. 

\subsubsection{Coupling of the apical oxygen ions to the in-plane orbitals}
Here, we describe the coupling of the apical oxygen ions to the copper and
in-plane oxygen ions. 
First, the hopping amplitude associated to the Cu:4s and its nearest neighbors
in ${\bf e}_z$ direction O$^{(a,b)}$:2p$_z$ orbitals is given by $t_{sp_z} =
\langle p_{z,{\bf m}}^{(a)}|\hat{T}|4s_{\bf i}\rangle$ and $-t_{sp_z} =
\langle p_{z,{\bf n}}^{(b)}|\hat{T}|4s_{\bf i}\rangle$ (as shown by the blue
arrows in Fig.~\ref{fig:dessin_2}). Second, symmetry yields $\langle p_{z,{\bf
    i} \pm r{\bf e}_z}^{(a,b)}|\hat{T}|3d_{x^2-y^2,{\bf i}}\rangle$ =
0. Hence, the couplings between the Cu orbitals and the 2p$_z^{(a,b)}$ apical
oxygen orbitals may be gathered in: 
\begin{equation}
 C_{k_z} = 
 \bordermatrix{~& \ket{p_z^{(a)}} & \ket{p_z^{(b)}} \cr
& 0 &  0  \cr
 & t_{sp_z}e^{irk_z} & -t_{sp_z}e^{-irk_z}   \cr}\,.
\end{equation}

In addition, the 2p$_z^{(a,b)}$ apical oxygen orbitals significantly hybridize
with the in-plane 2p$_{x,y}^{(X,Y)}$ oxygen ones. Indeed, with the distance
between a copper ion and an apical oxygen ion being d$_{\rm Cu-O_{ap}}$
$\simeq 0.64 a$ the distance between an apical oxygen ion and an in-plane
oxygen ion in the unit-cell is d$_{\rm O-O_{ap}}$ $\simeq 0.80a$. This is
comparable to the in-plane distance d$_{\rm O^{(X)}-O^{(Y)}}$ $\simeq 0.707 a$
in the unit-cell. Then, taking the point of view of apical oxygens, the
O$^{(a)}$:2p$_z^{(a)}$ orbital hybridizes with 2p$_x^{(X)}$ orbital along
$\boldsymbol{\delta}_{-}^{(\pm)} = \pm a {\bf e}_x/2 - d_{\rm Cu-O_{ap}}{\bf
  e}_z$ with the associated hopping integral $\mp t_{p_z} = \langle p_{z, {\bf
    m}}^{(a)}|\hat{T}|p_{x, {\bf m}
  +\boldsymbol{\delta}_{-}^{(\pm)}}^{(X)}\rangle$ and 2p$_y^{(Y)}$ along
$\boldsymbol{\delta}_{-}'^{(\pm)} = \pm a {\bf e}_y/2 - d_{\rm Cu-O_{ap}}{\bf
  e}_z$ with $\mp t_{p_z} = \langle p_{z, {\bf m}}^{(a)}|\hat{T}|p_{y, {\bf m}
  + \boldsymbol{\delta}_{-}'^{(\pm)} }^{(Y)}\rangle$. Besides,
O$^{(b)}$:2p$_z^{(b)}$ orbital also hybridizes with 2p$_x^{(X)}$ orbital along
$\boldsymbol{\delta}_{+}^{(\pm)} = \pm a{\bf e}_x + d_{\rm Cu-O_{ap}}{\bf
  e}_z$ with the associated hopping integral $\pm t_{p_z}$ = $\langle p_{z,
  {\bf n}}^{(b)}|\hat{T}|p_{x, {\bf n} +
  \boldsymbol{\delta}_{+}^{(\pm)}}^{(X)}\rangle$ and 2p$_y^{(Y)}$ along
$\boldsymbol{\delta}_{+}'^{(\pm)} = \pm a{\bf e}_y + d_{Cu-O_{ap}}{\bf e}_z$
with $\pm t_{p_z} = \langle p_{z, {\bf n}}^{(b)}|\hat{T}|p_{y, {\bf n} +
  \boldsymbol{\delta}_{+}'^{(\pm)}}\rangle$. These couplings are illustrated
in Fig.~\ref{fig:dessin_2} by green arrows. 

Furthermore, the coupling between the  2p$_z^{(a,b)}$ apical orbitals
(belonging to the nearest upper and lower CuO$_2$ layers) with in-plane
2p$_x^{(Y)}$ and 2p$_y^{(X)}$ orbitals should be considered, too. Below, the
distance between the current CuO$_2$ layer and the next-layer apical oxygen
ions is denoted by $v \equiv d_{\rm O_{ap}}^{\rm (next-layer)} = c/2 - d_{\rm
  Cu-O_{ap}} \simeq 1.1a$. 
Let us begin with O$^{(b)}:2p_z^{(b)}$ located in the upper layer at (a/2,
a/2, d$_{\rm O_{ap}}^{\rm (next-layer)}$). It hybridizes with the 2p$_x^{(Y)}$
orbital. The associated hopping amplitude is $\mp t_{p_z}''$ = $\langle
p_{z,\boldsymbol{\nu}_{+}}^{(b)}|\hat{T}|p_{x,\boldsymbol{\nu}_{+}-v{\bf e}_z
  \pm {\bf e}_x/2}^{(Y)}\rangle$. Here we introduced $\boldsymbol{\nu}_{\pm}$
= $a({\bf e}_x+{\bf e}_y)/2 \pm d_{\rm O_{ap}}^{\rm (next-layer)}{\bf e}_z$
the position of an O$^{(b)}$ (O$^{(a)}$) ion in the upper (lower) layer relative to the Cu
ion in the current layer. The next-layer $O^{(b)}$ also hybridizes with the
$2p_y^{(X)}$ orbital with the associated hopping amplitude $\mp t_{p_z}''$ =
$\langle
p_{z,\boldsymbol{\nu}_{+}}^{(b)}|\hat{T}|p_{y,\boldsymbol{\nu}_{+}-v{\bf e}_z
  \pm {\bf e}_y/2}^{(X)}\rangle$. Symmetry implies that the apical oxygen
O$^{(a)}$:2p$_z^{(a)}$ orbital belonging to the lower layer at (a/2, a/2,
-d$_{\rm O_{ap}}^{\rm (next-layer)}$) hybridizes (with opposite sign) with the
in-plane 2p$_x^{(Y)}$ and 2p$_y^{(X)}$ orbitals. Therefore, the associated
hopping integrals are $\pm t_{p_z}''$ = $\langle p_{z,
  \boldsymbol{\nu}_{-}}^{(a)}|\hat{T}|p_{x,\boldsymbol{\nu}_{-}+v{\bf e}_z \pm
  {\bf e}_x/2}^{(Y)}\rangle$ = $\langle p_{z,
  \boldsymbol{\nu}_{-}}^{(a)}|\hat{T}|p_{y,\boldsymbol{\nu}_{-}+v{\bf e}_z \pm
  {\bf e}_y/2}^{(X)}\rangle$. 
These couplings between in-plane 2p orbitals and out-of-plane 2p$_z^{(a,b)}$
orbitals may be gathered in: 
\begin{equation}
E_{\bf k} =
 \bordermatrix{~& \ket{p_z^{(a)}} & \ket{p_z^{(b)}} \cr
 & 2it_{p_z}p_xe^{irk_z} & -2it_{p_z}p_xe^{-irk_z}   \cr
 & 2it_{p_z}p_ye^{irk_z} & -2it_{p_z}p_ye^{-irk_z}   \cr
& -2it_{p_z}''p_ye^{-ivk_z} & 2it_{p_z}''p_ye^{ivk_z}   \cr
 & -2it_{p_z}''p_xe^{-ivk_z} & 2it_{p_z}''p_xe^{ivk_z}   \cr}\,.
\end{equation}
 
\subsubsection{Apical oxygens - Apical oxygens hopping amplitudes}
\begin{figure}[t!]
     \centering
     \includegraphics[width=1.0\columnwidth]{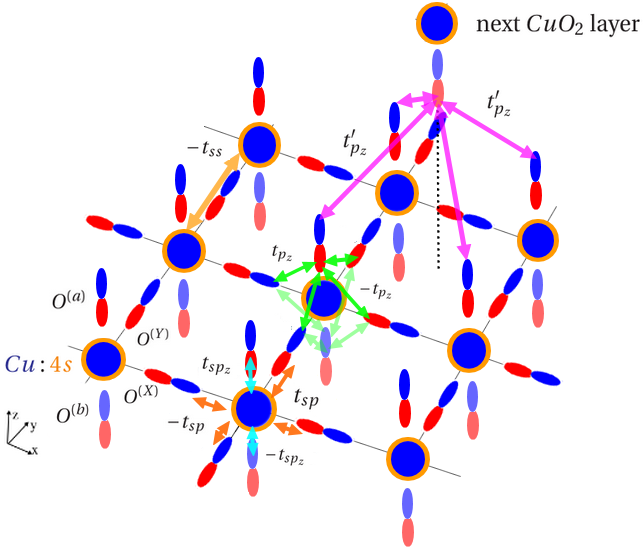}
     \caption{(Color online) Illustration of the body-centered tetragonal
       lattice of single-layer La-based cuprate involving the 2p$_{x,y}^{(X,Y)}$
       in-plane oxygen,  2p$_{z}^{(a,b)}$ out-of-plane apical oxygen, and Cu:4s
       orbitals. The inter-layer coupling is also illustrated.} 
     \label{fig:dessin_2}
\end{figure}

To complete the model we now proceed with the derivation of the tight-binding
Hamiltonian in the out-of-plane oxygen orbital subspace. Since neither the
2p$_x^{(a,b)}$ nor the 2p$_y^{(a,b)}$ orbitals couple to the 3d$_{x^2-y^2}$
and 4s copper orbitals we are left with the 2p$_z^{(a,b)}$ orbitals only. 
Further contributions to dispersion in the direction perpendicular to CuO$_2$
layers come from the 2p$_z$ orbitals of the apical oxygens. With the
body-centered tetragonal symmetry of the lattice, each apical oxygen O$^{(a)}$
(O$^{(b)}$) is surrounded by four nearest-neighbor apical oxygen O$^{(b)}$
(O$^{(a)}$) ions belonging to the upper (lower) next CuO$_2$ layer. For
example, we illustrate in Fig.~\ref{fig:dessin_2} (magenta arrows) the
coupling between the O$^{(a)}$ apical oxygen located at (0, 0, d$_{{\rm
    Cu-O_{ap}}}$) of the reference unit-cell with one of its apical nearest
neighbor O$^{(b)}$ located at (a/2, a/2, c/2-d$_{{\rm Cu-O_{ap}}}$) in the
unit-cell centered at (a/2, a/2, c/2). Since their distance is as small as
d$_{{\rm Cu-O_{ap}}}$, the associated hopping amplitude $t_{p_z}'$ has to be
taken into account. Concerning the apical oxygen O$^{(a)}$ located at ${\bf
  m}$, the four nearest-neighbor inter-plane apical oxygens O$^{(b)}$ are
located at ($\pm$a/2, $\pm$a/2, d$_{{\rm Cu-O_{ap}}}$ + d$_{\rm
  O^{(a)}-O^{(b)}}$) where $u\equiv d_{\rm O^{(a)}-O^{(b)}} = c/2 - 2d_{\rm
  Cu-O_{ap}} = 0.465a$ is the distance along the c-axis between two
inter-plane apical oxygen ions. Thus, we introduce the corresponding hopping
integrals $t_{p_z}'$ = $\langle p_{z,{\bf m}}^{(a)}|\hat{T}|p_{z,{\bf m} +
  a(\pm{\bf e}_x\pm{\bf e}_y)/2 + u{\bf e}_z }^{(b)}\rangle$. Symmetry
implies:  $t_{p_z}'$ = $\langle p_{z,{\bf n}}^{(b)}|\hat{T}|p_{z, {\bf n} +
  a(\pm{\bf e}_x\pm{\bf e}_y)/2 - u{\bf e}_z }^{(a)}\rangle$. The coupling
between O$^{(a)}$ and O$^{(b)}$ in the unit-cell is also taken into account
through $t_{p_z}'''$ = $\langle p_{z,{\bf m}}^{(a)}|\hat{T}|p_{z,{\bf
    n}}^{(b)}\rangle$. Note that the $\pi$-overlap between the
O$^{(a)}$:2p$_z$ orbitals - and between the O$^{(b)}$:2p$_z$ orbitals - is of
order $t_{\pi''}$ and may be neglected. Then, the matrix F$_{\bf k}$ which
accounts for the coupling between inter-layer apical oxygen orbitals is given
by: 
\begin{equation}
\resizebox{.89\columnwidth}{!} { \[
F_{\bf k} = 
 \bordermatrix{~& \ket{p_z^{(a)}} & \ket{p_z^{(b)}} \cr
& -\Delta_z & 4t_{p_z}'\pi_x\pi_ye^{iuk_z} + t_{p_z}'''e^{2irk_z}   \cr
& 4t_{p_z}'\pi_x\pi_ye^{-iuk_z}+t_{p_z}'''e^{-2irk_z}  & -\Delta_z   \cr}\]}
~
~
~
~
~
\label{eq.matrix_f}
\end{equation}
Let us note that all previously defined tight-binding parameters are taken
positive (the sign of the orbital lobes are taken care of through the adopted
sign convention). Symmetry implies that many seemingly large hopping
amplitudes actually vanish (see, e.~g.,
Eqs.~(\ref{eq.orb_vanish_1},\ref{eq.orb_vanish_2})). We neglected hybridizations
implying the 2p$_z^{(X,Y)}$ orbitals because they are smaller than the largest
retained ones, so that our model is an eight-band model that may hardly be
improved on in the tight-binding framework. 
Below, our model will be tested against the LDA results obtained by Markiewicz
$\textit{et al}$.\cite{marki05}, as well as against the one-band tight-binding
model they use to capture their numerical findings.

\section{Results and discussion}\label{sec:discus}
The purpose of this section is to establish an effective one-band model for
Cu:3d$_{x^2-y^2}$ band through the downfolding of the other seven
bands. Because of the relatively small value of the gap between the oxygen
bands and the copper one, one may infer that numerous relevant effective
hopping integrals will come into play. Our procedure will hence allow for
highlighting their origin in the multiband model, and for addressing what can
and cannot follow from the Emery model Eq.~(\ref{eq.emery_matrix}).

\begin{table*}[t]
\begin{andptabular}{X[5l]X[5l]X[5l]X[5l]X[5l]X[5l]X[5l]X[6l]}%
{Summary of the set of optimal tight-binding parameters in units of $t_{pd}$
  fitting the LDA dispersion of the conduction band (Fig.~\ref{fig:fig.lda}).} 
$\Delta_{pd,opt} = 3.5$ & $\Delta_{z,opt} = 2.6$ & $\Delta_{s,opt} = 6.5$ & $t_{\sigma,opt} = 0.95 $& $t_{\sigma',opt} = 0.13$ & $t_{\sigma'',opt} = 0.4$&$t_{sp,opt} = 1.3$  & $t_{\pi,opt} = 0.2375$  \\
$t_{ss,opt} = 0.40$ & $t_{ss,opt}' = 0.10$ & $t_{sp_z,opt} = 1.4$ & $t_{p_z,opt} = 0.95$ & $t_{p_z,opt}' = 0.45$ & $t_{p_z,opt}'' = 0.10$ & $t_{p_z,opt}''' = 0$ & $t_{\pi',opt} = 0.0325$
\label{tab.paramopt}
\end{andptabular}
\end{table*}

\subsection{Electronic structure and comparison to DFT}
\begin{figure}[t!]
     \centering
     \includegraphics[width=0.7\columnwidth]{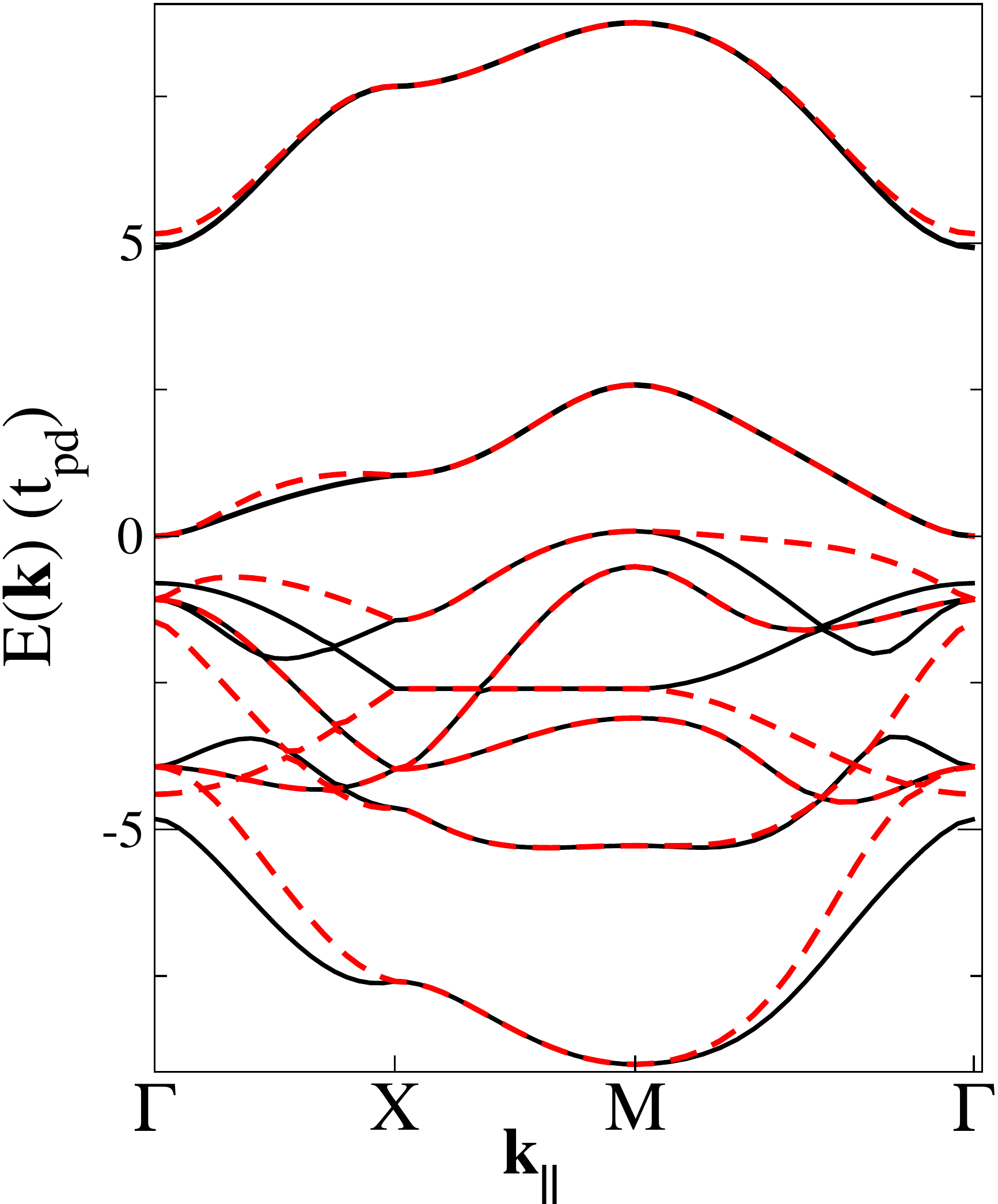}
     \caption{(Color online) Electronic structure along path 1 (full lines)
       and  path 2 (dashed lines) arising from the eight-band tight-binding
       model. The used tight-binding parameters are given in
       Table~\ref{tab.paramopt}.} 
     \label{fig:allbands}
\end{figure}

We now proceed with the dispersion of the bands resulting from the
tight-binding Hamiltonian Eq.~(\ref{eq.modelref}). In the following we focus
on the path 1 connecting the symmetry points $\Gamma$=(0, 0, 0), X=($\pi/a$,
0, 0) and M=($\pi/a$, $\pi/a$, 0) as well as on the path 2 connecting the
symmetry points Z=(0, 0, $2\pi/c$), R=($\pi/a$, 0, $2\pi/c$), and A=($\pi/a$,
$\pi/a$, $2\pi/c$). The electronic structure of the model is shown in
Fig.~\ref{fig:allbands}. The six lowest in energy bands have O:2p character:
the four in-plane ones are lower in energy than the two out-of-plane ones. Not
only do they show large dispersion along paths 1 and 2, but all six bands
display strong dependence on $k_z$ as well. Regarding the bands originating
from the Cu ions, the highest in energy one has a predominant Cu:4s character
while the second highest in energy one has predominant Cu:3d$_{x^2-y^2}$
character. On their own the former shows a weak ${\bf k}_\parallel$
dependence, while the latter is fully local. Hence they inheritate their
dispersion from their coupling to the oxygens. The so-obtained band widths
are smaller than for the oxygen based ones, but dispersion along the three
directions is obtained for both Cu-based bands. 

\begin{figure}[h!]
\begin{center}
\unitlength=0.21cm
\begin{picture}(40,50)
\put(3,25){\includegraphics*[width=0.83\columnwidth]{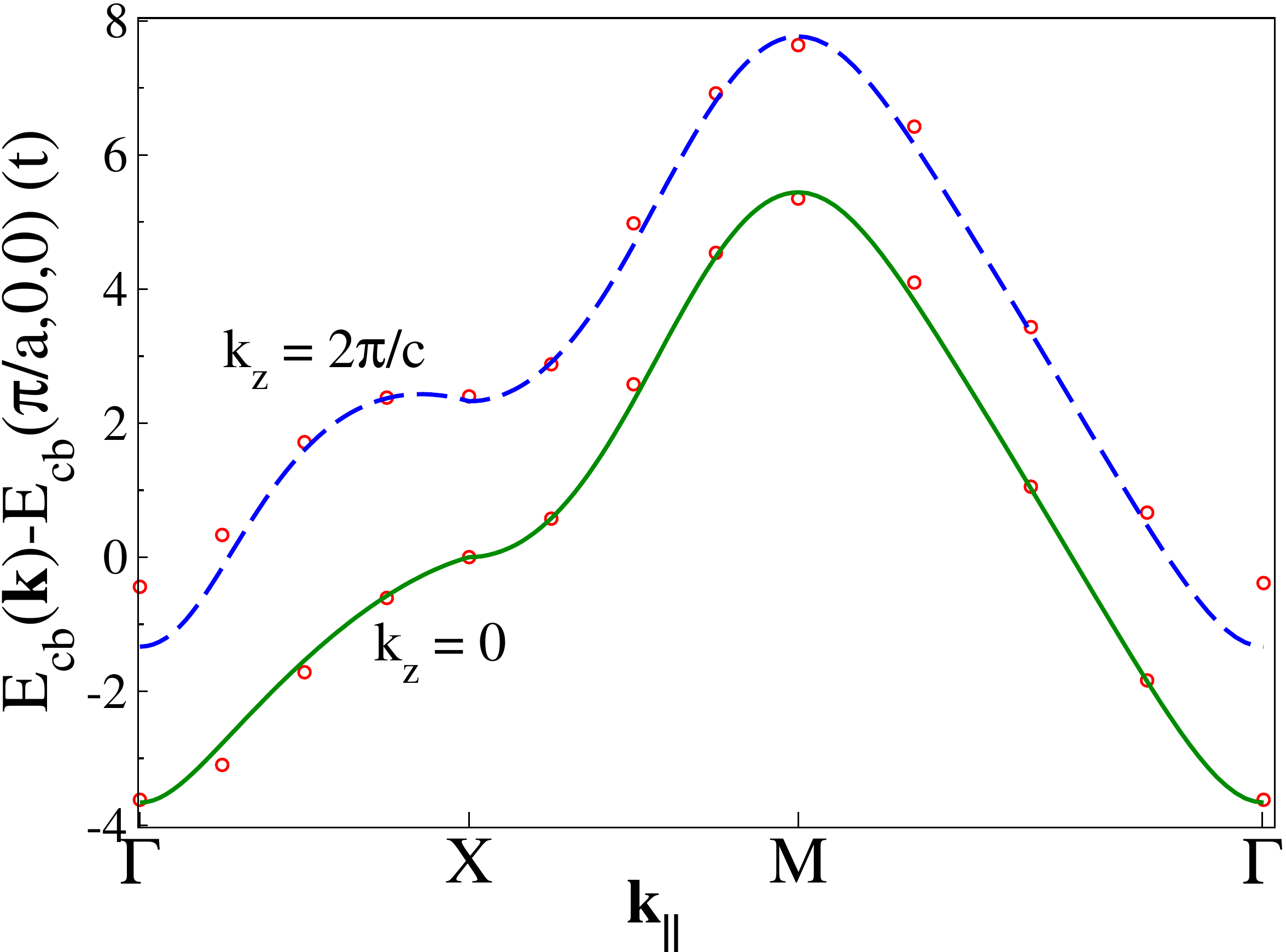}}
\put(2,0){\includegraphics*[width=0.85\columnwidth]{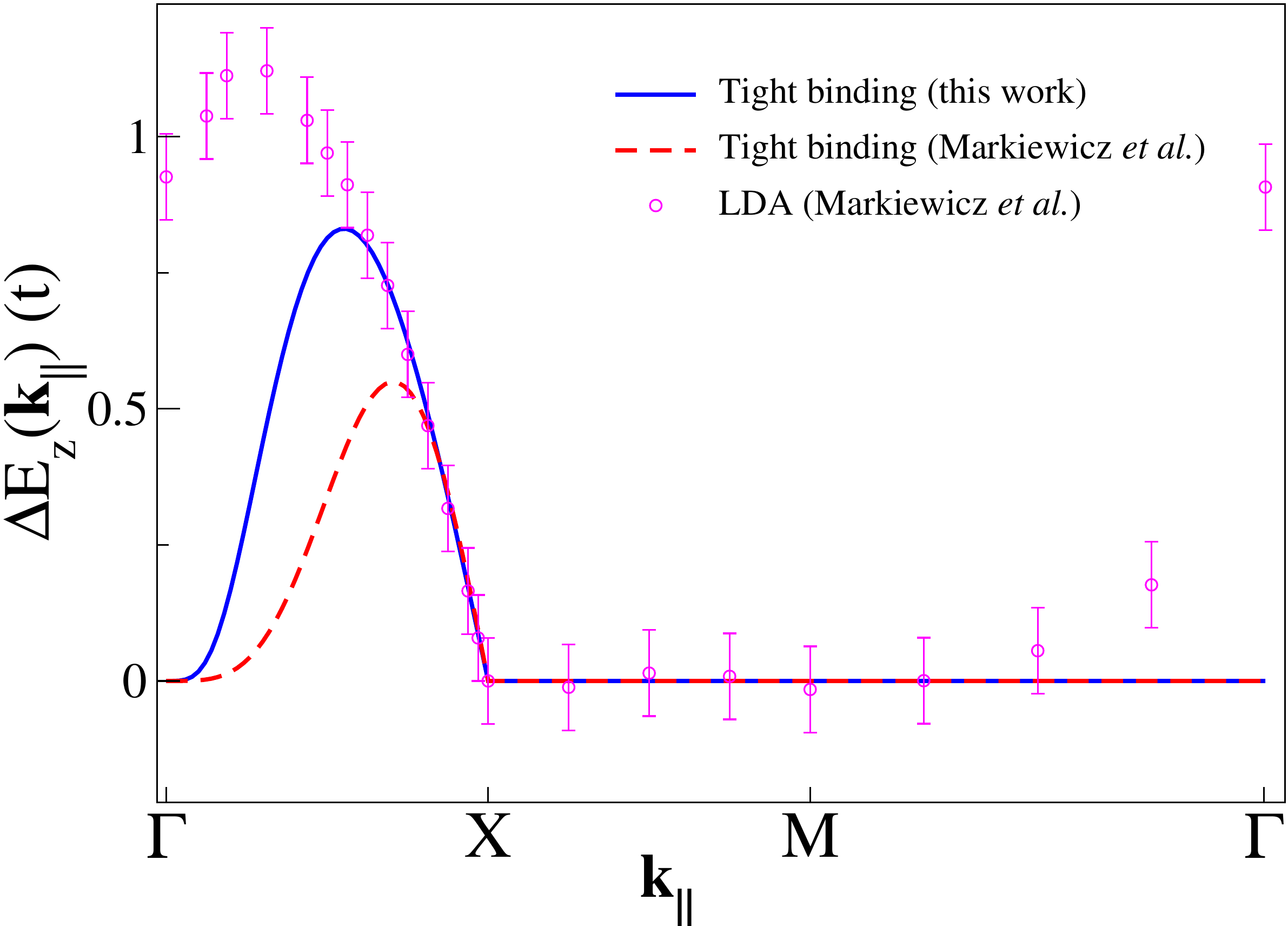}}
\put(33,47){ {\large (a)} } \put(33,22){ {\large (b)} }
\end{picture}
\end{center}
\caption {(Color online) (a) Comparison of the dispersion of the conduction
  band as obtained within DFT by Markiewicz $\textit{et al}$.\cite{marki05}
  (circles) to our tight-binding model. Here, $E_{cb}(\pi/a,0,0)$ defines
  the zero of energy and the $k_z = 2\pi/c$ line is shifted by 1~eV, for
  clarity. (b) Energy difference $\Delta E_z({\bf k}_\parallel)$
  Eq.~(\ref{eq.delta_ez}) as a function of ${\bf k}_\parallel$. The used
  tight-binding parameters are given in Table~\ref{tab.paramopt},
  together with $t/t_{pd} = 0.283$.}  
\label{fig:fig.lda}
\end{figure}

\begin{figure}[t!]
     \centering
     \includegraphics[width=0.83\columnwidth]{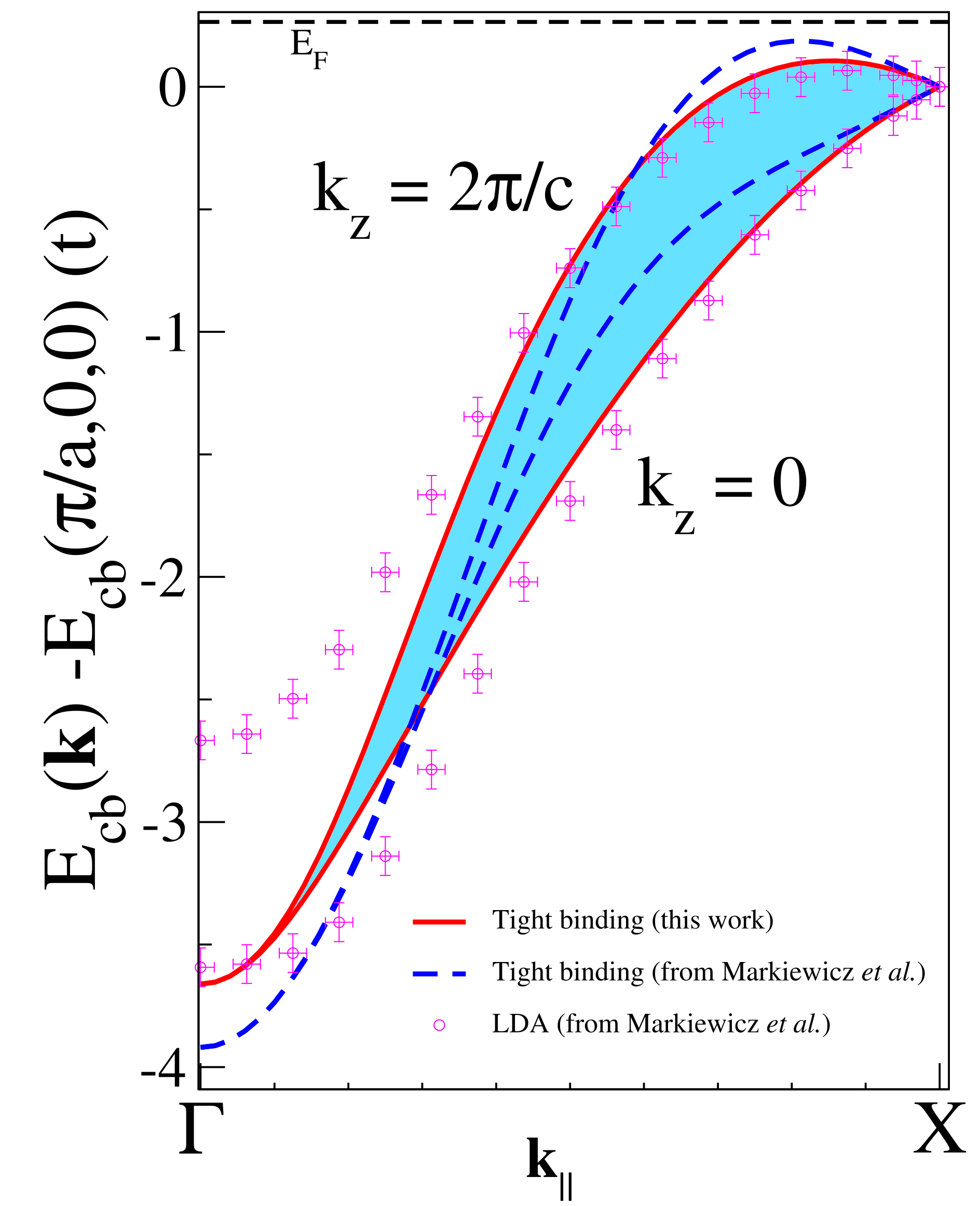}
     \caption{(Color online) Dependence on $k_z$ of the conduction band as a
       function of ${\bf k}_\parallel$ along $\Gamma$-X. The shaded blue
       region is bounded by the band for $k_z = 0$ and $k_z = 2\pi/c$. The
       used tight-binding parameters are given in Table~\ref{tab.paramopt},
       together with $t/t_{pd} = 0.283$.}  
     \label{fig.lda_2}
\end{figure}

Let us now focus on the Cu:3d$_{x^2-y^2}$ conduction band (the only one
crossing the Fermi level), of dispersion $E_{cb}({\bf k})$. As will be
discussed in 
sub-section~\ref{sec:numerical_approach}, the conduction band shows strong
sensitivity to the choice of the parameter set entering the Hamiltonian, not
only from the point of view of the band width, but also from the point of view
of the shape of the band. Hence we first stick to an optimal set of
tight-binding parameters (see Table~\ref{tab.paramopt}) which provides a good
fit of the conduction band as obtained from DFT by Markiewicz $\textit{et
  al.}$ for the LSCO compound\cite{marki05}. The dispersion of the conduction
band for $k_z$ = 0 and $k_z$ = $2\pi/c$ obtained from our model is compared to
a set of data points from LDA calculations \cite{marki05} in
Fig.~\ref{fig:fig.lda} and Fig.~\ref{fig.lda_2}. Regarding path 1 that
encompasses the symmetry lines $\Gamma$-X-M-$\Gamma$,
Figs.~\ref{fig:fig.lda}(a) and 5 demonstrate an almost perfect agreement
between our model and DFT calculations. A good agreement is also found along
path 2, at the exception of the vicinity of the Z point.
The energy difference: 
\begin{equation}
\Delta E_z({\bf k}_\parallel) = E_{cb}({\bf k}_\parallel,k_z = 2\pi/c)-E_{cb}({\bf k}_\parallel,k_z = 0)
\label{eq.delta_ez}
\end{equation}
is also plotted in Fig.~\ref{fig:fig.lda}(b). Amazingly no $k_z$-dispersion
arise above the X-M-$\Gamma$ symmetry line. It only appears above the
$\Gamma$-X symmetry line (anti-nodal direction). This feature is shared by
the Markiewicz 3D tight-binding model\cite{marki05}
(Eqs.~(\ref{eq.marki_eparall}, \ref{eq.marki_ez})) and is exhibited by
LDA calculations\cite{marki05,lindroo06}. It has also been observed in recent
ARPES experiments\cite{orio18}.  

A more accurate comparison of the calculated dispersion along $\Gamma$-X to
LDA is presented in Fig.~\ref{fig.lda_2}. Clearly, the dispersion yielded by
our model is in good agreement with the DFT results in the vicinity of
($\pi/a$, 0, $k_z$) that is close to the Fermi energy for a half-filled
band. The agreement remains good for ${\bf k}$ = ($\pi/2a$, 0, $k_z$) but
degrades when going towards the bottom of the band. In Fig.~\ref{fig.lda_2} we
also compare the tight-binding model set up by Markiewicz $\textit{et al.}$
\cite{marki05} (see Eqs.~(\ref{eq.marki_eparall}, \ref{eq.marki_ez})) to
their DFT results, in which case good agreement is obtained only in the very
vicinity of the ($\pi/a$, 0, $k_z$) symmetry line. In fact, for both models,
the dispersion of the LDA conduction band along $\Gamma$-Z is poorly accounted
for. At first glance this could be attributed to the neglected terms. Yet, the
largest (though small) of them do not bring dispersion of the conduction band
along $\Gamma$-Z, so that this rather points to a limitation of a
tight-binding description that neglects other Cu orbitals. Yet, when sticking
to the vicinity of half-filling, this primarily affects rather high energy
excitations, which turn out to be larger than the ones involving the top of the
(here neglected) 3d$_{3z^2-r^2}$ band \cite{kramer19}. 

\subsection{Analytical approach: microscopical origin of the effective
  hopping amplitudes from downfolding}

\begin{figure}[h!]
\begin{center}
\unitlength=0.23cm
\begin{picture}(36,44)
\put(7,20.5){\includegraphics[width=0.6\columnwidth]{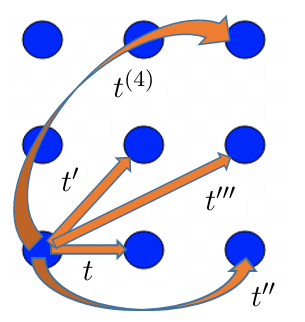}}
\put(6,-3.5){\includegraphics[width=0.6\columnwidth]{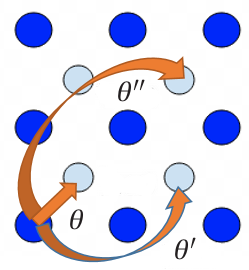}}
\put(30,42){ {\large (a)} } \put(30,17){ {\large (b)} }
\end{picture}
\end{center}
\caption {(Color online) Illustration of the copper ion positions on a BCT
  lattice and related effective hopping amplitudes. (a) The largest in-plane
  hopping integrals (between Cu ions) are denoted by $t$,$t'$,$t''$,$t'''$,
  and $t^{(4)}$. (b) The largest inter-plane hopping integrals are denoted by
  $\theta$, $\theta'$ and $\theta''$. The CuO$_2$ layers are shifted by (a/2,
  a/2).} 
\label{fig:tij}
\end{figure}

Below, we discuss the connection between hopping processes involved in our
model and within a one-band effective description of the conduction band
through a downfolding of the multiband Hamiltonian
Eq.~(\ref{eq.modelref}). Indeed, as previously seen in
Section~\ref{sec:motiv}, cuprates are often described within the Emery
model\cite{emer87,varm87} and it was suggested that this model can be reduced
to an equivalent effective single-band Mott-Hubbard system with the Zhang-Rice
singlet band playing the role of the lower Hubbard band \cite{rice88}. This
point of view supports the early proposal by P. W. Anderson \cite{anders87}
that essential physics of cuprates would be captured by a one-band
Hubbard-like model in which the kinetic part is made of the nearest neighbor
and the next-nearest neighbor integrals $t$ and $t'$ in addition to the local
repulsive interaction that favors electron localization. However, such a
description is only based on a square lattice and neglects the
body-centered tetragonal structure of single-layer La-based cuprates
responsible for inter-plane coupling. Here, focusing on the kinetic part of
the model, we follow this route by integrating out the oxygen and the Cu:4s
degrees of freedom in order to obtain a one-band effective model describing
the conduction band. The one-band Hamiltonian has the ${\bf k}$ dependence of
a Fourier series. Then, the copper band expression for the dispersion has an
in-plane part E$_\parallel({\bf k}_\parallel)$ and an inter-plane part
E$_z({\bf k})$, which, up to a constant ensuring $E(0)=0$, reads: 
\begin{equation}
E({\bf k}_\parallel,k_z) = E_\parallel(k_x,k_y) + E_z(k_x,k_y,k_z) =
\frac{1}{L}\sum_{i,j} t_{i,j}e^{i{\bf k}\cdot ({\bf R}_i-{\bf R}_j)}\,,
\label{eq.disptot}
\end{equation}
where the in-plane dispersion used in this paper is:
\begin{equation}
\begin{aligned}
E_\parallel&(k_x,k_y) = - 2t\left[\cos{(k_xa)} + \cos{(k_ya)}\right]\\
&- 4t'\cos{(k_xa)}\cos{(k_ya)} - 2t''\left[\cos{(2k_xa)}+\cos{(2k_ya)}\right]\\
&-4t'''\left[\cos{(2k_xa)}\cos{(k_ya)} + \cos{(2k_ya)}\cos{(k_xa)}\right]\\
&- 4t^{(4)}\cos{(2k_xa)}\cos{(2k_ya)}\\
&- 2t^{(5)}\left[\cos{(3k_xa)}+\cos{(3k_ya)}\right]\\
&- 4t^{(6)}\left[\cos{(3k_xa)}\cos{(k_ya)} + \cos{(3k_ya)}\cos{(k_xa)}\right]\\
&- 4t^{(7)}\left[\cos{(3k_xa)}\cos{(2k_ya)} + \cos{(3k_ya)}\cos{(2k_xa)}\right]\,.
\end{aligned}
\label{eq.dispinplane}
\end{equation}
Above, $t^{(n)}$=-$t_{i,j}$ denote the hopping integrals to the (n+1) nearest
neighbors on the copper lattice as illustrated in
Fig.~\ref{fig:tij}(a). Further smaller hopping amplitudes are
neglected. Furthermore, according to the BCT structure considered here (see
Fig.~\ref{fig:dessin_2}), inter-plane hopping amplitudes between two CuO$_2$
layers lead to the dispersion relation: 
\begin{equation}
\begin{aligned}
E_z&(k_x,k_y,k_z) = -8\pi_z(k_z)
\left\{\theta\cos{\left(\frac{k_xa}{2}\right)}\cos{\left(\frac{k_ya}{2}
    \right)}\right. \\  
&\!\!\!\!\!+\theta'\left[\cos{\left(\frac{3k_xa}{2}\right)}
\cos{\left(\frac{k_ya}{2}\right)} + 
\cos{\left(\frac{3k_ya}{2}\right)}\cos{\left(\frac{k_xa}{2}\right)}\right] \\
&\!\!\!\!\!+\theta''\cos{\left(\frac{3k_xa}{2}\right)}\cos{\left(\frac{3k_ya}{2}\right)} \\
&\!\!\!\!\!+\theta'''\left[\cos{\left(\frac{5k_xa}{2}\right)}
\cos{\left(\frac{k_ya}{2}\right)} + \cos{\left(\frac{5k_ya}{2}\right)}
\cos{\left(\frac{k_xa}{2}\right)}\right]\\
&\!\!\!\!\!+\theta^{(4)}\left[\!\cos{\left(\frac{5k_xa}{2}\right)}\!
\cos{\left(\frac{3k_ya}{2}\right)}+ \!
\cos{\left(\frac{5k_ya}{2}\right)}\!\cos{\left(\frac{3k_xa}{2}\right)}\!\right]\\
&\left. \!\!\!\!\!+\theta^{(5)}\cos{\left(\frac{5k_xa}{2} \right)}
\cos{\left(\frac{5k_ya}{2}\right)} \right\} \\
&\!\!\!\!\!-2t_{(0,0,c)}\cos{(k_zc)} \,.
\end{aligned}
\label{eq.disp_interplan}
\end{equation}
Here $\theta$, $\theta'$, $\theta''$, $\theta'''$, $\theta^{(4)}$, and
$\theta^{(5)}$ corresponds to the inter-plane hopping amplitudes onto copper
neighbors located in $(\pm\frac{a}{2}{\bf e}_x \pm\frac{a}{2}{\bf
  e}_y\pm\frac{c}{2}{\bf e}_z)$, $(\pm\frac{3a}{2}{\bf e}_x \pm\frac{a}{2}{\bf
  e}_y\pm\frac{c}{2}{\bf e}_z)$, $(\pm\frac{3a}{2}{\bf e}_x
\pm\frac{3a}{2}{\bf e}_y\pm\frac{c}{2}{\bf e}_z)$, $(\pm\frac{5a}{2}{\bf e}_x
\pm\frac{a}{2}{\bf e}_y\pm\frac{c}{2}{\bf e}_z)$, $(\pm\frac{5a}{2}{\bf e}_x
\pm\frac{3a}{2}{\bf e}_y\pm\frac{c}{2}{\bf e}_z)$, $(\pm\frac{5a}{2}{\bf e}_x
\pm\frac{5a}{2}{\bf e}_y\pm\frac{c}{2}{\bf e}_z)$, respectively. They are
depicted in Fig.~\ref{fig:tij}(b). Here $t_{(0,0,c)}$ corresponds to the
hopping amplitude to neighbors located in (0, 0, c). 

By using the Rayleigh-Schr\"odinger perturbation theory (PT) as encompassed in
Lindgren's notation to fourth order (see appendix), one obtains
the effective conduction band as: 
\begin{equation}
E_\parallel({\bf k}_\parallel)_{\rm eff} = E^{(2)}({\bf k}_\parallel)+E^{(3)}({\bf k}_\parallel)+E^{(4)}({\bf k}_\parallel)\,,
\end{equation}
since the first order contribution vanishes. Here 
$E^{(2)}$(${\bf k}_\parallel$), $E^{(3)}$(${\bf k}_\parallel$) and 
$E^{(4)}$(${\bf k}_\parallel$) are the second, third, and fourth order
contribution to the effective dispersion, respectively: 
\begin{equation}
E^{(2)}({\bf k}_\parallel) = -4t_{pd}^2\left(\frac{p_x^2}{\bar{\Delta}_{{\bf k}_\parallel}} + \frac{p_y^2}{\bar{\Delta}'_{{\bf k}_\parallel}}\right)
\label{eq.lind1}
\end{equation}
\begin{equation}
E^{(3)}({\bf k}_\parallel) = \frac{32t_{pd}^2t_{pp}p_x^2p_y^2}{\bar{\Delta}_{{\bf k}_\parallel}\bar{\Delta}'_{{\bf k}_\parallel}}
\label{eq.lind2}
\end{equation}
\begin{equation}
\begin{aligned}
E^{(4)}({\bf k}_\parallel) &= \frac{16t_{pd}^4}{\bar{\Delta}^2_{{\bf k}_\parallel}\bar{\Delta}'^2_{{\bf k}_\parallel}}\left(\frac{\bar{\Delta}'^2_{{\bf k}_\parallel}}{\bar{\Delta}_{{\bf k}_\parallel}}p_x^4+\frac{\bar{\Delta}^2_{{\bf k}_\parallel}}{\bar{\Delta}'_{{\bf k}_\parallel}}p_y^4 + p_x^2p_y^2\left(\bar{\Delta}_{{\bf k}_\parallel}+ \bar{\Delta}'_{{\bf k}_\parallel}\right)\right)\\&
-\frac{16t_{pd}^2t_{sp}^2}{\bar{\Delta}_{{\bf k}_\parallel}\bar{\Delta}'_{{\bf k}_\parallel}\tilde{\Delta}_{{\bf k}_\parallel}}\left(\frac{\bar{\Delta}'_{{\bf k}_\parallel}}{\bar{\Delta}_{{\bf k}_\parallel}}p_x^4+\frac{\bar{\Delta}_{{\bf k}_\parallel}}{\bar{\Delta}'_{{\bf k}_\parallel}}p_y^4 - 2p_x^2p_y^2\right)\\
&+\frac{32t_{pd}^2t_{p_z}^2}{\bar{\Delta}_{{\bf k}_\parallel}\bar{\Delta}'_{{\bf k}_\parallel}\Delta_{z}}\left(\frac{\bar{\Delta}'_{{\bf k}_\parallel}}{\bar{\Delta}_{{\bf k}_\parallel}}p_x^4+\frac{\bar{\Delta}_{{\bf k}_\parallel}}{\bar{\Delta}'_{{\bf k}_\parallel}}p_y^4 - 2p_x^2p_y^2\right)\\
&-\frac{64t_{pd}^2t_{pp}^2p_x^2p_y^2}{\bar{\Delta}_{{\bf k}_\parallel}\bar{\Delta}'_{{\bf k}_\parallel}}\left(\frac{p_x^2}{\bar{\Delta}_{{\bf k}_\parallel}} + \frac{p_y^2}{\bar{\Delta}'_{{\bf k}_\parallel}}\right) \\
&- \frac{64t_{pd}^2t_{pp}^{(2)^2}\pi_x^2\pi_y^2}{\bar{\Delta}_{{\bf k}_\parallel}\bar{\Delta}'_{{\bf k}_\parallel}}\left(\frac{\bar{\Delta}'_{{\bf k}_\parallel}}{\bar{\Delta}^2_{{\bf k}_\parallel}}p_x^2 + \frac{\bar{\Delta}_{{\bf k}_\parallel}}{\bar{\Delta}'^2_{{\bf k}_\parallel}}p_y^2\right)\\
&-\frac{16t_{pd}^2t_{\sigma''}^{2}p_{2x}^2p_{2y}^2}{\bar{\Delta}_{{\bf k}_\parallel}\bar{\Delta}'_{{\bf k}_\parallel}}\left(\frac{p_x^2}{\bar{\Delta}_{{\bf k}_\parallel}} + \frac{p_y^2}{\bar{\Delta}'_{{\bf k}_\parallel}}\right) \\
&-
\frac{64t_{pd}^2t_{\sigma''}t_{pp}^{(2)}p_xp_yp_{2x}p_{2y}\pi_x\pi_y}{\bar{\Delta}^2_{{\bf
      k}_\parallel}\bar{\Delta}'^2_{{\bf
      k}_\parallel}}\left(\bar{\Delta}_{{\bf k}_\parallel}+\bar{\Delta}'_{{\bf
      k}_\parallel}\right)\,.
\end{aligned}
\label{eq.lind3}
\end{equation}
No $k_z$ dependence arises at this order in PT.
\begin{figure}[h!]
\begin{center}
\unitlength=0.18cm
\begin{picture}(40,85)
\put(0.75,55){\includegraphics[width=0.87\columnwidth]{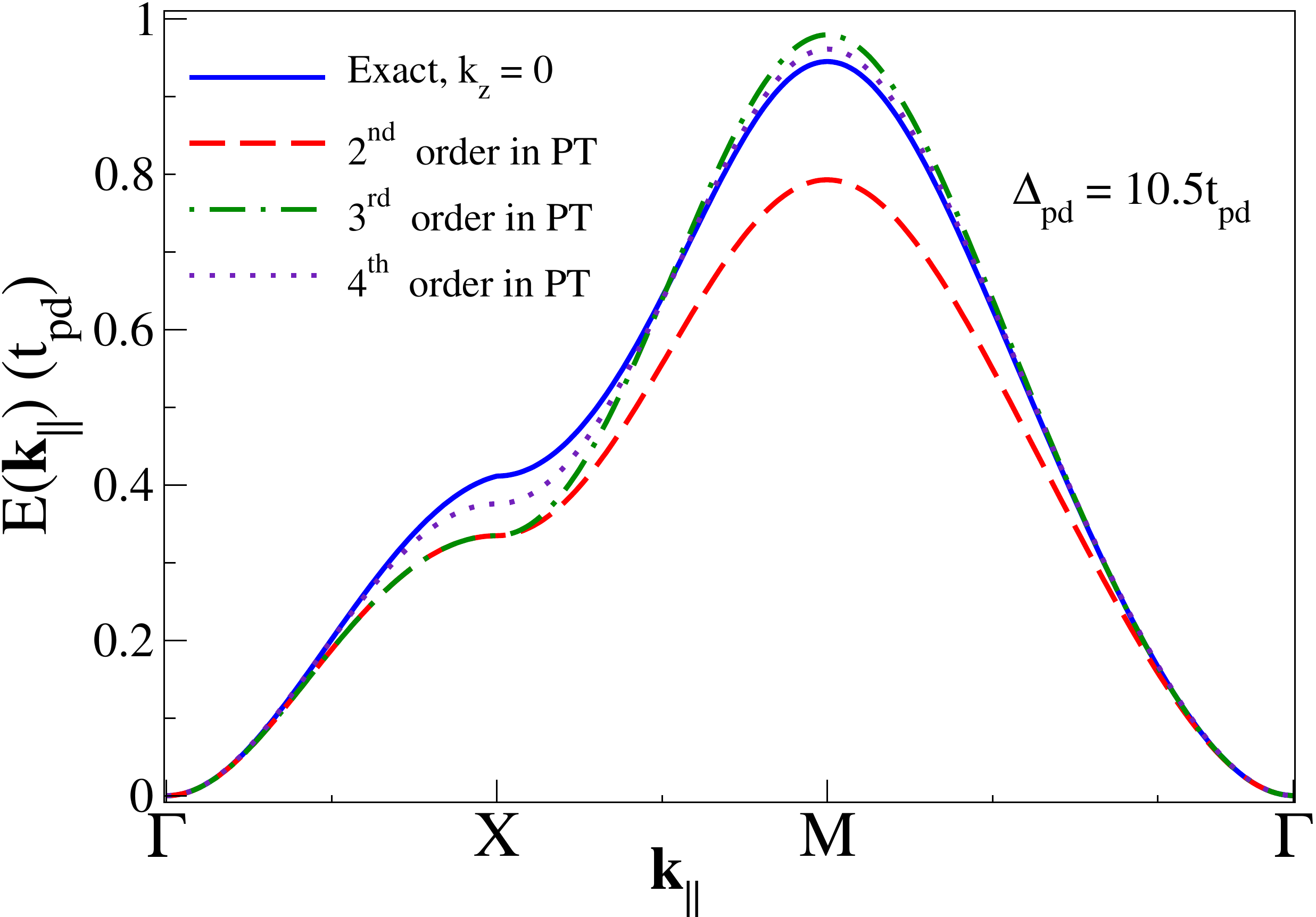}}
\put(1,25.5){\includegraphics[width=0.86\columnwidth]{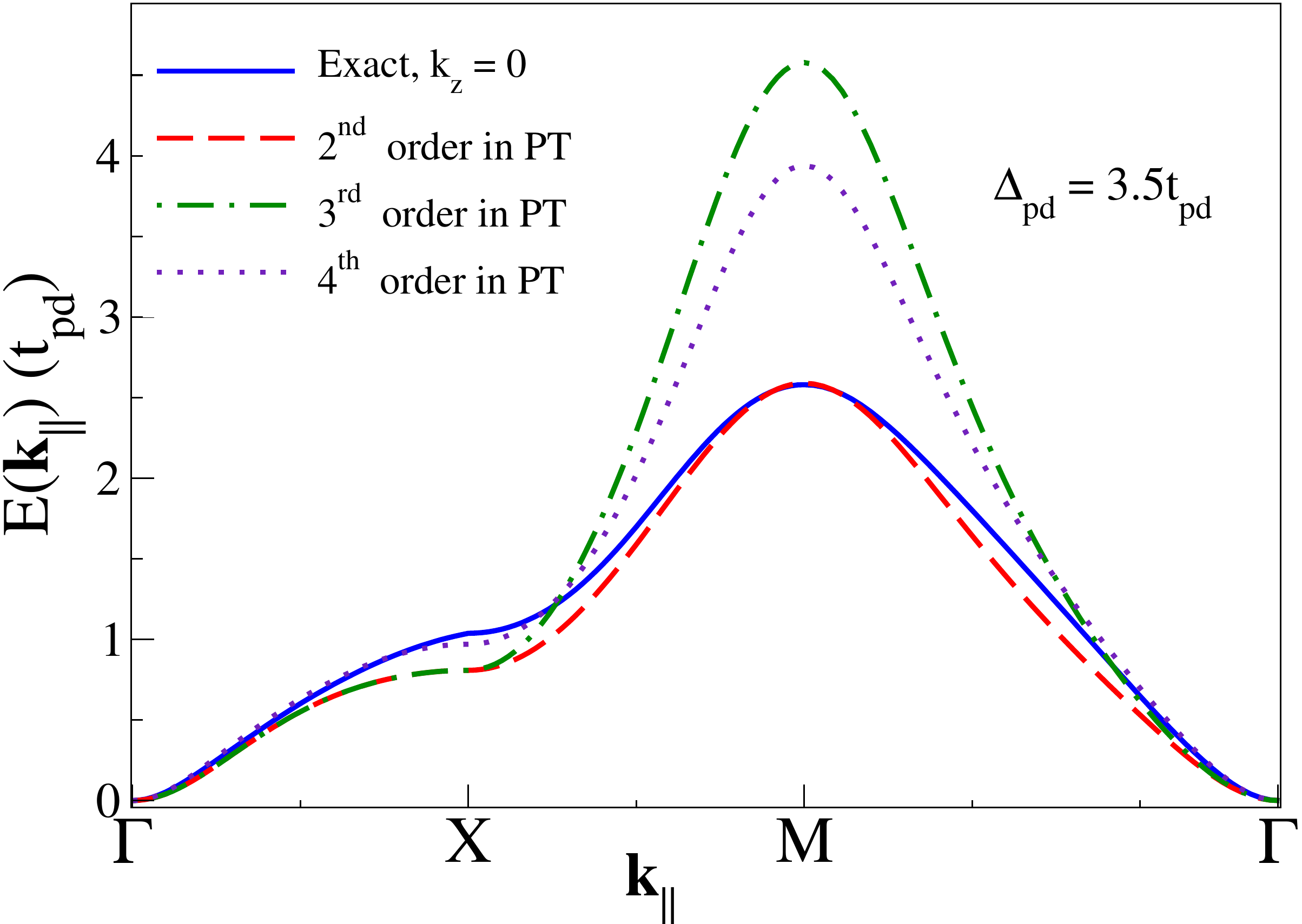}}
\put(1,-4){\includegraphics[width=0.86\columnwidth]{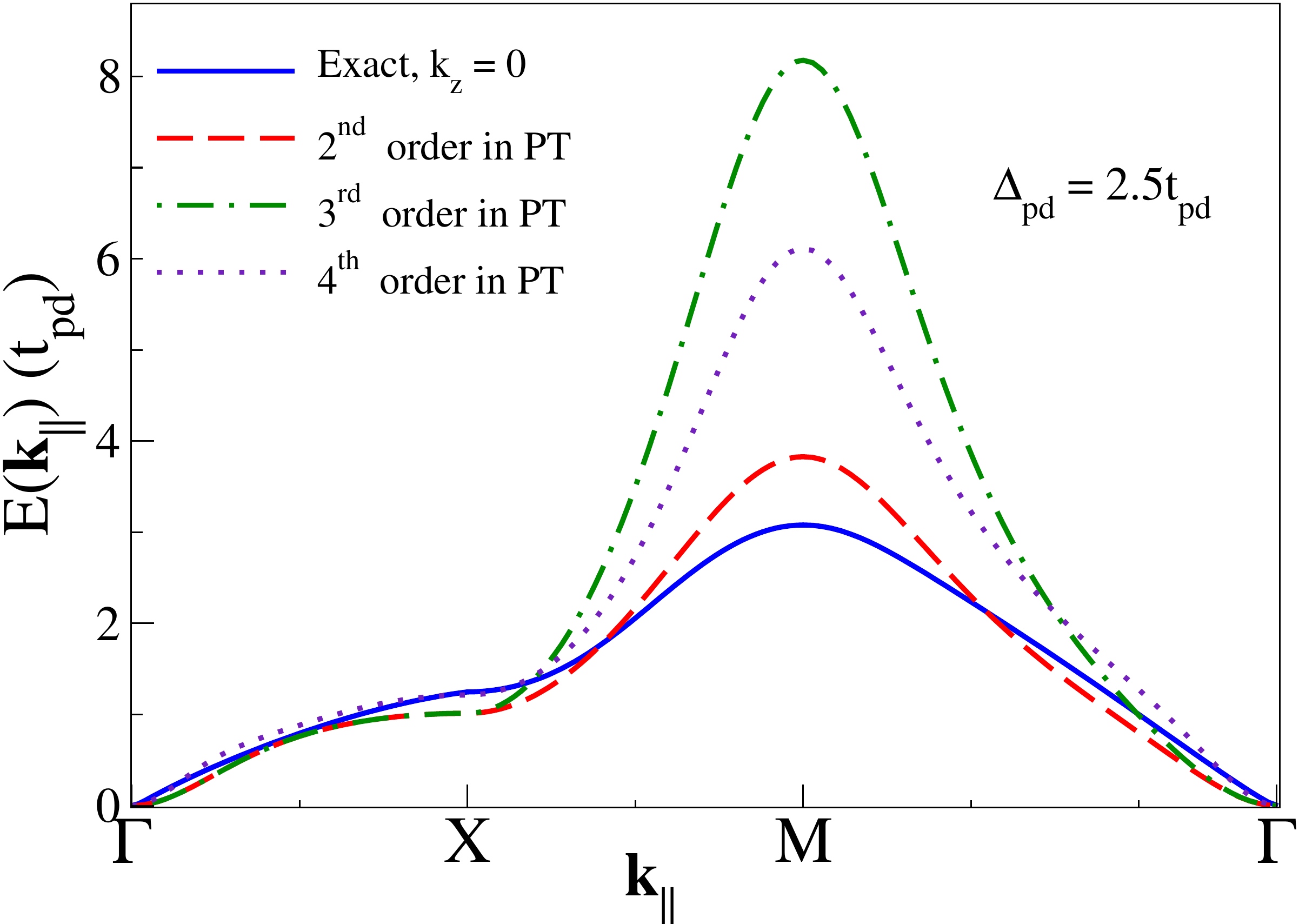}}
\put(37,80){ {\large (a)} } \put(37,51){ {\large (b)} }\put(37,22){ {\large (c)} }
\end{picture}
\end{center}
\caption {(Color online) Comparison between the exact conduction band with the
  effective one obtained from perturbation theory for different values of the
  charge transfer gap. The used tight-binding parameters are given in
  Table~\ref{tab.paramopt}.}  
\label{fig:complindgren}
\end{figure}
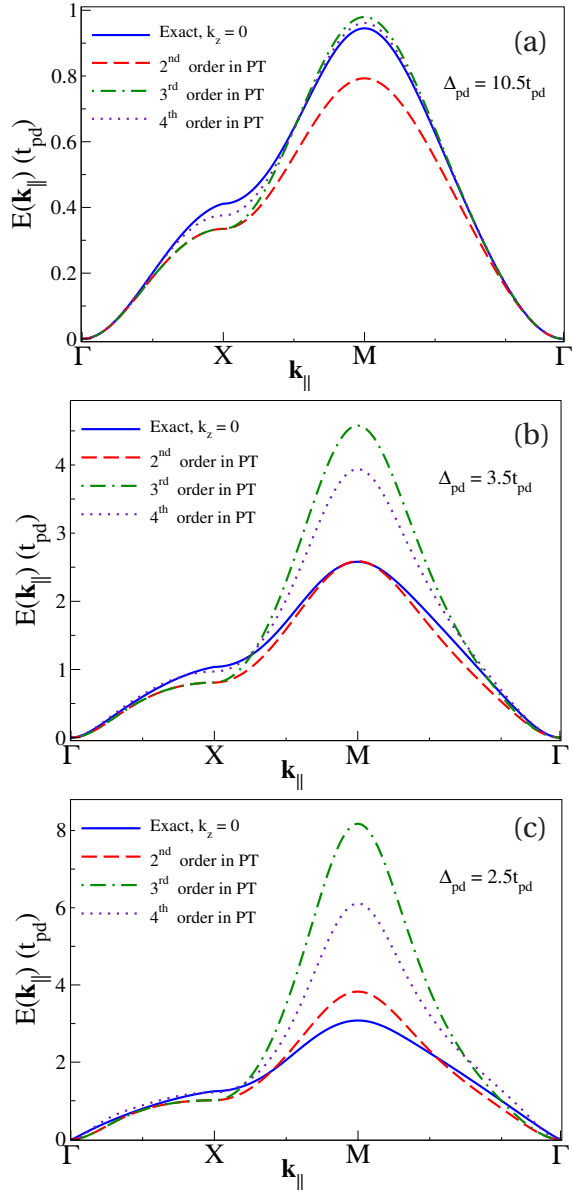

\begin{figure}[t!]
     \centering
     \includegraphics[width=0.95\columnwidth]{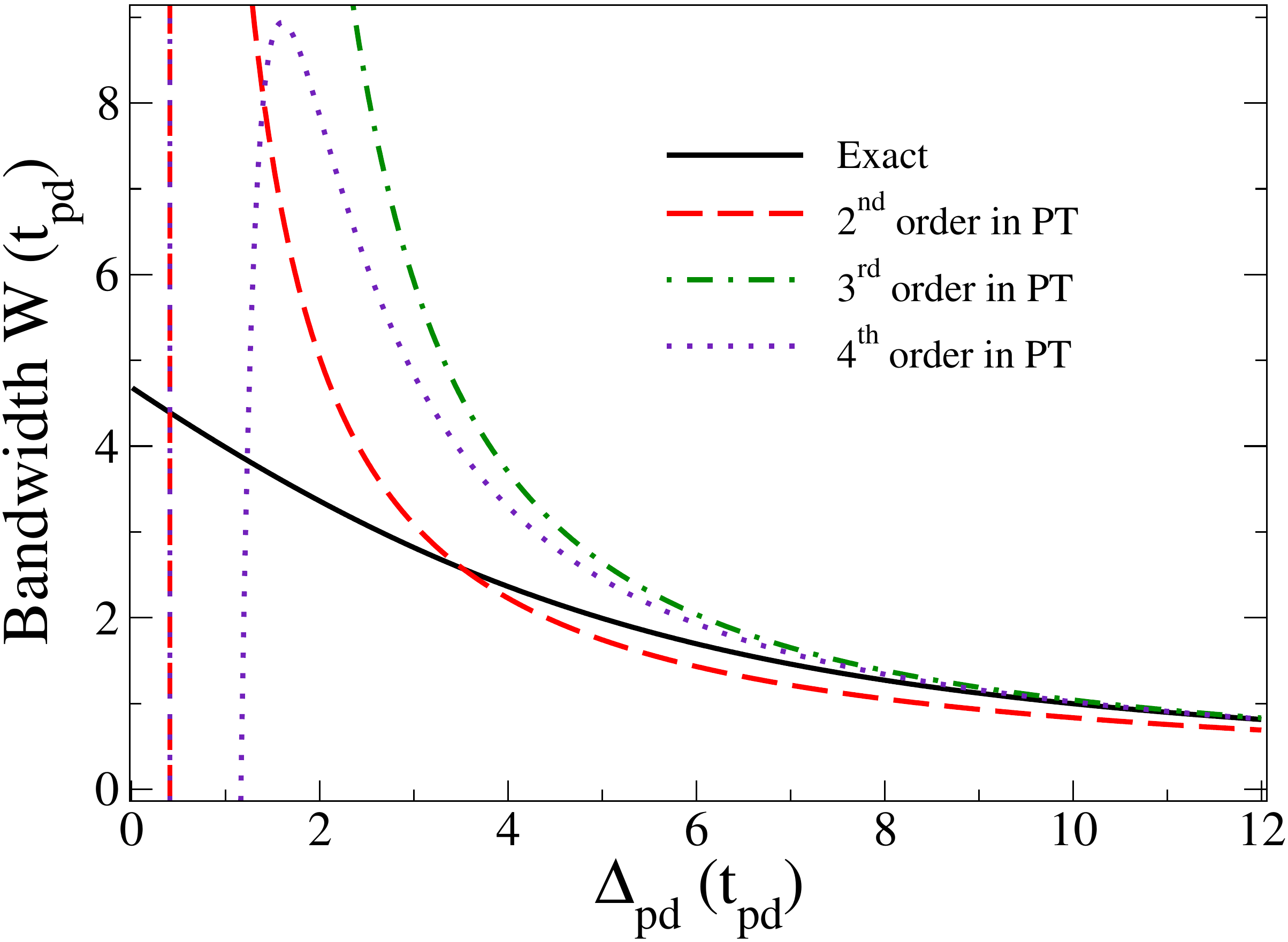}
     \caption{(Color online) The exact bandwidth of the conduction band is
       compared to different orders in perturbation theory. A good
       approximation of the bandwidth is obtained for $\Delta_{pd}\gtrsim
       5t_{pd}$. The used tight-binding parameters are given in
       Table~\ref{tab.paramopt}.} 
     \label{fig:complindgren_2}
\end{figure}
In Fig.~\ref{fig:complindgren} we compare the in-plane dispersion of the
$3d_{x^2-y^2}$ band obtained from numerical diagonalization to the
perturbative results obtained to second, third and fourth orders. Rapid
convergence for all ${\bf k}_\parallel$ values is reached for large values of
$\Delta_{pd}$ (Fig.~\ref{fig:complindgren}(a)). For instance, for $\Delta_{pd}
= 10.5t_{pd}$, the dispersion obtained to fourth order in perturbation theory
is found to almost perfectly reproduce the dispersion obtained by
diagonalizing the eight-band Hamiltonian Eq.~(\ref{eq:modeltot}). Yet, this good agreement gets
gradually lost when reducing the charge transfer gap. Indeed, for $\Delta_{pd}
= 3.5t_{pd}$, which is a broadly accepted value for LSCO
\cite{macma87,dagot94}, the effective dispersion to second order agrees better
to exact diagonalization than the one obtained to third order. Going to fourth
order only yields a small improvement (see Fig.~\ref{fig:complindgren}(b)). No
good agreement could be obtained for $\Delta_{pd} = 2.5t_{pd}$ (see
Fig.~\ref{fig:complindgren}(c)). Furthermore, we show in
Fig.~\ref{fig:complindgren_2} that the bandwidth is correctly recovered for
large values $\Delta_{pd} > 8t_{pd}$, only. Thus, for realistic values of the
charge transfer gap the convergence of the perturbative approach to the exact
conduction band is at best very slow. Such a difficulty is not specific to our
model, but arises as well when tackling the Emery model
Eq.~(\ref{eq.emery_matrix}), and no accurate dispersion relation for realistic
values of $\Delta_{pd}$ and $t_{pp}$ can be obtained. Therefore, either higher
orders in perturbation theory are needed to improve the approximation of the
conduction band, or the perturbation theory breaks down altogether. In the
former case, this implies that smaller hopping integrals become relevant and
yield new hopping processes (for example, $t_{\sigma''}$ or $t_{p_z}'$)  which
are essential to the $3d_{x^2-y^2}$ band dispersion. This idea is crucial in
order to explain the size of the longer-ranged in-plane and inter-plane
hopping amplitudes $t''$, $t'''$ and $t_{z}$ invoked to fit the experimental
(ARPES) and LDA Fermi surfaces as reported in Tables ~\ref{tab:final_1} and
\ref{tab:final_2}. Indeed, as can be seen in Eqs.~(\ref{eq.lind1},
\ref{eq.lind2}, \ref{eq.lind3}), the $k_z$-dispersion along $\Gamma$-X is not
captured by the fourth order. In fact, one needs to go to the fifth order to
obtain inter-plane coupling. It follows from contributions to
$\mathcal{\hat{H}}_{\rm eff}^{(5)}$ given in the appendix. We expand this term
according to the matrix elements of our model
Eq.~(\ref{eq.modelref}). Inter-plane contributions contained in
Eq.~(\ref{eq.lindorder4}) to the effective dispersion of the conduction band
are given by: 
\begin{equation}
\begin{split}
E_{z,{\rm eff}}({\bf k}) =\:\:\:\:\:\:\:\:\:\:\:\:\:\:\:\:\:\:\:\:\:\:\:\:\:\:\:\:\:\:\:\:\:\:\:\:\:\:\:\:\:\:\:\:\:\:\:\:\:\:\:\:\:\:\:\:\:\:\:\:\:\:\:\:\:\:\:\:\:\:\:\:\:\:\:\:\:\:\:\:\:\:\:\:\:\:\:\:\:\:\:\:\:\:\:\:\:\:\:\:\:\:\:\:\:\:\:\:\:\:\:\:\:\:\:\:\:\\
 -128t_{pd}^2t_{p_z}^2t_{p_z}'\pi_x\pi_y\pi_z\left(\frac{p_x^4}{\Delta_{z}^2\bar{\Delta}^2_{{\bf k}_\parallel}} + \frac{p_y^4}{\Delta_{z}^2\bar{\Delta}'^2_{{\bf k}_\parallel}} -\frac{2p_x^2p_y^2}{\Delta_{z}^2\bar{\Delta}'_{{\bf k}_\parallel}\bar{\Delta}_{{\bf k}_\parallel}}\right) \\
 -256t_{pd}^2t_{p_z}t_{p_z}''t_{pp}^{(2)}\pi_x\pi_y\pi_z\left[\frac{p_x^4}{\Delta_{z}\bar{\Delta}^3_{{\bf k}_\parallel}}\right. + \frac{p_y^4}{\Delta_{z}\bar{\Delta}'^3_{{\bf k}_\parallel}}\\
 -\frac{p_x^2p_y^2}{\Delta_{z}\bar{\Delta}_{{\bf k}_\parallel}\bar{\Delta}'_{{\bf k}_\parallel}}\left.\left(\frac{1}{\bar{\Delta}_{{\bf k}_\parallel}} + \frac{1}{\bar{\Delta}'_{{\bf k}_\parallel}}\right)\right]\,.
\end{split}
\label{eq.interplan_lind}
\end{equation}
\begin{table*}[t]
\begin{andptabular}{X[2c]X[2c]X[2c]X[2c]X[2c]X[2c]}%
{In-plane tight-binding parameters set determined from LDA calculations or
  ARPES data and compared to the ones from the Emery model ($\Delta_{pd} =
  3.5t_{pd}$, $t_{pp} = 0.6t_{pd}$) and this work (parameters given in
  Table~\ref{tab.paramopt}).}%
In-plane & $t$ & $t'/t$ & $t''/t$ & $t'''/t$ & $t^{(4)}/t$\\
ARPES (Ref.\cite{marki05}) & 0.25 (eV) & -0.09 & 0.07 & 0.105 & $\cdot$ \\
ARPES (Ref.\cite{norman07}) & 0.195 (eV) & -0.095 & 0.075 & 0.09 & 0.02\\
LDA (Ref.\cite{marki05}) & 0.43 (eV) & -0.09 & 0.07 & 0.08 & $\cdot$\\
Emery model & 0.29 ($t_{pd}$) & -0.11 & 0.05 & -0.0056 & -0.0003\\
This work & 0.28 ($t_{pd}$) & -0.136 & 0.068 & 0.061 & -0.017
\label{tab:final_1}
\end{andptabular}
\end{table*}

\begin{figure}[b!]
     \centering
     \includegraphics[width=0.95\columnwidth]{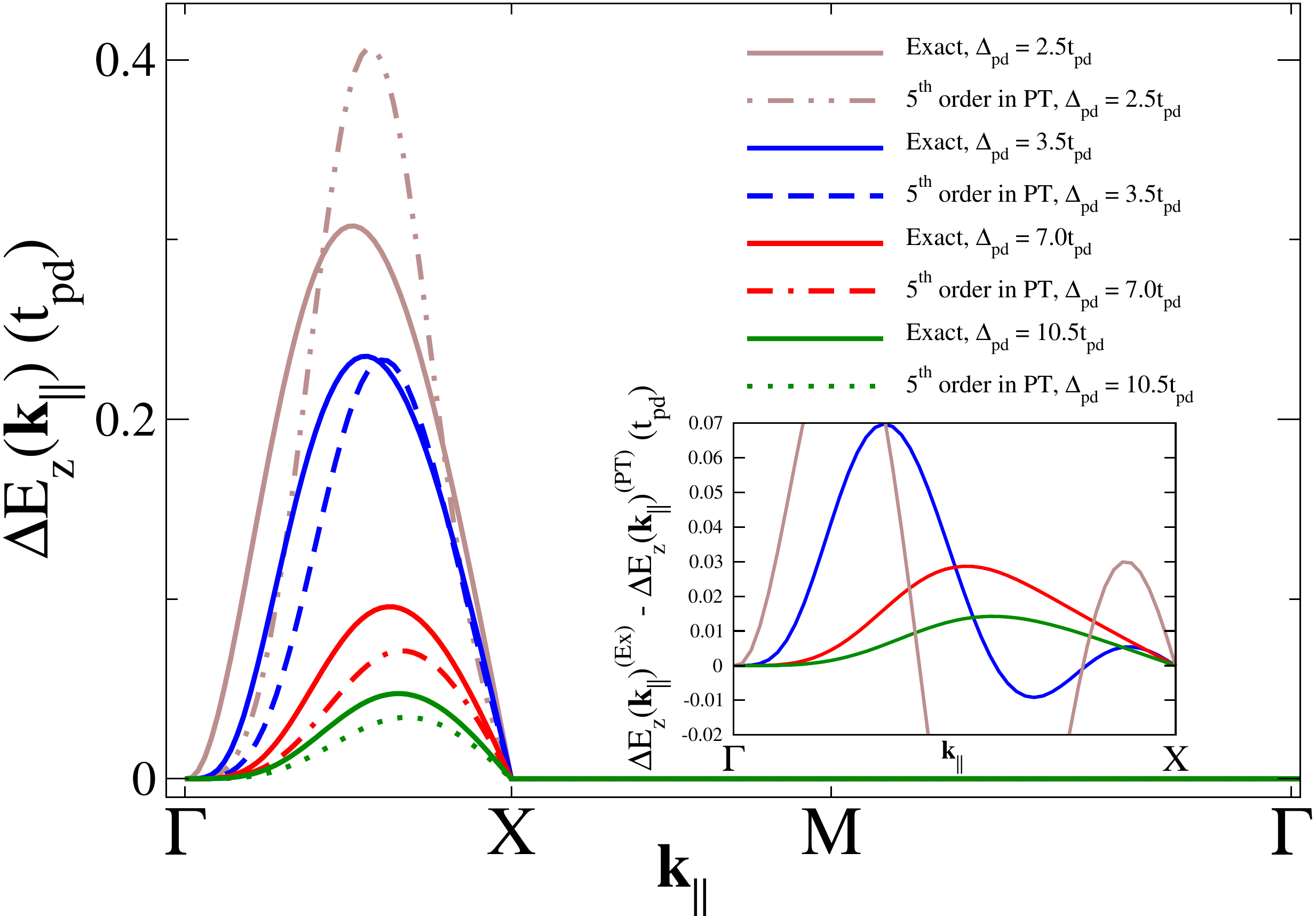}
     \caption{(Color online) The exact $\Delta E_z$(${\bf k}_\parallel$)
       Eq.~$(\ref{eq.delta_ez})$ is compared to the one determined from the
       fifth order in perturbation theory for different values of
       $\Delta_{pd}$. The used tight-binding parameters are given in
       Table~\ref{tab.paramopt}.} 
     \label{fig:deltaezcomp}
\end{figure}
In Fig.~\ref{fig:deltaezcomp}, we plot the energy difference $\Delta
E_z$(${\bf k}_\parallel$) Eq.~$(\ref{eq.delta_ez})$ along path 1, obtained
both with numerical diagonalization and with the effective dispersion
$E_{z,{\rm eff}}({\bf k})$ (Eq.~\ref{eq.interplan_lind}). For small value of
$\Delta_{pd}$ (e.~g. $\Delta_{pd} = 2.5t_{pd}$), $\Delta E_z$(${\bf
  k}_\parallel$) from PT poorly compares to the exact one. Increasing
$\Delta_{pd}$ yields a slightly better agreement but convergence to the exact
value is at best slow, as already observed in the case of the in-plane
dispersion (Fig.~\ref{fig:complindgren}). For realistic $\Delta_{pd} =
3.5t_{pd}$, the $k_z$-dependence of the dispersion is not well captured.  

Therefore, the Rayleigh-Schr\"odinger perturbation theory is not suitable to
properly describe the dispersion of the conduction band. Indeed, since the
first contribution to the $k_z$-dependence of the conduction band arises at
fifth order in PT, good converging behavior of the latter is mandatory. Yet,
Figs.~\ref{fig:complindgren_2} and \ref{fig:deltaezcomp} show that this
happens for unrealistically large values of $\Delta_{pd}/t_{pd}$,
only. However, this approach yields important qualitative informations about
how hopping processes emerging from the coupling between oxygen and copper
orbitals contribute to the dispersion of the conduction band when the
high-energy degrees freedom are integrated out. Indeed, with the charge
transfer gap larger than all hopping integrals involved in the model and
neglecting $t_{\pi'}$, a first order expansion of
Eq.~(\ref{eq.interplan_lind}) in power of $1/\Delta_{pd}$
(e.~g. $1/\Delta_{pd}(1-t_{\alpha}/\Delta_{pd}) 
\simeq (1/\Delta_{pd})(1+t_{\alpha}/\Delta_{pd})$)
leads to: 
\begin{equation}
\begin{aligned}
E_{z,{\rm eff}}({\bf k}) =& -2t_z\pi_x\pi_y\pi_z\left[\cos{(k_xa)}-\cos{(k_ya)}\right]^2\\
&-\frac{2t_z'}{\Delta_{pd}}\pi_x\pi_y\pi_z\left[\cos{(k_xa)}-\cos{(k_ya)}\right]^2\times\\&\left[t_{\sigma''}\cos{(k_xa)}\cos{(k_ya)}{\phantom{\frac{1}{1}}}\right. \\
&\left.-\frac{t_{\sigma'}}{2}\left(\cos{(k_xa)}+\cos{(k_ya)} - 1\right) \right]\:,
\end{aligned}
\label{eq.Eznous_marki}
\end{equation} 
with
\begin{equation}
\begin{aligned}
t_z &= \frac{16t_{pd}^2t_{p_z}}{\Delta_{z}\Delta_{pd}^2}\left(\frac{t_{p_z}t_{p_z}'}{\Delta_{z}} + \frac{2t_{p_z}''t_{pp}^{(2)}}{\Delta_{pd}}\right) \\
t_z' &= \frac{64t_{pd}^2t_{p_z}}{\Delta_{z}\Delta_{pd}^2}\left(\frac{t_{p_z}t_{p_z}'}{\Delta_{z}} + \frac{3t_{p_z}''t_{pp}^{(2)}}{\Delta_{pd}}\right)\:\:.
\end{aligned}
\label{eq.t_zpr}
\end{equation}

\begin{table*}[t]
\begin{andptabular}{X[2c]X[2c]X[2c]X[2c]X[2c]X[2c]X[2c]}%
{Comparison of inter-layer tight-binding parameters to the ones determined from LDA.}%
Inter-plane & $\theta/t$& $\theta'/t$& $\theta''/t$& $\theta'''/t$&$\theta^{(4)}/t$&$\theta^{(5)}/t$\\
LDA (Ref.\cite{marki05}) & 0.015 & -0.0075 & -0.015 & 0.0075 & $\cdot$ & $\cdot$\\
This work & 0.0285 & -0.0070 & -0.0224 & 0.0068 & -0.0054 & -0.0049
\label{tab:final_2}
\end{andptabular}
\end{table*}
Hence the dispersion along $k_z$ predominantly follows from hopping processes
$\sim t_{pd}^2t_{p_z}^2t_{p_z}'/\Delta_{pd}^2\Delta_{z}^2$ and $\sim
t_{pd}^2t_{p_z} t_{p_z}''t_{pp}^{(2)}/\Delta_{pd}^3\Delta_{z}$. Therefore, the
Cu:4s orbital 
is not involved in the leading order contributions to the interlayer hopping
as proposed in Refs.\cite{ander95,xiang96,pavar01}. Instead, this role is
devoted to in-plane and apical oxygens. Strikingly, we show in
Eq.~(\ref{eq.Eznous_marki}) that the dominant term, proportional to $t_z$, is
exactly the inter-layer part of the phenomenological formula proposed by
Markiewicz $\textit{et al.}$\cite{marki05} in order to fit LDA results for
LSCO Eq.~(\ref{eq.marki_ez}) while the second term enriches it. The former
term was also used in Ref.~\cite{orio18} to fit the three-dimensional Fermi
surface of overdoped La-based cuprates. The factor $\gamma \equiv
\left[\cos{(k_xa)}-\cos{(k_ya)}\right]$ of d-wave symmetry suppresses
dispersion along $k_z$ above the high-symmetry line $\Gamma$-M observed in our
model (see Fig.~\ref{fig:fig.lda}). We may note that band structures
calculated for bi-layers Bi-2212 \cite{mattheiss88,marki05}, Tl-2201
\cite{mishonov10}, tri-layer Tl-2223 and four-layer Tl-2234  \cite{angil99}
revealed that the dominant inter-layer hopping exhibits a
$\gamma^2$-modulation, too. In addition, Chakravarty $\textit{et al.}$
\cite{chakrav93} assumed the same $\gamma^2$-modulation of the
inter-layer hopping term which thus plays an important role in the inter-layer
pair tunneling mechanism for boosting T$_c$. In our model, $\gamma$ arises
from the layer-to-layer hybridization between O:2p$_z$ apical oxygen orbitals
through the in-plane O:2p$_{x,y}$ ones, see Eq.~(\ref{eq.Eznous_marki}). This
leads to virtual processes 
involving hopping from, e.~g., an O:2p$_x$ orbital to a nearest O:2p$_z$
orbital, then to a nearest layer O:2p$_z$ orbital, and then to an O:2p$_{x,y}$
orbital. 

However, this perturbative treatment yields a tight-binding model different
from the one of Ref.~\cite{marki05} (Eqs.~(\ref{eq.marki_eparall},
\ref{eq.marki_ez})) as it contains other longer-ranged hopping terms
included in the contribution proportional to $t_z'$
(Eq.~(\ref{eq.t_zpr})). Let us now remark that Eq.~(\ref{eq.marki_ez}) may be
recast in the form of Eq.~(\ref{eq.disp_interplan}), since: 
\begin{equation}
\begin{aligned}
-2t_z\pi_z\pi_x\pi_y\left[\cos{(k_xa)}-\cos{(k_ya)}\right]^2 =\\
 -2t_z\pi_z\left(\frac{1}{2}\cos{\left(\frac{k_xa}{2}\right)}\cos{\left(\frac{k_ya}{2}\right)}\right. \\
-\frac{1}{4}\left[\cos{\left(\frac{3k_xa}{2}\right)}\cos{\left(\frac{k_ya}{2}\right)} + \cos{\left(\frac{3k_ya}{2}\right)}\cos{\left(\frac{k_xa}{2}\right)}\right]\\
-\frac{1}{2}\cos{\left(\frac{3k_xa}{2}\right)}\cos{\left(\frac{3k_ya}{2}\right)}\\
+\frac{1}{4}\left[\cos{\left(\frac{5k_xa}{2}\right)}\cos{\left(\frac{k_ya}{2}\right)} + \left.\cos{\left(\frac{5k_ya}{2}\right)}\cos{\left(\frac{k_xa}{2}\right)}\right]\right)\,.
\end{aligned}
\end{equation}
Hence Eq.~(\ref{eq.marki_ez}) corresponds to Eq.~(\ref{eq.disp_interplan})
under the assumption $t_z = 8\theta = -16\theta' = -8\theta'' = 16\theta'''$,
which may not be justified within in our model, unless all contributions to
the out-of-plane hopping amplitudes are neglected, but the leading one. In our
model, the multiplicity of processes involved in $t_z'$ is higher than those
involved in $t_z$, see Eq.~(\ref{eq.t_zpr}). In addition, the in-plane
$t_{\sigma''}$ and $t_{\sigma'}$ 
integrals involved in the term proportional to $t_z'$ significantly contribute
to the effective inter-plane hopping integrals. Therefore, the implicit
relation between $t_z$, $\theta$, $\theta'$, $\theta''$ and $\theta'''$
involved in the formula Eq.~(\ref{eq.marki_ez}) used in
Ref.~\cite{marki05,orio18} is broken in our model. Indeed, by expanding
Eq.~(\ref{eq.Eznous_marki}) on the basis of
Eq.~(\ref{eq.disp_interplan}), we explicit the microscopical hopping processes
contribution to the inter-plane effective hopping integrals $\theta$,
$\theta'$, $\theta''$ and $\theta'''$. We obtain: 
\begin{equation}
\begin{aligned}
\theta &=
\frac{2t_{pd}^2t_{p_z}}{\Delta_{pd}^2\Delta_{z}}
\left[\frac{t_{p_z}t_{p_z}'}{\Delta_{z}}
\left(1-\frac{t_{\sigma''}-t_{\sigma'}}{\Delta_{pd}}\right) \right. \\ 
&\left.+ \frac{2t_{p_z}''t_{pp}^{(2)}}{\Delta_{pd}}
\left(1-\frac{3(t_{\sigma''}-t_{\sigma'})}{2\Delta_{pd}}\right)\right]\\
\theta' &= \frac{-t_{pd}^2t_{p_z}}{\Delta_{pd}^2\Delta_{z}}
\left[\frac{t_{p_z}t_{p_z}'}{\Delta_{z}}
\left(1-\frac{2(t_{\sigma''}-t_{\sigma'})}{\Delta_{pd}}\right) \right.\\
&\left.+ \frac{2t_{p_z}''t_{pp}^{(2)}}{\Delta_{pd}}
\left(1-\frac{3(t_{\sigma''}-t_{\sigma'})}{2\Delta_{pd}}\right)\right]\\
\theta'' &= \frac{-2t_{pd}^2t_{p_z}}{\Delta_{pd}^2\Delta_{z}}
\left[\frac{t_{p_z}t_{p_z}'}{\Delta_{z}}
\left(1-\frac{2t_{\sigma''}-t_{\sigma'}}{\Delta_{pd}}\right)\right. \\
&\left.+ \frac{2t_{p_z}''t_{pp}^{(2)}}{\Delta_{pd}}
\left(1-\frac{3(2t_{\sigma''}-t_{\sigma'})}{2\Delta_{pd}}\right)\right]\\
\theta''' &= \frac{t_{pd}^2t_{p_z}}{\Delta_{pd}^2\Delta_{z}}
\left[\frac{t_{p_z}t_{p_z}'}{\Delta_{z}}
\left(1-\frac{(3t_{\sigma''}-2t_{\sigma'})}{\Delta_{pd}}\right)\right. \\
&\left.+ \frac{2t_{p_z}''t_{pp}^{(2)}}{\Delta_{pd}}
\left(1-\frac{3(3t_{\sigma''}-2t_{\sigma'})}{2\Delta_{pd}}\right)\right]\,.
\label{eq.hoppingparam_effect_interp}
\end{aligned}
\end{equation}
These are the main inter-plane effective hopping amplitudes. As a matter of
fact, the nearest neighbor $\theta$ and the second nearest neighbor $\theta''$
are the dominating inter-plane hopping integrals. Their amplitudes are very
close to one another, while $\theta'$ is smaller by a factor close to 2,
only. Despite the needed fifth perturbative order, inter-plane hopping
processes must be taken into account since  $\Delta_{z}$ < $\Delta_{pd}$
\cite{jang15}. The underlying microscopic mechanism implies hopping from the
Cu:3d$_{x^2-y^2}$ orbital to an in-plane oxygen one, then to an apical oxygen
one, then to an apical oxygen one belonging to the next CuO$_2$ layer, then to
an in-plane oxygen one in this layer, and finally to the Cu:3d$_{x^2-y^2}$ one
in this layer. The various $\theta$'s are proportional, to leading order, to
$t_{pd}^2t_{p_z}^2t_{p_z}'/\Delta_{pd}^2\Delta_{z}^2$. Furthermore, the
leading corrections may also significantly contribute and become important for
small values of $\Delta_{z}$ and $\Delta_{pd}$, which is the relevant
experimental situation. Besides, hopping processes through the direct coupling
between in-plane oxygen orbitals and next-layer apical oxygen orbital of order
$t_{pd}^2t_{p_z}t_{p_z}''t_{pp}^{(2)}/\Delta_{pd}^3\Delta_{z}$ contribute as
well. In addition, in-plane hopping integrals $t_{\sigma'}$ and $t_{\sigma''}$
trigger longer-ranged hopping processes and are contributing to the anisotropy
of the inter-plane effective hopping integrals. 

Furthermore, a similar study to the in-plane effective dispersion can be
performed. Fourth order in PT suffices to obtain the leading
contributions. Then, by linearizing Eq.~(\ref{eq.heff}) and performing a
Taylor expansion as done for Eq.~(\ref{eq.interplan_lind}), one obtains the
in-plane effective hopping integrals as: 
\begin{equation}
\begin{aligned}
t &= t_{E}  + \frac{t_{pd}^2}{\Delta_{pd}^2}(t_{\sigma''}-2t_{\sigma'}) + \frac{t_{pd}^2}{\Delta_{pd}^3}\left[4t_{pp}(2t_{\sigma''}\right.-3t_{\sigma'})\\
&+2t_{\sigma''}^2+\left.3t_{\sigma'}^2 - 2t_{pp}^{(2)^2} - 4t_{\sigma'}t_{\sigma''}\right]\\
t' &= t_{E}' + \frac{2t_{pd}^2}{\Delta_{pd}^2}\left(\frac{2t_{p_z}^2}{\Delta_{z}} - \frac{t_{sp}^2}{\Delta_{s}} - t_{\sigma''}\right) + \frac{2t_{pd}^2}{\Delta_{pd}^3}\left[3t_{\sigma'}\right.t_{\sigma''}\\
& - \left.4t_{pp}(t_{\sigma''}-t_{\sigma'})\right] \\
 t'' &= t_{E}'' + \frac{2t_{pd}^2}{\Delta_{pd}^2}\left(-\frac{t_{p_z}^2}{\Delta_{z}} + \frac{t_{sp}^2}{2\Delta_{s}} + \frac{t_{\sigma'}}{2}\right)+ \frac{2t_{pd}^2}{\Delta_{pd}^3}\left[t_{pp}^{(2)^2}\right. \\
&-\left.2t_{pp}(t_{\sigma''}-t_{\sigma'}) + t_{\sigma''}t_{pp}^{(2)}-t_{\sigma'}^2\right]\\
t''' &= t_{E}''' + \frac{t_{pd}^2t_{\sigma''}}{2\Delta_{pd}^2} + \frac{t_{pd}^2}{\Delta_{pd}^3}\left[2t_{pp}(t_{\sigma''}-t_{\sigma'})\right.+\frac{t_{\sigma''}^2}{2}\\
&\left.+ t_{pp}^{(2)^2}- 2t_{\sigma'}t_{\sigma''} \right]\,.
\end{aligned}
\label{eq.hoppingparam_effect_inp}
\end{equation}
Where $t_{E}^{(i)}$ are the hopping integrals arising when downfolding the Emery model Eq.~(\ref{eq.emery_matrix}) via the same perturbative treatment:
\begin{equation}
\begin{aligned}
t_{E} &= \frac{t_{pd}^2}{\Delta_{pd}} + \frac{4t_{pd}^2t_{pp}}{\Delta_{pd}^2} + \frac{14t_{pd}^2t_{pp}^2}{\Delta_{pd}^3} - \frac{8t_{pd}^4}{\Delta_{pd}^3}\\
t_{E}' &= -\frac{2t_{pd}^2t_{pp}}{\Delta_{pd}^2} - \frac{8t_{pd}^2t_{pp}^2}{\Delta_{pd}^3} + \frac{2t_{pd}^4}{\Delta_{pd}^3}\\
t_{E}'' &= -\frac{2t_{pd}^2t_{pp}^2}{\Delta_{pd}^3} + \frac{t_{pd}^4}{\Delta_{pd}^3}\\
t_{E}''' &=  \frac{t_{pd}^2t_{pp}^2}{\Delta_{pd}^3}\,.
\end{aligned}
\label{eq.hoppingparam_emer}
\end{equation}
Not only does our model capture the in-plane d-p and O$^{(X)}$-O$^{(Y)}$
hopping processes involved in the Emery Model Eq.~(\ref{eq.emery_matrix}), but
the purpose of Eq.~(\ref{eq.hoppingparam_effect_inp}) is to show how the
effective hopping parameters entailed in the Emery model are modified within
our model. Indeed, it turns out that the added O-O hopping integrals
$t_{\sigma'}$ and $t_{\sigma''}$ modify the copper lattice effective hopping
parameters. For example, the leading contribution to $t''$ follows from
$t_{\sigma'}$. It is proportional to $t_{\sigma'}(t_{pd}/\Delta_{pd})^2$ while
in the Emery model $t''$ is proportional to
$(t_{pp}^2/\Delta_{pd})(t_{pd}/\Delta_{pd})^2$ (to leading order). Similarly,
the main contribution to the third nearest neighbor hopping amplitude $t'''$
is no longer governed by $(t_{pp}^2/\Delta_{pd})(t_{pd}/\Delta_{pd})^2$ as in
the Emery model, but by $t_{\sigma''}(t_{pd}/\Delta_{pd})^2$. Furthermore,
$t_{\sigma''}$ also yields a leading order contribution to $t'$. In addition,
we find $t_{\sigma'}$ and $t_{\sigma''}$ to produce longer-ranged hopping amplitudes
of order $1/\Delta_{pd}^3$  until $t^{(7)}$. Moreover,
hopping processes involving apical oxygen ions and Cu:4s orbitals yield
sub-leading order contributions to $t'$ and $t''$. Indeed, hopping processes
involving the $t_{p_z}$ hopping integral between in-plane oxygen and apical
oxygen orbitals ($\sim t_{pd}^2t_{p_z}^2/\Delta_{pd}^2\Delta_{z}$) strongly
reduce the amplitude of $t'$ and $t''$. This is compatible with the recent
experimental observation that $t'$ decreases when the apical oxygen ions are
brought closer to the basal plane \cite{peng17}, which further supports the
empirical correlation between $t'$ and d$_{\rm Cu-O_{ap}}$ found by Pavarini
$\textit{et al}$.\cite{pavar01}. Moreover, we show that $t'$ and $t''$ only
are sensitive to the apical oxygen ions, while $t'''$ is not at fourth order in
perturbation theory Eq.~(\ref{eq.hoppingparam_effect_inp}). Besides, as
already shown in Ref.\cite{pavar01}, the $t_{sp}$ hopping integral resulting
from the hybridization between Cu:4s and in-plane oxygen orbitals enhances
$t'$ and, to a lesser extent, $t''$ through hopping processes ($\sim
t_{pd}^2t_{sp}^2/\Delta_{pd}^2\Delta_{s}$). 

\subsection{Numerical approach: role of microscopical
  parameters}\label{sec:numerical_approach}

The previous section shows that the perturbative approach is suitable to
interprete the different leading order hopping processes involved in the
multiband model. Yet, this description in terms of higher order superexchange
processes is not accurate enough to quantitatively provide the effective
dispersion of the conduction band. A better alternative consists in applying
the Fourier transform of the conduction band obtained via numerical
diagonalization. This quantitative approach allows us to find the numerical
values of the in-plane and inter-plane hopping integrals which perfectly fit
the conduction band and, therefore, provides a realistic one-band effective
model. For instance, we obtain the hopping parameter $\theta$ as:
\begin{equation}
\theta = \int \frac{d{\bf k}}{(2\pi)^3} 
e^{i{\bf k}\cdot(\frac{a}{2}({\bf e_x} + {\bf e_y}) + \frac{c}{2}{\bf e_z})}
E_{cb}({\bf k})
\label{eq:43}
\end{equation}
and accordingly for the other ones.

The numerically obtained 3d$_{x^2-y^2}$ band is very sensitive to the choice
of the tight-binding parameters of the model introduced in
Section~\ref{sec:model}. Then we have found a set of optimal parameters
yielding a good fit of the LDA conduction band (see Figs.~\ref{fig:fig.lda}
and \ref{fig.lda_2}). The parameters of the 8-band model are expressed in
$t_{pd}$ unit where $t_{pd}\simeq 1.2-1.5$ eV
\cite{macma87,emer87,mila88,hyberts89,kampf94,dagot94,dopf92,pavar01,kent08}. 
Some of them are well known through LDA calculations and we fix them to their
typical value. We accordingly set the optimal energy gaps between the
3d$_{x^2-y^2}$ band and the other band as: $\Delta_{pd,opt} =
3.5t_{pd}$\cite{macma87,kampf94,jang15}, $\Delta_{s,opt} =
6.5t_{pd}$\cite{ander95,pavar01}, and $\Delta_{z,opt} =
2.6t_{pd}$\cite{jang15,weber10}. Concerning typical values for the hopping
amplitudes, we set: $t_{sp,opt} = 1.3t_{pd}$\cite{ander95,pavar01}, $t_\pi$ =
$t_\sigma/4$ and $t_{\pi'}$ = $t_{\sigma'}/4$ according to
Refs.~\cite{ebrah14,harris99}. The uncertainty on $t_{\sigma}$, $t_{\sigma'}$
and $t_{\sigma''}$ is larger. Here we choose $t_{\sigma,opt} = 0.95t_{pd}$
since t$_{pp}$ = $(t_{\sigma}+t_{\pi})/2 \simeq 0.6t_{pd}$ which is the
typical admitted value of the O$^{(X)}$-O$^{(Y)}$ hopping matrix element in
the Emery model \cite{dagot94,kampf94,weber10}. Since the distance between the
involved oxygen ions increases: $t_{\sigma,opt} > t_{\sigma',opt}$,
$t_{\sigma'',opt}$. Then, we set $t_{\sigma'',opt} = 0.4t_{pd}$ according to
Ref.\cite{kent08} in which a sizeable hopping amplitude is determined between
next-nearest neighbors O$^{(\beta)}$-O$^{(\beta)}$ ($\beta$ = X or Y) oxygens
from first-principle calculations for La-based cuprates. However, the hopping
amplitude between nearest neighbors O$^{(\beta)}$-O$^{(\beta)}$ oxygens was
determined to be small \cite{kent08}. Accordingly, we set
$t_{\sigma',opt} = 0.13t_{pd}$. Besides, we set $t_{ss,opt} = 0.4t_{pd}$ and
$t_{ss,opt}' = 0.1t_{pd}$ according to the inter-atomic distance between the copper
atoms. Concerning hopping integrals involving apical oxygens, no consensus
about their values has been reached in the literature. All these unknown
in-plane and out-of-plane hopping amplitudes are determined in order to
optimize the ${\bf k}_\parallel$ dispersion of the calculated conduction band
as well as the effect of the $k_z$-dispersion in order to provide a better
comparison with LDA (Figs.~\ref{fig:fig.lda} and \ref{fig.lda_2}). We found
optimal values: $t_{sp_z,opt} = 1.4t_{pd}$, $t_{p_z,opt} = 0.95t_{pd}$,
$t_{p_z,opt}' = 0.45t_{pd}$, $t_{p_z,opt}'' = 0.1t_{pd}$, and $t_{p_z,opt}'''$
= 0. These values are consistent with the different inter-atomic distances
involving apical oxygens which are detailed in Section~\ref{sec:model}. The
set of optimal parameter is reported in Table~\ref{tab.paramopt}.

\begin{figure}[h!]
\begin{center}
\unitlength=0.22cm
\begin{picture}(38,44)
\put(1,20.5){\includegraphics[width=0.86\columnwidth]{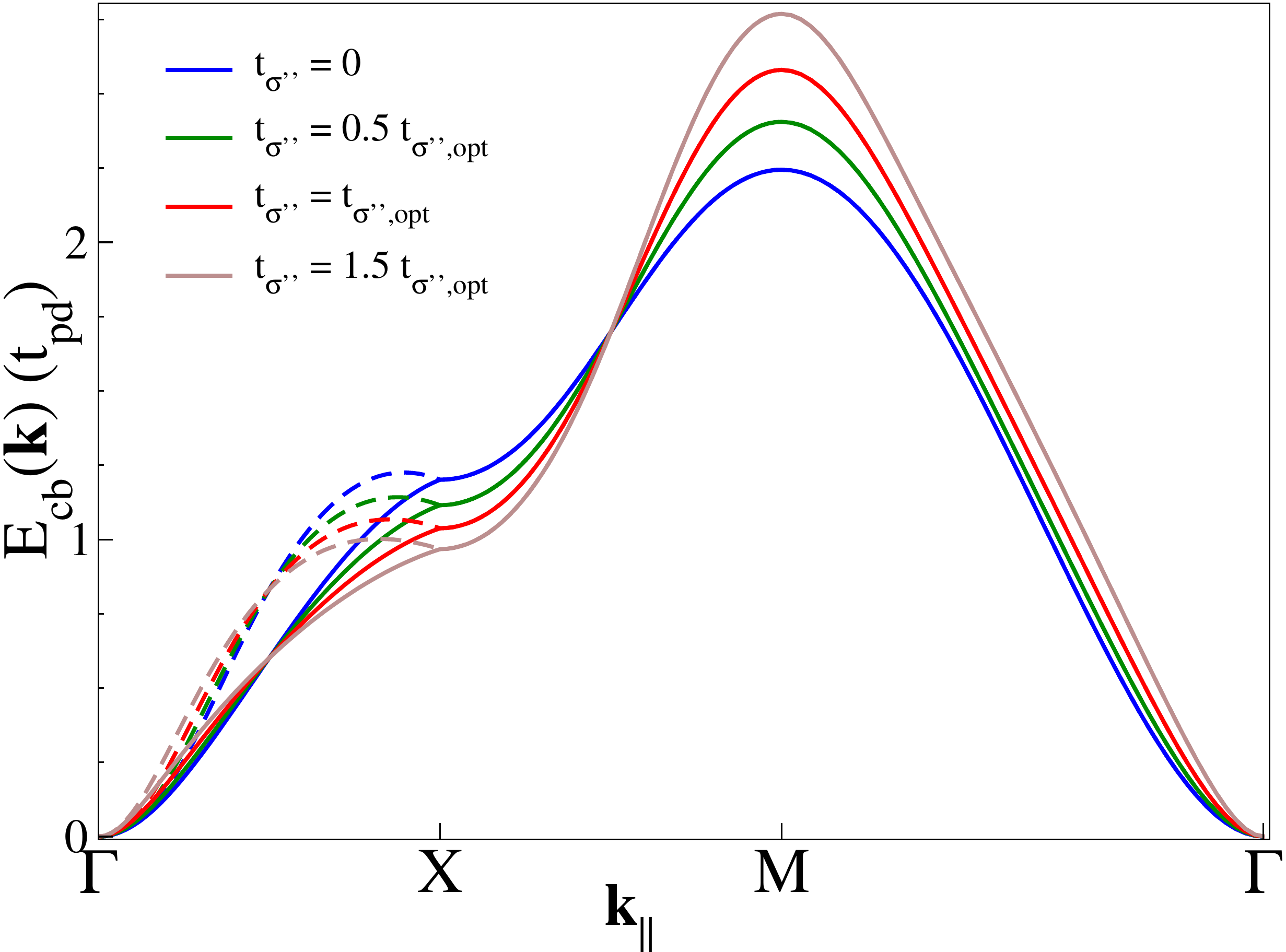}}
\put(0,-3.5){\includegraphics[width=0.875\columnwidth]{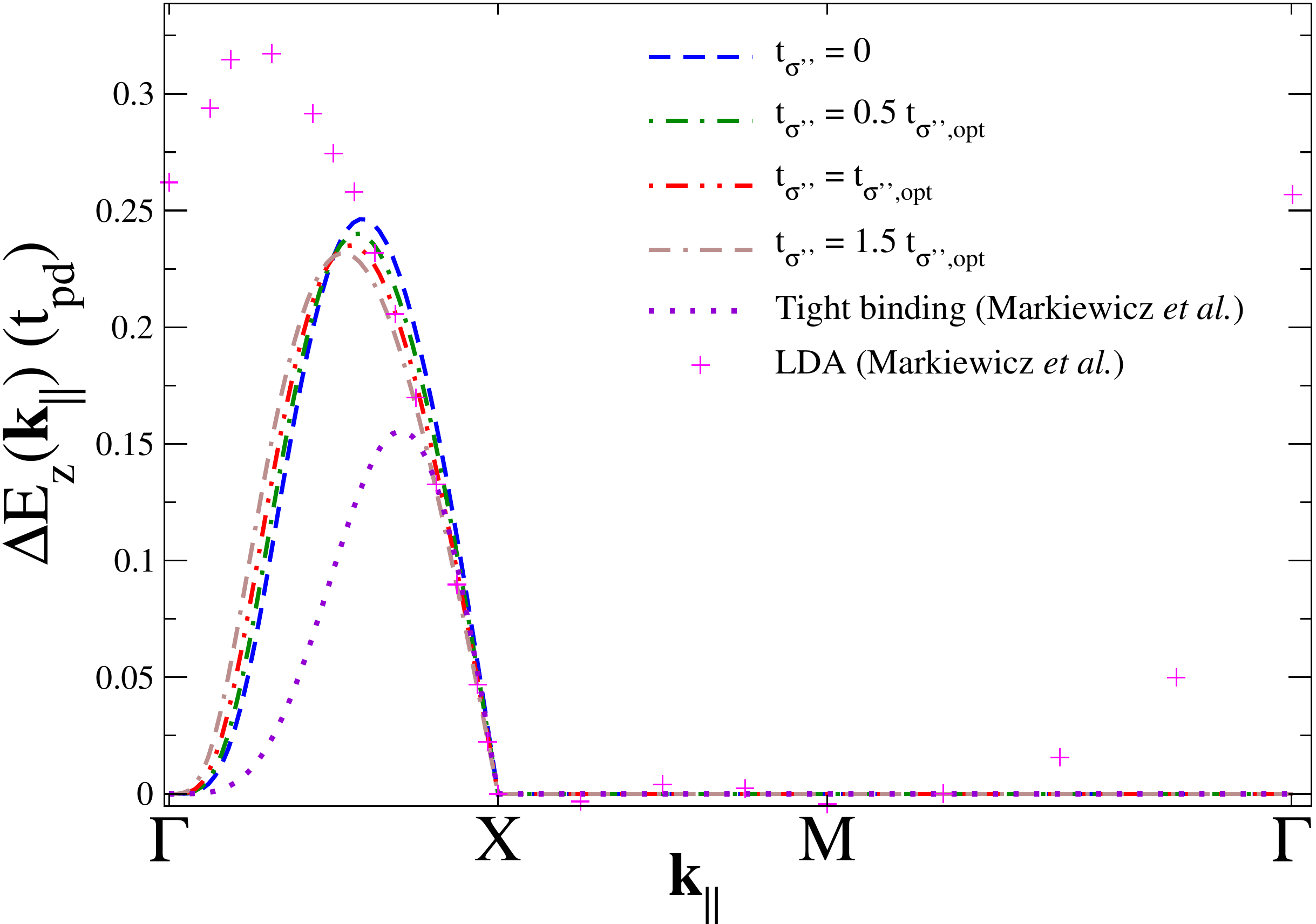}}
\put(30,42){ {\large (a)} } \put(30,17){ {\large (b)} }
\end{picture}
\end{center}
\caption {(Color online) (a) Dispersion of the conduction band along path 1
  (solid lines) and path 2 (dashed lines) with $t_{\sigma''}$ varied around
  its optimal value. (b) $\Delta E_z({\bf k}_\parallel)$ with $t_{\sigma''}$
  varied around its optimal value. The other used tight-binding parameters are
  given in Table~\ref{tab.paramopt}.} 
\label{fig:tsigdiag}
\end{figure}
\begin{table*}
\begin{andptabular}{X[6c]X[5c]X[5c]X[5c]X[5c]X[5c]X[5c]X[5c]X[5c]}%
{Dependence of the main in-plane and inter-plane effective hopping amplitudes
  on $t_{\sigma''}$ expressed in units of its optimal value. The other used
  tight-binding parameters are given in Table~\ref{tab.paramopt}.}%
$t_{\sigma''}/t_{\sigma'',opt}$ & $t/t_{pd}$ & $t'/t$ & $t''/t$ & $t'''/t$ & $t^{(4)}/t$ & $\theta/t$ & $\theta'/t$& $\theta''/t$\\
0 & 0.267 & 0.038 & 0.014 & 0.029& 0.015 & 0.0303 & -0.0089 & -0.0258\\
0.5 & 0.275 & -0.052 & 0.038 & 0.045 & 0.003 & 0.0293 & -0.0079 & -0.0241\\
1 & 0.283 & -0.136 & 0.068 & 0.061 & -0.017 & 0.0285 & -0.0069 & -0.0224\\
1.5 & 0.291 & -0.222 & 0.103 & 0.074 & -0.049 & 0.0279 & -0.0059 & -0.0209
\label{tab:tsigdiag_1}
\end{andptabular}
\end{table*}

It turns out that some tight-binding parameters have stronger impact on the
dispersion of the conduction band than others. In fact the unknown hopping
integrals $t_{\sigma'}$, $t_{\pi'}$, $t_{ss}$, $t_{ss'}$ and $t_{sp_z}$
modifies weakly the conduction band and are, therefore non-critical. In
contrast, $t_{p_z}$, $t_{p_z}'$ and $t_{\sigma''}$ strongly affect the
3$d_{x^2-y^2}$ dispersion because they give birth to leading order hopping
processes (see the perturbative expansion
Eqs.~(\ref{eq.hoppingparam_effect_interp},\ref{eq.hoppingparam_effect_inp})). 
Indeed, Fig.~\ref{fig:tsigdiag}(a) shows  
the strong influence of the $t_{\sigma''}$ hopping integral around its optimal
value on the dispersion along path 1 of the Brillouin zone. Increasing
$t_{\sigma''}$ reduces the energy of the band at X whereas the energy value at
M (the bandwidth) is strongly increased. Hence the entire dispersion along
$\Gamma$-X-M-$\Gamma$ is modified. However, the dispersion along $k_z$ of
width $\Delta E_z({\bf k}_\parallel)$ shown in Fig.~\ref{fig:tsigdiag}(b) is
largest above the $\Gamma$-X line. It is hardly influenced by
$t_{\sigma''}$. Table~\ref{tab:tsigdiag_1} shows how the in-plane and the
inter-plane microscopic hopping parameters $t^{(i)}/t$ are modified by
$t_{\sigma''}$. When $t_{\sigma''}$ is increased, then $t$ is moderately
increased in agreement with Eq.~(\ref{eq.hoppingparam_effect_inp}). However,
$t'/t$ and $t'''/t$ strongly vary when $t_{\sigma''}$ is modified around its
optimal value. 
Indeed, as shown in the perturbative treatment
Eq.~(\ref{eq.hoppingparam_effect_inp}), leading order hopping processes, $\sim
-t_{\sigma''}t_{pd}^2/\Delta_{pd}^2$, contribute to $t'$. This indicates that
$t'$ may get either increasingly negative when increasing $t_{\sigma''}$ from
$t_{\sigma'',opt}$, or possibly positive  when decreasing $t_{\sigma''}$ from
$t_{\sigma'',opt}$. This trend is confirmed by the exact calculation. Likewise,
the decrease of $t'''$ when decreasing $t_{\sigma''}$ from
$t_{\sigma'',opt}$ predicted by perturbation theory is realized by the exact
calculation, whereas $t''$ is weakly modified by $t_{\sigma''}$ since the
leading order hopping processes, $\sim t_{\sigma'}t_{pd}^2/\Delta_{pd}^2$,
follow from $t_{\sigma'}$.
Besides, all inter-plane hopping parameters depend
on $t_{\sigma''}$ as shown in the perturbative treatment
Eq.~(\ref{eq.hoppingparam_effect_interp}). However their magnitudes weakly
decrease when increasing $t_{\sigma''}$ (see Table~\ref{tab:tsigdiag_1}). 
 
\begin{figure}[h!]
\begin{center}
\unitlength=0.22cm
\begin{picture}(38,44)
\put(1,20.5){\includegraphics[width=0.86\columnwidth]{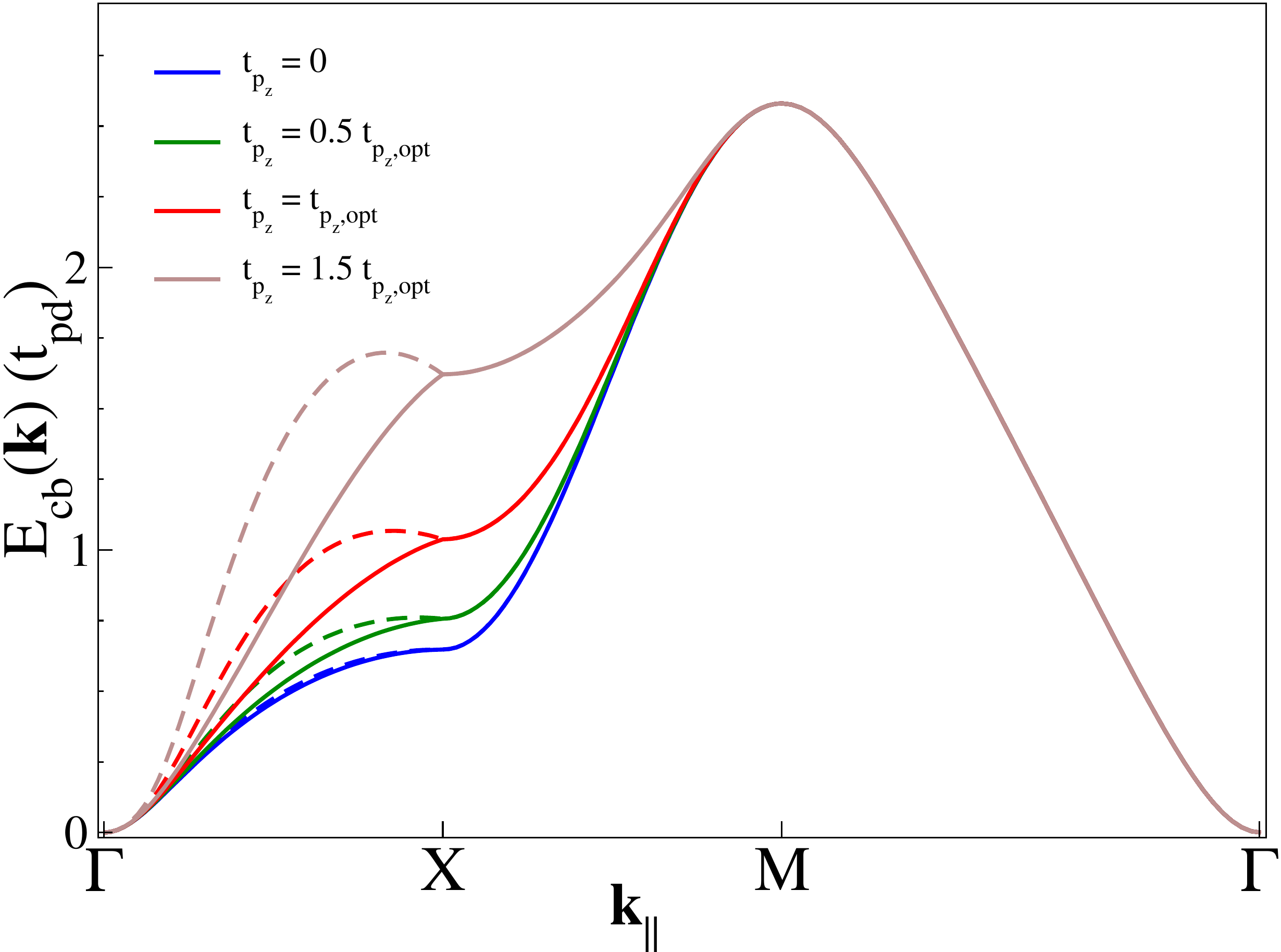}}
\put(0,-3.5){\includegraphics[width=0.875\columnwidth]{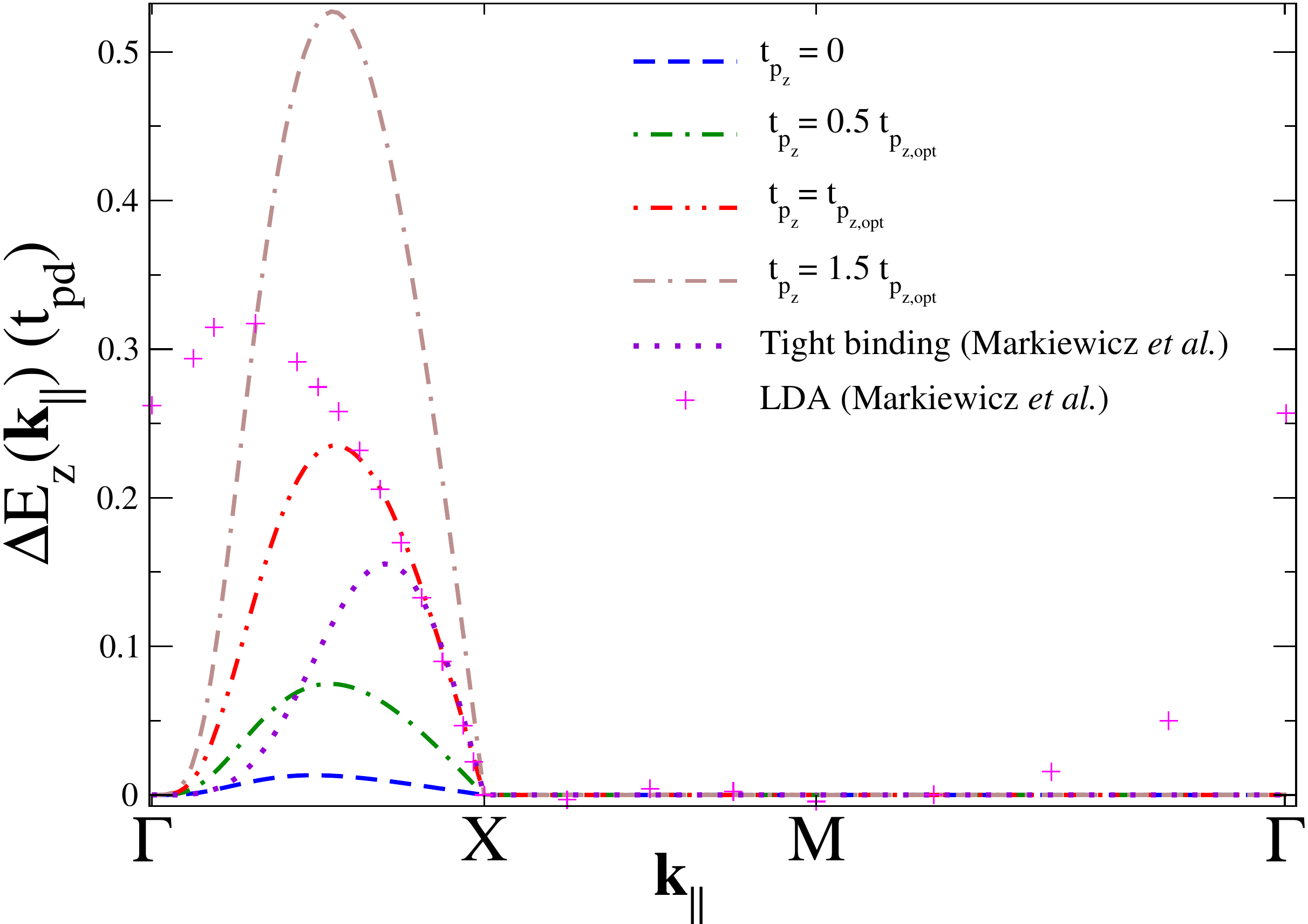}}
\put(30,42){ {\large (a)} } \put(30,17){ {\large (b)} }
\end{picture}
\end{center}
\caption {(Color online) (a) Dispersion of the conduction band along path 1
  (solid lines) and path 2 (dashed lines) with $t_{p_z}$ varied around its
  optimal value. (b) $\Delta E_z ({\bf k}_\parallel)$ with $t_{p_z}$ varied
  around its optimal value. The other used tight-binding parameters are given
  in Table~\ref{tab.paramopt}.} 
\label{fig:tpz}
\end{figure}

\begin{table*}
\begin{andptabular}{X[6c]X[5c]X[5c]X[5c]X[5c]X[5c]X[5c]X[5c]X[5c]}%
{Dependence of the main in-plane and inter-plane effective hopping amplitudes
  on $t_{p_z}$ expressed in units of its optimal value. The other used
  tight-binding parameters are given in Table~\ref{tab.paramopt}.}%
$t_{p_z}/t_{p_z,opt}$ & $t/t_{pd}$ & $t'/t$ & $t''/t$ & $t'''/t$ & $t^{(4)}/t$ & $\theta/t$& $\theta'/t$& $\theta''/t$\\
0 & 0.295 & -0.302 & 0.148 & 0.027 & -0.043 & 0.0014 & -0.0002 & -0.0010\\
0.5 & 0.292 & -0.258 & 0.126 & 0.037 & -0.036 & 0.0082 & -0.0017 & -0.0065\\
1 & 0.283 & -0.136 & 0.068 & 0.061 & -0.017 & 0.0285 & -0.0069 & -0.0224\\
1.5 & 0.273 & 0.168 & -0.039 & 0.086 & -0.049 & 0.0857 & -0.0267 & -0.0508
\label{tab:tpz_1}
\end{andptabular}
\end{table*}

Concerning the hopping integral $t_{p_z}$ accounting for the coupling between
the 2p in-plane oxygen orbitals and the 2$p_z$ apical oxygen orbitals, its
impact on the dispersion of the conduction band is shown in
Fig.~\ref{fig:tpz}. When $t_{p_z}=0$, there is only very small $k_z$
dispersion above $\Gamma$-X. It originates from the terms $\sim
t_{p_z}''t_{sp_z}$, which lead to high order inter-plane hopping processes
(higher than five) and non-vanishing inter-plane hopping integrals reported in
Table~\ref{tab:tpz_1}. An increase in $t_{p_z}$ influences the dispersion of
the conduction band. Yet, regarding the symmetry lines, this increase is not
limited to $\Gamma$-M.
As shown in Fig.~\ref{fig:tpz}(b), the
difference in energy of the dispersion along $\Gamma$-X and Z-R strongly
increases with $t_{p_z}$, in contrast to the bandwidth that is not
affected. Table~\ref{tab:tpz_1} shows that the presence of apical oxygens has
a strong impact on the in-plane and inter-plane microscopic hopping
parameters. Indeed, $t'/t$ is drastically reduced, in agreement with the
perturbative result Eq.~(\ref{eq.hoppingparam_effect_inp}). In addition,
Eqs.~(\ref{eq.hoppingparam_effect_inp}, \ref{eq.hoppingparam_emer})
reveal that Emery processes to leading order $\sim
-t_{pp}t_{pd}^2/\Delta_{pd}^2$ are in competition with the out-of-plane
processes $\sim t_{p_z}^2t_{pd}^2/\Delta_{pd}^2\Delta_{z}$. These numerical
observations are in good agreement with first-principle calculations by
Pavarini $\textit{et al.}$ \cite{pavar01} showing that the higher Cu-apical
oxygen distance is correlated to the higher $|t'/t|$ as experimentally
supported \cite{yoshid06,yoshid07,peng17}. In a similar way, $t''/t$ is
decreased by increasing $t_{p_z}$, in agreement with
Eq.~(\ref{eq.hoppingparam_effect_inp}), too. Surprisingly, $t'''/t$ is
significantly enhanced when $t_{p_z}$ is increased whereas no apical hopping
processes are seen in the perturbative expansion up to fourth
(Eq.~(\ref{eq.hoppingparam_effect_inp})), and even to fifth order. In fact,
one needs to compute the sixth order to unravel the origin the rising of
$t'''/t$ with apical oxygen couplings and the interesting term is given in
the appendix by Eq.~(\ref{eq.fifth}). 
\begin{figure}[t!]
     \centering
     \includegraphics[width=0.75\columnwidth]{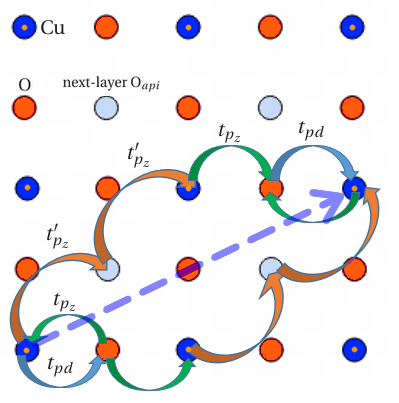}
     \caption{(Color online) Illustration of non-negligible inter-layer
       hopping processes contributing to $t''' \sim
       t_{pd}t_{p_z}t_{p_z}'^2t_{p_z}t_{pd}/\Delta_{pd}^2\Delta_{z}^3$ and
       indicated by dashed arrow. Orange circles denote the in-plane oxygens,
       blue circles denote the copper sites above which the apical oxygens are
       located (orange dots), whereas shaded blue circles denote the apical
       oxygens located below the next-layer copper sites.} 
     \label{fig:ttier_process}
\end{figure}
$\mathcal{\hat{H}}_{\rm eff}^{(6)}$ contains non-negligible hopping processes which are
contributing to $t''' \sim
t_{pd}^2t_{p_z}^2t_{p_z}'^2/\Delta_{pd}^2\Delta_{z}^3$ and are originated from
the inter-layer coupling (see Fig.~\ref{fig:ttier_process}). Besides,
inter-plane hopping integrals are naturally enhanced by increasing
$t_{p_z}$. As seen in Eq.~(\ref{eq.hoppingparam_effect_interp}), inter-plane
hopping processes $\sim t_{pd}^2t_{p_z}^2t_{p_z}'/\Delta_{pd}^2\Delta_{z}^2$
are leading orders for $\theta$ and $\theta''$ with opposite sign as
numerically observed in Table~\ref{tab:tpz_1}. 

\begin{figure}[h!]
\begin{center}
\unitlength=0.22cm
\begin{picture}(38,44)
\put(1,20.5){\includegraphics[width=0.86\columnwidth]{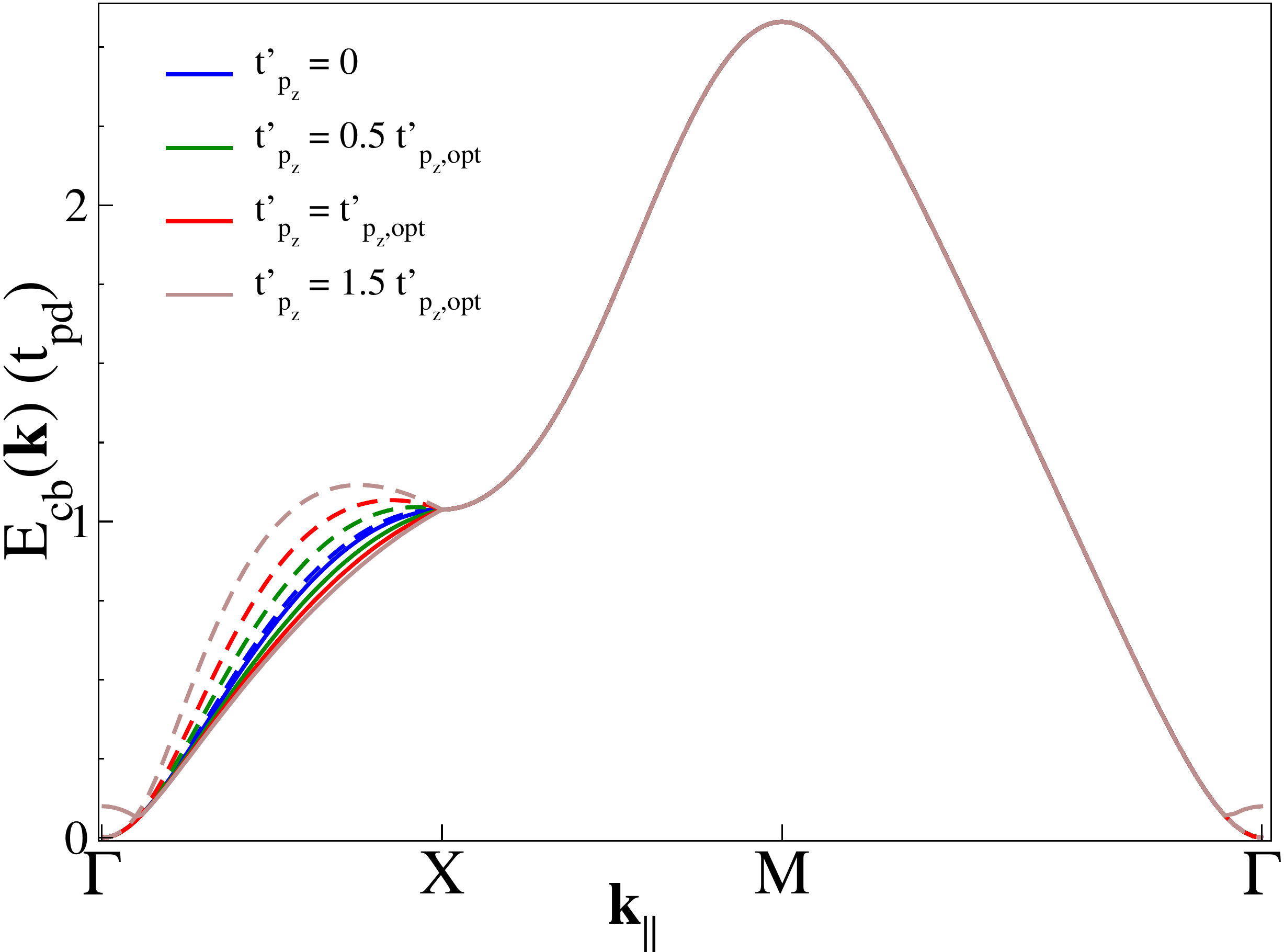}}
\put(0,-3.5){\includegraphics[width=0.875\columnwidth]{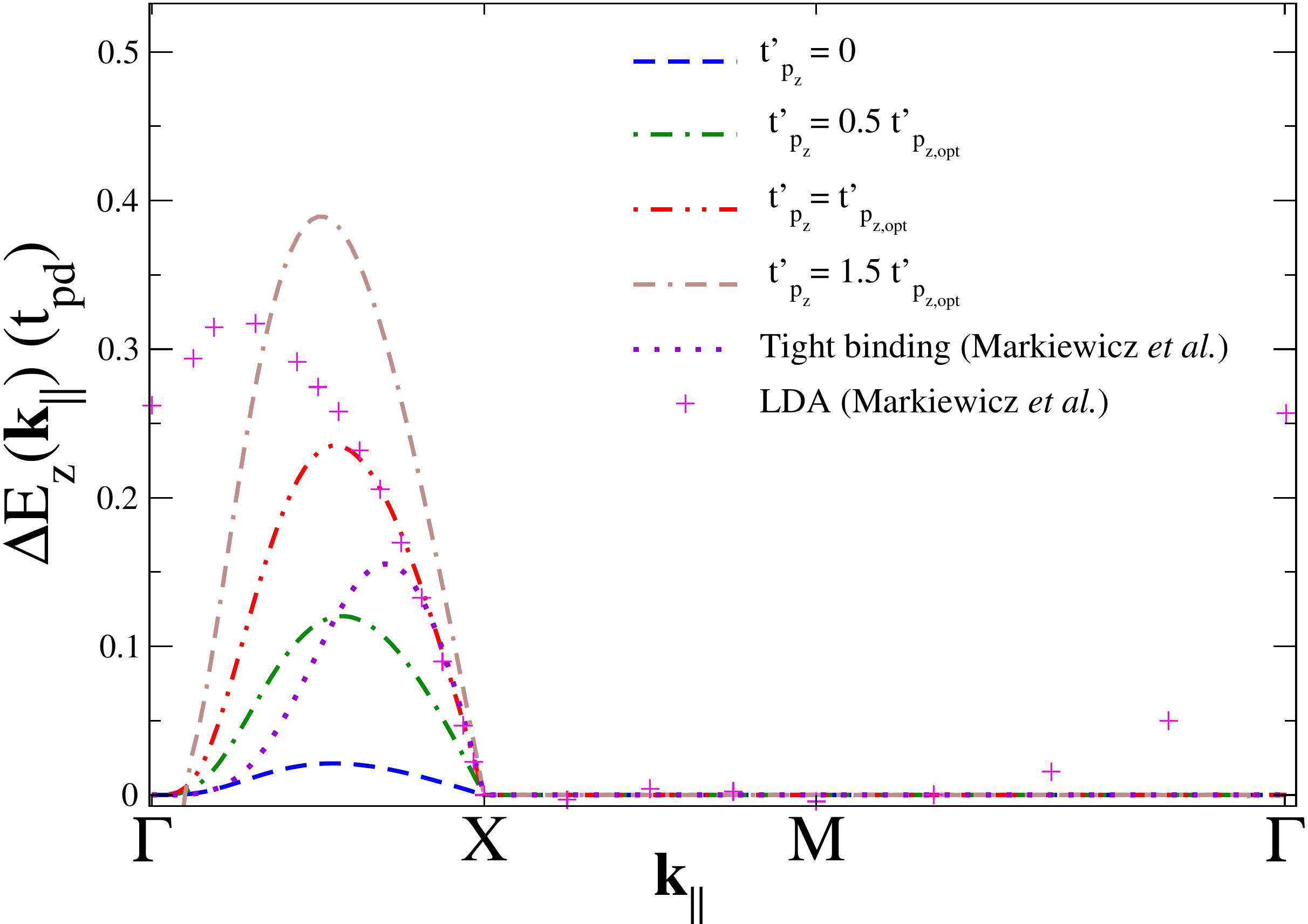}}
\put(30,42){ {\large (a)} } \put(30,17){ {\large (b)} }
\end{picture}
\end{center}
\caption {(Color online) (a) Dispersion of the conduction band along path 1 (solid lines) and path 2 (dashed lines) with $t_{p_z}'$ varied around its optimal value. (b) $\Delta E_z({\bf k}_\parallel)$ with $t_{p_z}'$ varied around its optimal value. The other used tight-binding parameters are given in Table~\ref{tab.paramopt}.}
\label{fig:tpz_pr}
\end{figure}

\begin{table*}
\begin{andptabular}{X[6c]X[5c]X[5c]X[5c]X[5c]X[5c]X[5c]X[5c]X[5c]}%
{Dependence of the main in-plane and inter-plane effective hopping amplitudes on $t_{p_z}'$ expressed in units of its optimal value. The other used tight-binding parameters are given in Table~\ref{tab.paramopt}.}%
$t_{p_z}'/t_{p_z,opt}'$ & $t/t_{pd}$ & $t'/t$ & $t''/t$ & $t'''/t$ &  $t^{(4)}/t$ & $\theta/t$& $\theta'/t$& $\theta''/t$\\
0 & 0.287 & -0.136 & 0.067 & 0.053 & -0.026 & 0.0026 & -0.0007 & -0.0019\\
0.5 & 0.286 & -0.136 & 0.067 & 0.055 & -0.024 & 0.0141 & -0.0039 & -0.0115\\
1 & 0.283 & -0.136 & 0.068 & 0.061 & -0.017 & 0.0285 & -0.0069 & -0.0224\\
1.5 & 0.277 & -0.143 & 0.068 & 0.072 & -0.0007 & 0.0508 & -0.0094 & -0.0364
\label{tab:tpz_pr_1}
\end{andptabular}
\end{table*}

Concerning the hopping integral $t_{p_z}'$ accounting for the coupling between
inter-layer apical oxygens, Fig.~\ref{fig:tpz_pr} shows how it modifies the
dispersion of the conduction band. In agreement with Eq.~(\ref{eq.matrix_f}),
$t_{p_z}'$ does not affect the dispersion along X-M-$\Gamma$ as expected from
perturbation theory Eq.~(\ref{eq.Eznous_marki}). When $t_{p_z}'$ = 0, there is
naturally no inter-plane coupling and the result is similar to the case
$t_{p_z}$ = 0. When $t_{p_z}'$ is increased, the splitting between $k_z$ = 0
and $k_z = 2\pi/c$ emerges along $\Gamma$-X and crosses the optimal
value. Table~\ref{tab:tpz_pr_1} shows that in-plane hopping parameters are
essentially independent of $t_{p_z}'$, yet with the exception of $t'''/t$ (due
to sixth order inter-layer hopping processes Eq.~(\ref{eq.fifth}) illustrated
in Fig.~\ref{fig:ttier_process}). Similarly to the previously examined
$t_{p_z}$ case, $t_{p_z}'$ yields hopping processes which essentially affect
the diagonal inter-plane hopping integrals $\theta$ and $\theta''$, with
opposite sign as numerically observed in Table~\ref{tab:tpz_pr_1}. 

\begin{figure}[t!]
     \centering
     \includegraphics[width=0.95\columnwidth]{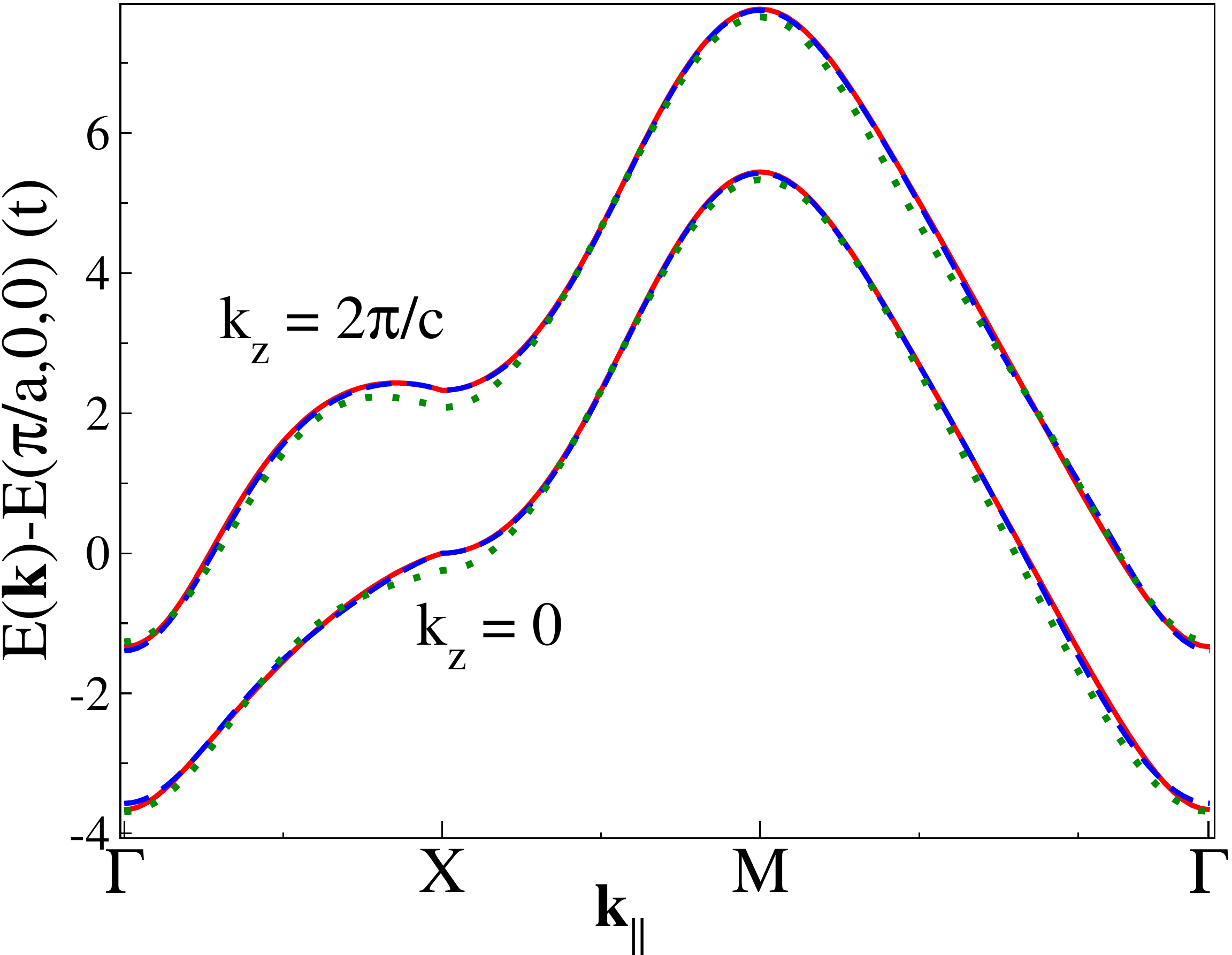}
     \caption{(Color online) The exact dispersion of the conduction band along
       path 1and path 2 (red full lines) is compared to the effective one-band
       dispersion. i) With all hopping amplitudes retained (blue dashed
       lines), and ii) with $t$, $t'$, $t''$, $t'''$, $\theta$, and 
       $\theta''$ retained, only (green dotted lines). The
       used tight-binding parameters are given in
       Table~\ref{tab.paramopt}, together with $t/t_{pd} = 0.283$.}  
     \label{fig:disp_effcomp_ourmod}
\end{figure}

\subsection{Numerical approach: optimal parameters}\label{sec:opt_params}

Having clarified the role of all tight-binding parameters entering
Eq.~(\ref{eq.modelref}) we now set them to their optimal value. After
diagonalizing the Hamiltonian matrix Eq.~(\ref{eq.matrixmodel}), we apply the
Fourier transform of the Cu:3d$_{x^2-y^2}$ band and we obtain the numerical
value of the microscopic hopping parameters which ultimately parameterize the
conduction band through the one-band effective dispersion
Eqs.~(\ref{eq.dispinplane}) and (\ref{eq.disp_interplan}). In-plane hopping
parameters are given by:  $t'/t = -0.1364$, $t''/t = 0.0677$, $t'''/t =
0.0608$, $t^{(4)}/t = -0.0166$, $t^{(5)}/t = -0.0017$, $t^{(6)}/t = 0.0125$,
$t^{(7)}/t = 0.0071$. Inter-plane hopping parameters are given by: $\theta/t =
0.0285$, $\theta'/t = -0.0070$, $\theta''/t = -0.0224$, $\theta'''/t =
0.0068$, $\theta^{(4)}/t = -0.0052$, $\theta^{(5)}/t = -0.0047$ and
t$_{\left(0,0,c\right)}/t = -0.0007$. Fig.~\ref{fig:disp_effcomp_ourmod} shows
the almost perfect fit of the one-band effective dispersion of the conduction
band Eq.~(\ref{eq.disptot}). Simplifying the effective model by retaining the
largest in-plane ($t'/t$, $t''/t$, and $t'''/t$) and out-of-plane ($\theta/t$,
and $\theta''/t$) hopping integrals yields a good fit, too, yet with a small
discrepancy along $\Gamma$-X. This approximation should nevertheless be
sufficient for further purposes. 

\subsection{Fermi surface and density of states}
\begin{figure}[h!]
\begin{center}
\unitlength=0.210cm
\begin{picture}(38,84)
\put(3,55){\includegraphics[width=0.75\columnwidth]{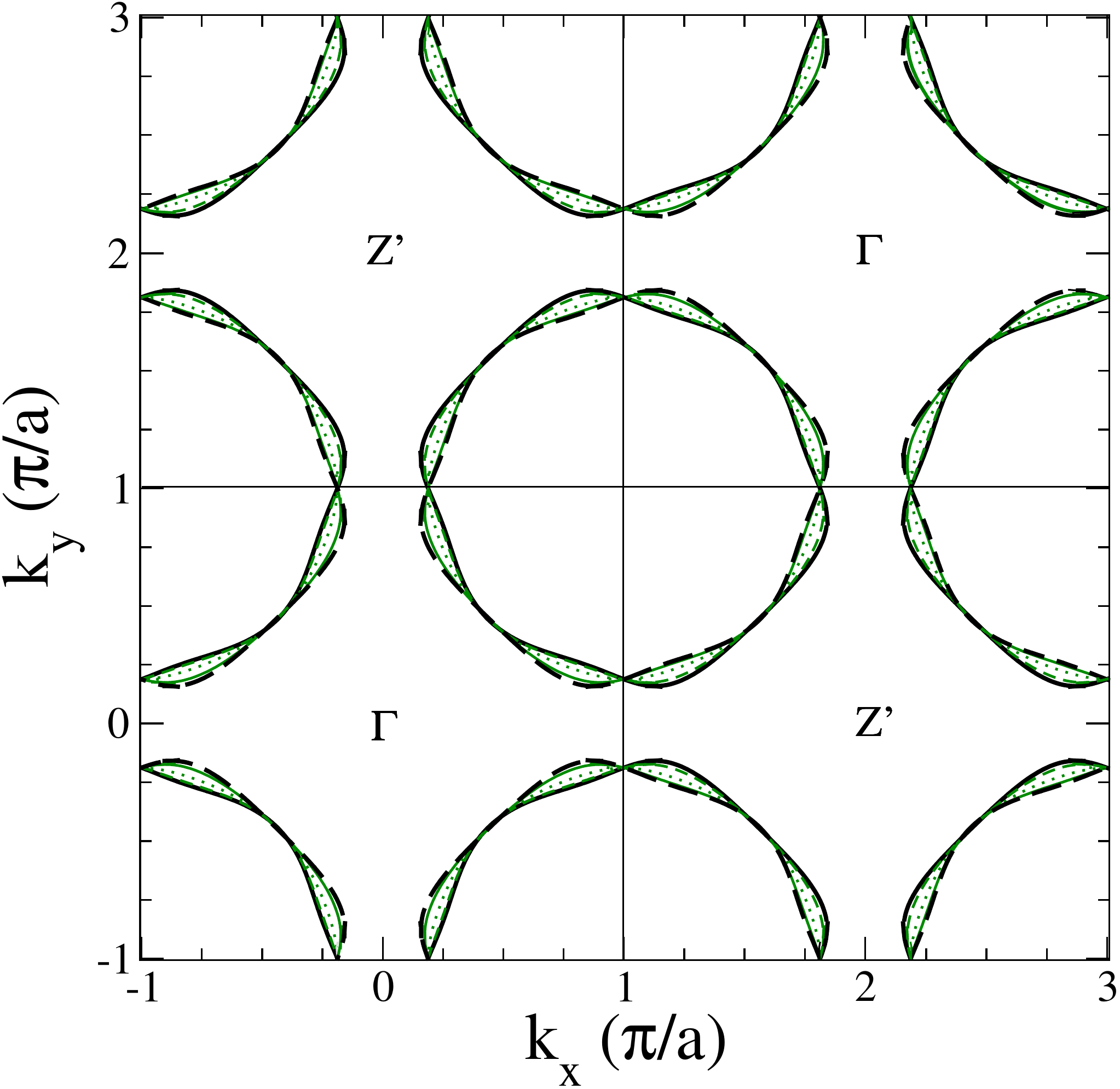}}
\put(3,25.5){\includegraphics[width=0.75\columnwidth]{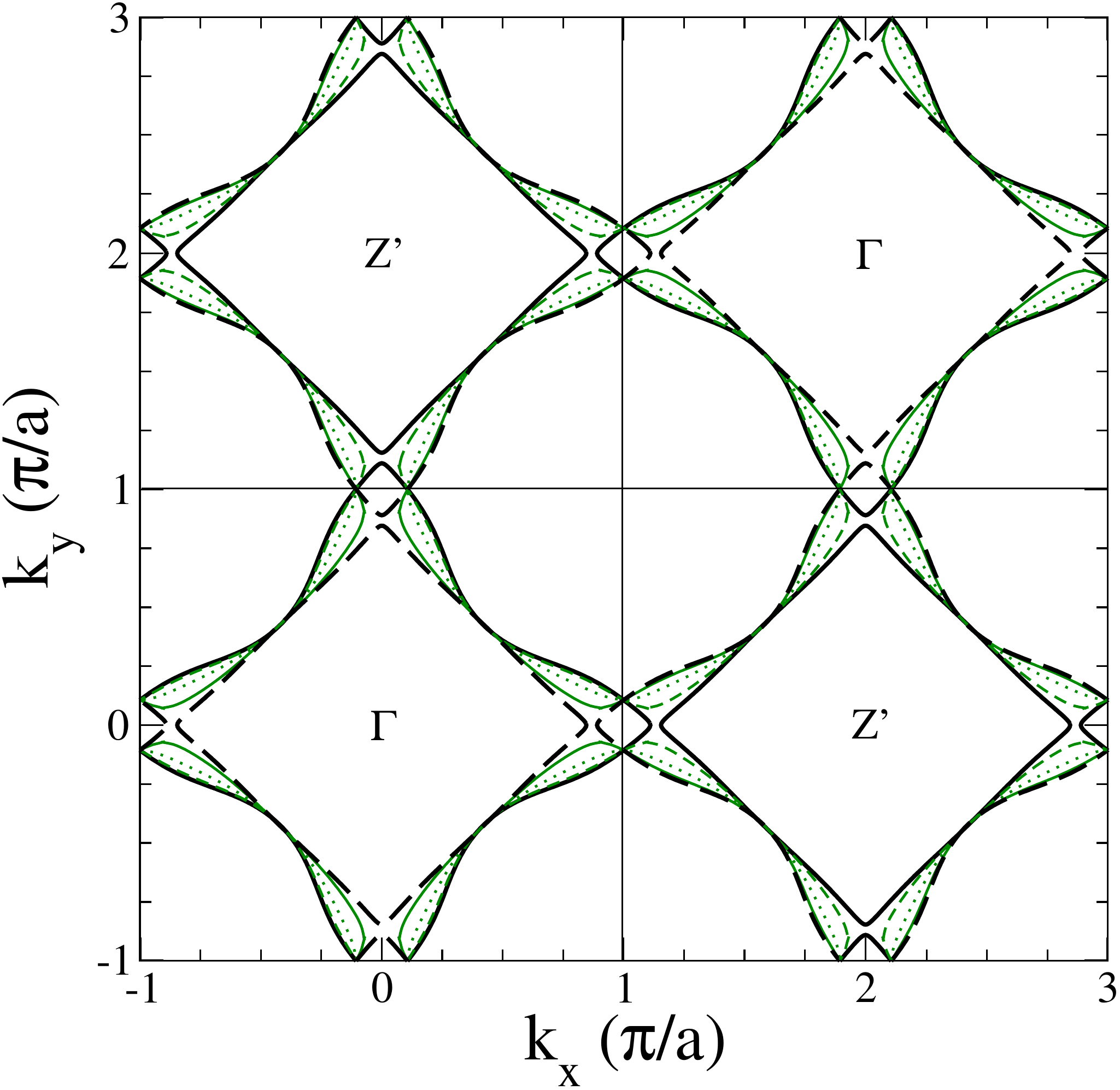}}
\put(3,-4){\includegraphics[width=0.75\columnwidth]{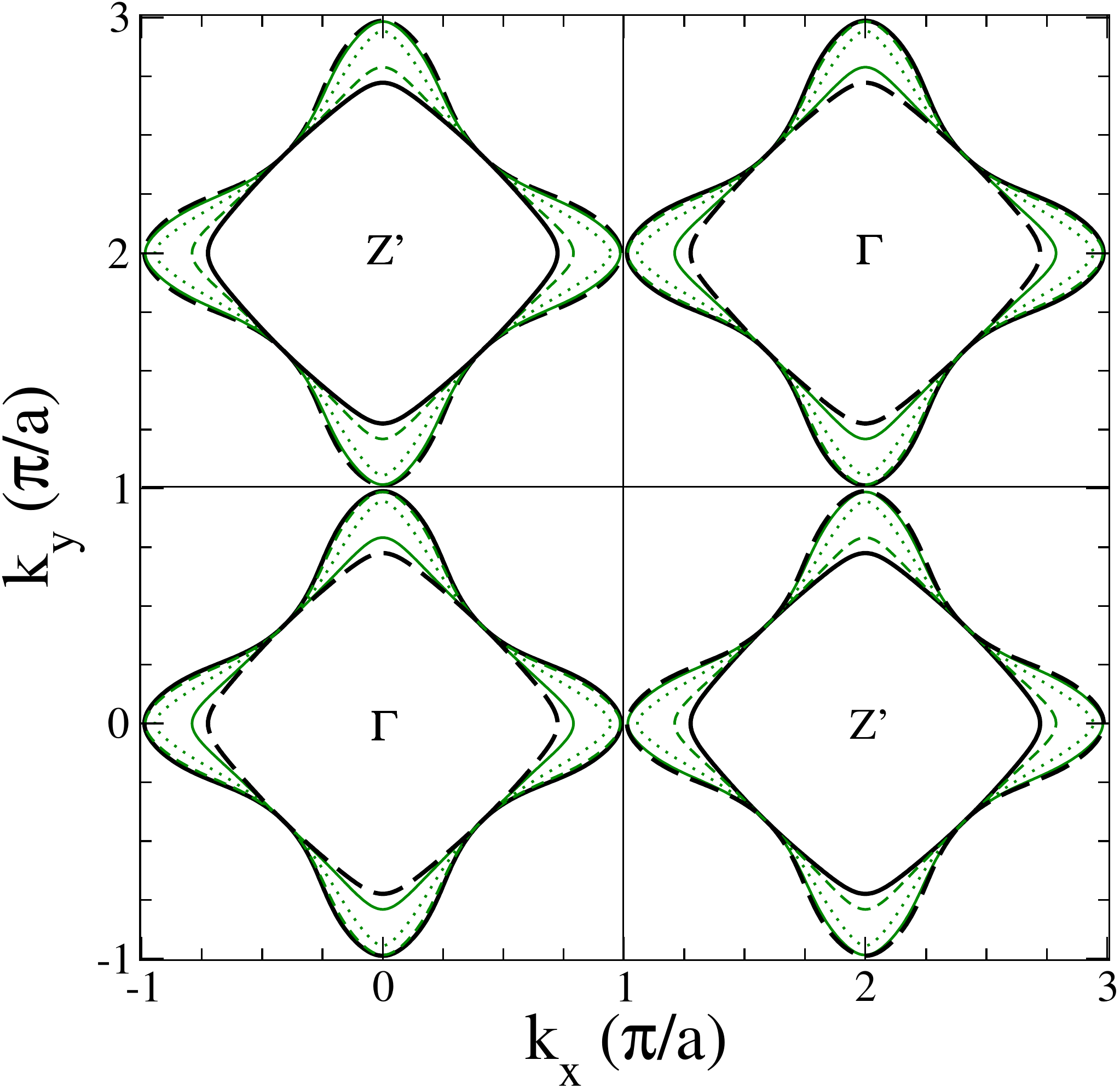}}
\put(30,60){ {\large (a)} } \put(30,30){ {\large (b)} }\put(30,1){ {\large (c)} }
\end{picture}
\end{center}
\caption {(Color online) Projection of the Fermi surfaces on the basal plane
  for several representative values of $k_z$ and electron density. (a) n = 1.0
  (half-filling), (b) n = 0.875 (underdoped), and (c) n = 0.78
  (overdoped). Parameter: $k_z = 0$ (black full line), $k_z = \pi/2c$ (green
  dashed line), $k_z = \pi/c$ (green dotted line), $k_z = 3\pi/2c$ (green full
  line), and $k_z = 2\pi/c$ (black dashed line). The used tight-binding
  parameters are given in Table~\ref{tab.paramopt}.}  
\label{fig:fermisur_kzvar}
\end{figure}
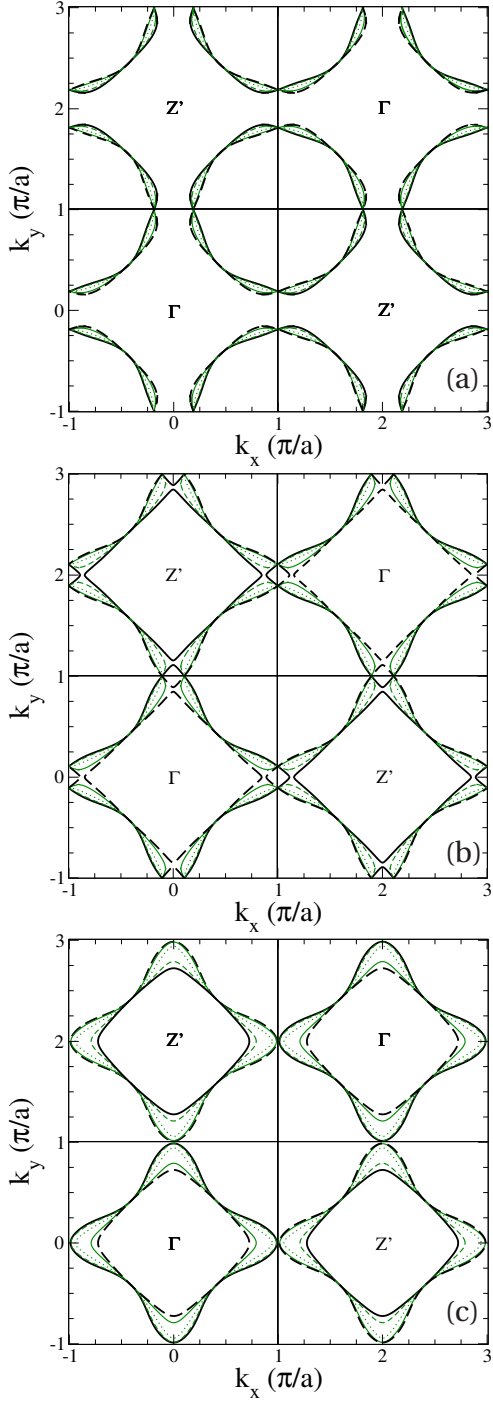

Let us now turn to the Fermi surfaces following from our model. In
Fig.~\ref{fig:fermisur_kzvar} we plot projections of the 3D Fermi surface onto
the ${\bf k}_\parallel$-plane for three important density values: half-filling
(n = 1), an underdoped case (n = 0.875) and an overdoped one (n = 0.78). These
projections are plotted for several positive values of $k_z$ as they depend on
|$k_z$|, only. At half-filling (Fig.~\ref{fig:fermisur_kzvar}(a)), we obtain
hole-like, cylindrical Fermi surfaces, for all values of $k_z$. They are open
in $k_z$-direction and closed in the ${\bf k}_\parallel$ plane. In fact, in
this case, $k_z$ has very little influence on the projected Fermi surface (PFS)
in general, and virtually none for ${\bf k}_\parallel$ along $\Gamma$-M since
in that case, E(${\bf k}$) very weakly depends on $k_z$. We further notice
that the hopping amplitudes beyond $t$ no longer lead to a nested Fermi
surface. 

In the 1/8 hole doped case, Fig.~\ref{fig:fermisur_kzvar}(b), the Fermi
surface retains its closed cylindrical shape. Astonishingly enough, the PFS
around $\Gamma$ (Z') for $k_z = 2\pi/c$ (0) are nested, with a nesting vector
${\bf Q} \simeq (0.84,0.84)\pi/a$. One might then infer that our model is most
prone to develop an incommensurate magnetic instability in this case, which
might result in the formation of a stripe order (a combination of
charge-density-wave and spin-density-wave modulations) that has been reported
in Eu-LSCO \cite{fink11}, Nd-LSCO \cite{tranqua95,tranqua11} and LBCO
\cite{hucker11,tranqua11}. Besides, the influence of $k_z$ on the PFSs is
stronger than at half-filling. Indeed, for $k_z = 0$, the PFSs splits into two
closed pieces: a smaller nested one centered around the Z'-point and a larger
one centered around the $\Gamma$-point that even goes beyond the X and Y
points of the square lattice. This piece has both hole and electron characters,
depending on whether ${\bf k}_\parallel$ is rather on the nodal directions, or
not. For $k_z = 2\pi/c$ the roles of $\Gamma$ and Z' are exchanged. For $k_z$
close to $\pi/c$, the two pieces join and the PFS shows hole-like character,
only. Let us stress that the PFS for $k_z=0$ is in good agreement with 2D
Fermi surface map experimentally obtained via ARPES \cite{yoshid06,yoshid07}
and in LDA for hole doped LSCO cuprate \cite{marki05,lindroo06}. 

The PFSs consist of two pieces for all values of $k_z$. For small values of
$k_z$ a smaller electron-like piece is centered around Z', while the larger
one is centered around the $\Gamma$-point. Again, this piece has both hole and
electron character, depending on the direction of ${\bf k}_\parallel$. For
large values of $k_z$ the roles of $\Gamma$ and Z' are again exchanged. In
fact, such split PFSs, together with the staggering of their size and straight
and rounded shapes have been recently observed in ARPES experiment on
overdoped ($\delta = 22\%$) LSCO \cite{orio18}. Since these features may
neither be accounted for on the square lattice nor on the cubic one, they
originate from the BCT structure. Furthermore it gives rise to peculiar large
momentum-low energy excitations that are absent in simpler models. 

Truly, these PFSs may equally well be obtained using the conduction band
arising from the diagonalization of the Hamiltonian Eq.~(\ref{eq.matrixmodel})
or with its tight-binding form Eqs.~(\ref{eq.dispinplane}) and
(\ref{eq.disp_interplan}), in a broad density range around half-filling. Given
the rather involved form of the full tight-binding model it is tempting to
neglected the smallest hopping amplitudes. This leads to retain nearest
neighbor hopping amplitudes, $t$ and $\theta$, followed by $t'$, $t''$,
$t'''$, $\theta'$, $\theta''$, and $\theta'''$, only (see
Table~\ref{tab:final_1} and Table~\ref{tab:final_2}). This approximation
suffices to reproduce the Fermi surfaces with a high accuracy in all
considered cases. One may then further consider neglecting $\theta'$ and
$\theta'''$ as well, since the so obtained conduction band is in very good
agreement with the exact one (see Fig.~\ref{fig:disp_effcomp_ourmod}).  
At half-filling, the resulting PFSs are in excellent agreement with the
exact ones. Yet, already from $\delta = 1/8$ on, the agreement degrades and,
e.~g., the (in-plane) PFSs no longer close around $\Gamma$ and $Z'$, but around
$M$. Yet, it only takes a small increase in doping to recover this feature. 
Finally, further simplifications such as neglecting $\theta''$ as well, result
into an even poorer account of the conduction band. This may hardly come as a
surprise given the above discussed slow convergence of the perturbative
calculation of the effective model, which itself follows from the relatively
small value of the charge transfer gap, and which unavoidably generates a
larger number of non-negligible hopping amplitudes. 

\begin{figure}[t!]
     \centering
     \includegraphics[width=0.95\columnwidth]{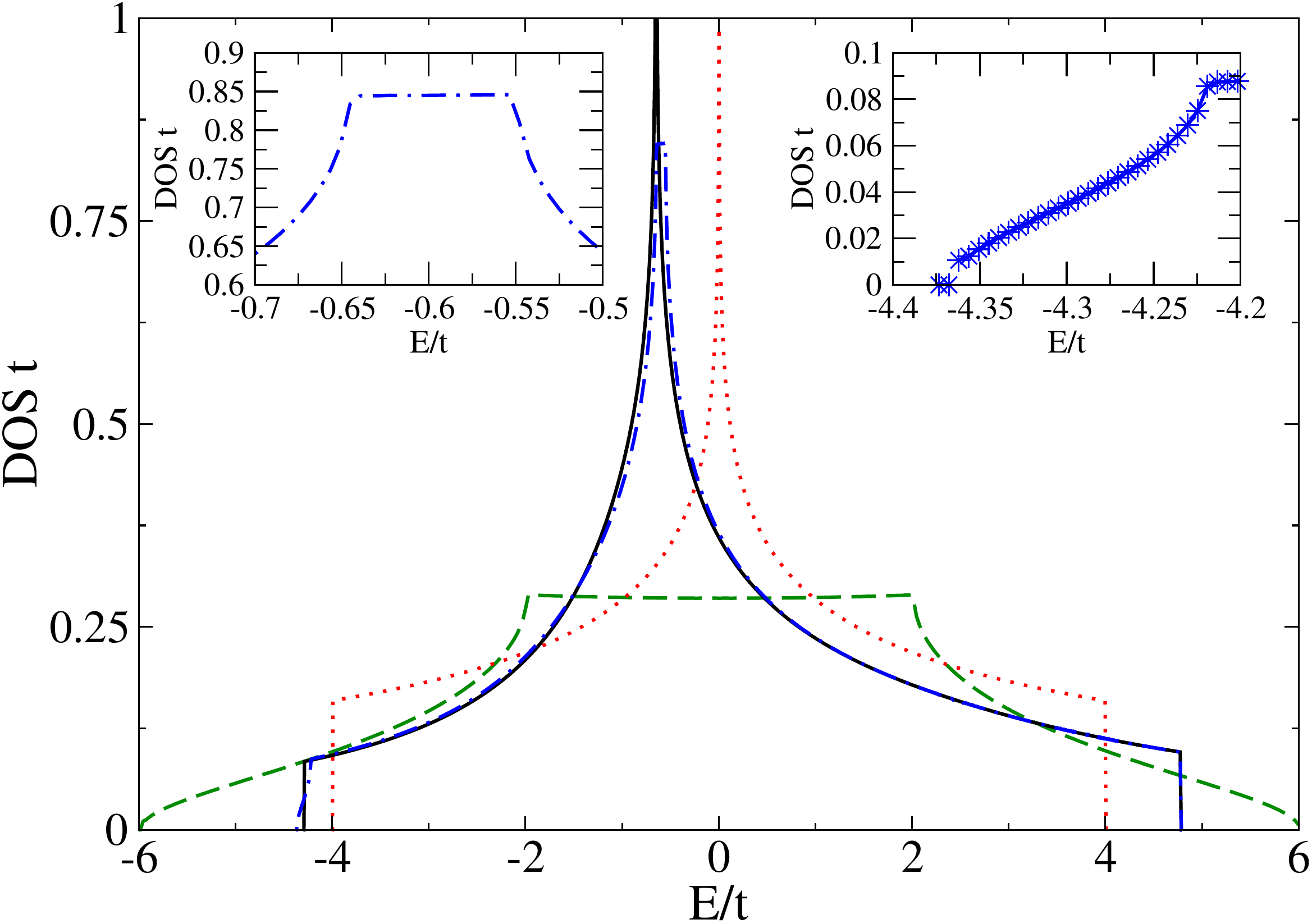}
     \caption{(Color online) The density of states of: our effective 3D model
       Eq.~(\ref{eq.disptot}) (blue dashed-dotted line), the corresponding 2D
       model (black solid line), the tight-binding model to nearest neighbor
       only on the square (dotted red line) and cubic (green dashed line)
       lattices. Right inset: the Van Hove singularity of our model close to
       the bottom of the band. Left inset: the plateau exhibited by our model
       and surrounded by two Van Hove singularities in the vicinity of the
       middle of the band.} 
     \label{fig:dos}
\end{figure}

The density of states of our model is shown in Fig.~\ref{fig:dos}, where it is
compared to the simpler model where inter-plane hopping is set to
zero. Comparison to nearest neighbor only hopping on both the square and the
cubic lattices is performed, too. Interestingly, our three-dimensional model
inherits features from both cases. Indeed, the density of states at the bottom
of the band and at the top of the band is finite, as in the 2D case. Yet, it
now exhibits a plateau surrounded by two Van Hove singularities in the
vicinity of the middle of the band as for the cubic lattice. However, the
plateau centered around $-0.6 t$ is of considerably narrower width, as a
consequence of the anisotropy of the model. Expressed in terms of the charge
carrier density, the plateau extends from (1-$\delta$) = 0.81 to (1-$\delta$)
= 0.89. Furthermore, one notices an additional Van Hove singularity for small
density; it does not exhibit any counterpart for large density, as is expected
for a particle-hole symmetric model. Finally, the 3D case remains strikingly
close to the 2D one. This behavior finds its explanation when expanding
Eq.~(\ref{eq.dispinplane}) around the bottom of the band that is located at
the $\Gamma$-point. One finds $E_{cb}({\bf k}) = \epsilon_0 +
\epsilon_1(k_x^2+k_y^2)$ where the $k_z$-dependence is missing. A similar
behavior is found in the vicinity of the top of the band. The lack of
$k_z$-dependence above the symmetry lines X-M-$\Gamma$ contributes to this
feature, too.

\section{Summary and conclusion}\label{sec:conclusion}

Summarizing, we have re-analyzed the conduction band relevant to single-layer
La-based superconducting cup\-rates in the tight-binding framework. We have
shown that it naturally emerges from an eight-band model involving two copper
orbitals and six oxygen orbitals. As a consequence of their strong mutual
hybridization, longer-ranged tight-binding parameters were taken into
account. In particular, we obtained that the retained apical oxygen orbitals
are not only crucial to the dispersion of the band perpendicular to the basal
plane, but also significantly renormalize the in-plane dispersion as compared
to the Emery model. This allowed us to accurately reproduce the DFT results of
Markiewicz $\textit{et al}$.\cite{marki05}, as well as to shed light on the
peculiar $k_z$-dependence of the conduction band. 

We then proceeded to the determination of the parameters entering the
tight-binding Hamiltonian characterizing the conduction band. We first applied
the Rayleigh-Schr\"odinger perturbation theory in order to unravel the
microscopic processes contained in the eight-band model that determine the
various hopping parameters entering the effective low-energy model. We have
shown that the model extends the Emery model on two aspects since, on one
hand, accounting for apical oxygens significantly affects the in-plane hopping
amplitudes (see Table~\ref{tab:final_1}), and, on the other hand, it gives
birth to out-of-plane dispersion. It turns out that perturbation theory to
fifth order is mandatory to obtain non-vanishing layer to layer hopping that
we found to be primarily governed by apical oxygen orbitals. This extends the
3D model proposed by Markiewicz $\textit{et al}$.\cite{marki05} (see
Eq.~(\ref{eq.Eznous_marki})). Yet, because of the relatively small value of the
charge transfer gap, the perturbative approach lacks accuracy. 
 
We then overcame this difficulty by directly computing the hopping parameters
through the Fourier transform of the numerically obtained conduction
band Eq.~(\ref{eq:43}). This leads to longer ranged hopping amplitudes that slowly decay with
distance, and one needs to take many of them into account in order to
accurately reproduce the conduction band and the Fermi surfaces. As revealed
by Table~\ref{tab:final_1} the so obtained main in-plane hopping amplitudes
are in very 
good agreement with the ones used to fit the Fermi surfaces obtained with
ARPES experiments on La-based cuprates \cite{yoshid06,yoshid07}. In
particular, we found $t'/t\simeq-0.1$ and $t''/t'\simeq-0.5$ and we obtained,
in addition, $(|t'|+|t''|)/t\simeq 0.2$ as reported for overdoped
La$_{1.78}$Sr$_{0.22}$CuO$_4$\cite{yoshid06,yoshid07,chang13}. Furthermore, we
found $t''' \simeq t''$, which was empirically used to model ARPES data or DFT
calculations \cite{marki05,norman07}. Regarding inter-layer couplings (see
Table~\ref{tab:final_2}), we found the magnitude of $\theta'$ to be smaller
than the one of $\theta$ and $\theta''$. Furthermore, the relation between
$\theta$, $\theta'$, $\theta''$, and $\theta'''$ assumed in
Eq.~(\ref{eq.marki_ez})\cite{marki05} could not be recovered, so that the
leading contribution to the dispersion perpendicular to the layers is indeed
given by Eq.~(\ref{eq.Eznous_marki}). In the doped case our model yields
peculiar Fermi surfaces which projections on the basal plane alternate in
size and shape. Since this may not arise in tight-binding models on square or
cubic lattices this may be seen as a signature of the body-centered tetragonal
structure that is at the heart of this work. It is compatible with recent
ARPES measurements of the 3D Fermi surface of overdoped LSCO
\cite{orio18}. Finally, it would be of interest to carry out the corresponding
analysis to other cuprate families, or to highly anisotropic oxides such as
PdCoO$_2$ \cite{hicks12,daou15}, which structure entails shifted layers such
as Bi-2212 \cite{marki05} or Tl-2201 \cite{pickett92,peets07}. We also
unraveled that simplifying Eqs.~(\ref{eq.dispinplane}) and
(\ref{eq.disp_interplan}) down to a model that only entails the in-plane $t$,
$t'$, $t''$, and $t'''$ as well as the out-of-plane $\theta$, and $\theta''$
hopping amplitudes (given in Tables~\ref{tab:final_1} and \ref{tab:final_2})
yields a reasonable description of the conduction band and should therefore
suffice for future purposes. For instance, it would be of interest to
understand how the found peculiar form of nesting and interaction generate
magnetic or pairing instabilities within a Hubbard model on the BCT lattice.
Work along this line is in progress.

{\it Acknowledgments.}\, 
The authors acknowledge the financial support of the Minist\`ere de la
Recherche, the R\'egion Normandie and the French Agence Nationale de la
Recherche (ANR), through the program Investissements d'Avenir
(ANR-10-LABX-09-01) and LabEx EMC3. The authors gratefully acknowledge
G. A. Sawatsky, A. F. Santander-Syro, A. M. Ole\'s, V. H. Dao and O. Juillet
for inspiring discussions. 

\appendix
\section{Lindgren's formulation of Rayleigh-Schr\"odinger perturbation theory}
The above Eqs.~(\ref{eq.lind1}, \ref{eq.lind2}, \ref{eq.lind3},
\ref{eq.interplan_lind}) have been derived using the Rayleigh-Schr\"odinger 
perturbation theory. More specifically, we made use of Lindgren's formulation
\cite{lindgren74} to set up an effective tight-binding Hamiltonian and to put
forward the microscopic origin of the various hopping amplitudes. Below, we
summarize the main steps leading to an effective Hamiltonian at a given order
in perturbation theory. 
The Hamiltonian $\hat{H}$ Eqs.~(\ref{eq:modeltot},
  \ref{eq.matrixmodel}) is separated into its diagonal part
  that plays the role of the unperturbed Hamiltonian, $ \sum_{{\bf k},\sigma}
  \mathcal{\hat{H}}_{0,\sigma}({\bf k}) \equiv \sum_{{\bf k},\sigma} \sum_{\nu} 
\epsilon_\nu({\bf k}) \hat{\Psi}_{{\bf k},\sigma,\nu}^\dagger \hat{\Psi}_{{\bf k},\sigma, \nu}^{\phantom{\dagger}}$, and its off-diagonal part this is
  considered as the perturbation 
$ \sum_{{\bf k},\sigma} \mathcal{\hat{H}}_{1,\sigma}({\bf k})$
\begin{equation}
\mathcal{\hat{H}} = \sum_{{\bf k},\sigma}\left(\mathcal{\hat{H}}_{0,\sigma}({\bf k}) +
  \mathcal{\hat{H}}_{1,\sigma}({\bf k})\right)\:. 
\end{equation} 
Usually, perturbation theory is applied to problems lacking
  an exact solution. In our case the latter has been found, but in numerical
  form, only, and our goal is not to recover it. Instead, it is to shed light
  on both how the perturbation generates 
  the dispersion of the conduction band, and on the microscopical origin of
  the various hopping amplitudes entering the effective one-band model.
In Lindgren's approach \cite{lindgren74}, the effective Hamiltonian acting on
the low energy sector of the Hilbert space, here spanned by $|d_{{\bf
    k},\sigma}\rangle$, is expressed in terms of a wave operator
$\hat{\Omega}({\bf k})$ \cite{moller45} as: 
\begin{equation}
\begin{split}
\mathcal{\hat{H}}_{\rm eff}^{(i)} = \sum_{{\bf k}}|d_{\bf k}\rangle\langle d_{\bf k}|\mathcal{\hat{H}}_1({\bf k})\left(\hat{\Omega}^{(0)}({\bf k}) + \hat{\Omega}^{(1)}({\bf k}) + \hat{\Omega}^{(2)}({\bf k})+...\right.\\
+\left.\hat{\Omega}^{(i-1)}({\bf k})\right)\:\:,
\end{split}
\label{eq.append_heff}
\end{equation}
where the dull index $\sigma$ is omitted for simplicity. Having measured all energies relative to $\epsilon_d$ forces the zeroth order contribution to vanish. To zeroth order the wave operator is simply the projection operator onto the low-energy sector of our model:
\begin{equation}
\hat{\Omega}^{(0)}({\bf k}) =  |d_{\bf k}\rangle\langle d_{\bf k}|\:.
\end{equation}
Starting from Schr\"odinger's equation, Lindgren obtained a recursion formula for $\hat{\Omega}^{(l)}({\bf k})$:
\begin{equation}
\begin{aligned}
[\hat{\Omega}^{(l)}({\bf k}),\mathcal{\hat{H}}_0({\bf k})] =& (1-|d_{\bf k}\rangle\langle d_{\bf k}|)\mathcal{\hat{H}}_1({\bf k})\hat{\Omega}^{(l-1)}({\bf k})\\
 &- \sum_{m=1}^{l-1}\hat{\Omega}^{(l-m)}({\bf k})\mathcal{\hat{H}}_1({\bf k})\hat{\Omega}^{(m-1)}({\bf k})\:.
\end{aligned}
\end{equation}
The lowest orders are then explicitly obtained as: 
\begin{equation}
\hat{\Omega}^{(1)}({\bf k}) = \sum_{\nu}|\nu_{\bf k}\rangle\langle d_{\bf k} |\frac{\langle \nu_{\bf k}|\mathcal{\hat{H}}_1({\bf k})|d_{\bf k}\rangle}{\epsilon_d-\epsilon_\nu({\bf k})}
\end{equation}
\begin{equation}
\hat{\Omega}^{(2)}({\bf k}) = \sum_{\nu}|\nu_{\bf k}\rangle\langle d_{\bf k} |\frac{\langle \nu_{\bf k}|\mathcal{\hat{H}}_1({\bf k})\hat{\Omega}^{(1)}({\bf k})-\hat{\Omega}^{(1)}({\bf k})\mathcal{\hat{H}}_1({\bf k})|d_{\bf k}\rangle}{\epsilon_d-\epsilon_\nu({\bf k})}
\end{equation}
\begin{equation}
\begin{aligned}
&\hat{\Omega}^{(3)}({\bf k}) = \sum_{\nu}|\nu_{\bf k}\rangle\langle d_{\bf k}|\times\\
&\frac{\langle \nu_{\bf k}|\mathcal{\hat{H}}_1({\bf k})\hat{\Omega}^{(2)}({\bf k})
\!-\!\hat{\Omega}^{(1)}({\bf k})\mathcal{\hat{H}}_1({\bf k})\hat{\Omega}^{(1)}({\bf k})
\!-\!\hat{\Omega}^{(2)}({\bf k})\mathcal{\hat{H}}_1({\bf k})|d_{\bf k}\rangle}{\epsilon_d-\epsilon_\nu({\bf k})}
\end{aligned}
\end{equation}
\begin{equation}
\begin{aligned}
&\hat{\Omega}^{(4)}({\bf k}) = \sum_{\nu}|\nu_{\bf k}\rangle\langle d_{\bf k}|\\
&\times\frac{\langle \nu_{\bf k}|\left[\mathcal{\hat{H}}_1({\bf k})\hat{\Omega}^{(3)}({\bf k})-\hat{\Omega}^{(1)}({\bf k})\mathcal{\hat{H}}_1({\bf k})\hat{\Omega}^{(2)}({\bf k})\right.}{\epsilon_d-\epsilon_\nu({\bf k})}\\
&-\frac{\left.\hat{\Omega}^{(2)}({\bf k})\mathcal{\hat{H}}_1({\bf k})\hat{\Omega}^{(1)}({\bf k})+\hat{\Omega}^{(3)}({\bf k})\mathcal{\hat{H}}_1({\bf k})\right]|d_{\bf k}\rangle}{\epsilon_d-\epsilon_\nu({\bf k})}\,,
\end{aligned}
\end{equation}
where we have kept (the here vanishing) $\epsilon_d$, for clarity. Hence
\begin{equation}
\begin{aligned}
\mathcal{\hat{H}}_{\rm eff}^{(4)} &= \sum_{{\bf k}}|d_{\bf k}\rangle\langle d_{\bf
  k}|\left[\epsilon_d+\sum_{\nu}\frac{|h_{\nu}({\bf
      k})|^2}{\epsilon_d-\epsilon_\nu({\bf k})} \right.\\
&+ \sum_{\nu,\nu'}
\frac{h_\nu({\bf k}) g_{\nu,\nu'}({\bf k}) h_{\nu'}({\bf k})}
{(\epsilon_d - \epsilon_\nu({\bf k}))
(\epsilon_d - \epsilon_{\nu'}({\bf k}))}\\
&+\sum_{\nu,\nu',\nu''}\frac{h_{\nu}({\bf k})g_{\nu,\nu'}({\bf k})
g_{\nu',\nu''}({\bf k})h_{\nu''}^*({\bf k})}{(\epsilon_d - \epsilon_\nu({\bf
  k}))
(\epsilon_d - \epsilon_{\nu'}({\bf k}))(\epsilon_d-\epsilon_{\nu''}({\bf k}))}\\
 &- \left.\sum_{\nu,\nu'}\frac{|h_\nu({\bf k})|^2|h_{\nu'}({\bf k})|^2}{(\epsilon_d-\epsilon_{\nu'}({\bf k}))^2(\epsilon_d-\epsilon_{\nu'}({\bf k}))} \right]\,,
\end{aligned}
\label{eq.heff}
\end{equation}
with
\begin{equation}
\begin{aligned}
\mathcal{\hat{H}}_{1,d,\nu} &= \sum_{{\bf k}}t_{pd}\left(h_{3}({\bf k})|d_{\bf k}\rangle\langle p_{x,{\bf k}}^{(X)}| + h_{4}({\bf k})|d_{\bf k}\rangle\langle p_{y,{\bf k}}^{(Y)}|\right)\\
\mathcal{\hat{H}}_{1,\nu,\nu'} &= \sum_{{\bf k}}\sum_{\nu,\nu'}g_{\nu,\nu'}({\bf k})|\nu_{\bf k}\rangle\langle \nu_{\bf k}'|\:,
\end{aligned}
\end{equation}
where $h_{3}({\bf k}) = 2it_{pd}p_x$ and $h_{4}({\bf k}) = -2it_{pd}p_y$. The
matrix elements $\langle\nu_{\bf k}|\mathcal{\hat{H}}_1({\bf k})|\nu'_{\bf k} \rangle$ are denoted by $g_{\nu,\nu'}({\bf k})$.
Regarding the inter-layer coupling given by Eq.~(\ref{eq.matrix_f}), it turns out to appear at fifth order, from the contributions to $\mathcal{\hat{H}}_{\rm eff}^{(5)}$ reading:
\begin{equation}
\sum_{\nu,\nu',\nu'',\nu'''}\frac{h_{\nu}({\bf k})g_{\nu,\nu'}({\bf k})g_{\nu',\nu''}({\bf k})g_{\nu'',\nu'''}({\bf k})h_{\nu'''}^*({\bf k})}{(\epsilon_d-\epsilon_\nu({\bf k}))(\epsilon_d-\epsilon_{\nu'}({\bf k}))(\epsilon_d-\epsilon_{\nu''}({\bf k}))(\epsilon_d-\epsilon_{\nu'''}({\bf k}))}\:.
\label{eq.lindorder4}
\end{equation}
Besides, inter-layer couplings may significantly contribute to in-plane hopping amplitudes. This is especially relevant to $t'''$, from contribution to $\mathcal{\hat{H}}_{\rm eff}^{(6)}$ of the form:
\begin{equation}
\begin{aligned}
\sum_{\nu,\nu',\nu''} &
\frac{h_{\nu}({\bf k})g_{\nu,\nu'}({\bf k})g_{\nu',\nu''}({\bf k})}
{(\epsilon_d-\epsilon_\nu({\bf k}))(\epsilon_d-\epsilon_{\nu'}({\bf k}))
(\epsilon_d-\epsilon_{\nu''}({\bf k}))} \\
\times \sum_{\nu''',\nu''''} &
\frac{g_{\nu'',\nu'''}({\bf k})g_{\nu''',\nu''''}({\bf k})h_{\nu''''}^*({\bf k})}
{(\epsilon_d-\epsilon_{\nu'''}({\bf k}))(\epsilon_d-\epsilon_{\nu''''}({\bf k}))}
\,.
\end{aligned}
\label{eq.fifth}
\end{equation}

\end{document}